\documentclass[fleqn,usenatbib]{mnras}

\usepackage{newtxtext,newtxmath}
\usepackage[T1]{fontenc}
\usepackage{ae,aecompl}

\usepackage{graphicx}	% Including figure files
\usepackage{amsmath}	% Advanced maths commands
\usepackage{amssymb}	% Extra maths symbols
\usepackage{breakurl}
\usepackage{gensymb}    % For the degree symbol

\newcommand{\etal}{{et al}\/.}
\newcommand{\casa}{\textsc{casa}\,}
\newcommand{\casaviewer}{\textsc{casaviewer}\,}
\newcommand{\brats}{\textsc{brats}\,}
\newcommand{\python}{\textsc{python}\,}

\title[Unveiling the cause of HyMoRS]{Unveiling the cause of hybrid morphology radio sources (HyMoRS)}

\author[J.J.~Harwood \etal]{Jeremy J.\ Harwood$^{1}$\thanks{E-mail: jeremy.harwood@physics.org}, Tessa Vernstrom$^{2}$ and Andra Stroe$^{3}$\thanks{Clay Fellow}
\\$^{1}$Centre for Astrophysics Research, School of Physics, Astronomy and Mathematics, University of Hertfordshire, College Lane,\\ Hatfield, Hertfordshire AL10 9AB, UK
\\$^{2}$CSIRO Astronomy $\&$ Space Science, Kensington, Perth 6151, Australia
\\$^{3}$Center for Astrophysics | Harvard \& Smithsonian, 60 Garden St, Cambridge, MA 02138, USA
}

\date{Accepted XXX. Received YYY; in original form ZZZ}

\pubyear{2019}

\begin{document}
\label{firstpage}
\pagerange{\pageref{firstpage}--\pageref{lastpage}}
\maketitle

\graphicspath{{./images/}}

\begin{abstract}

Hybrid morphology radio sources (HyMoRS) are a rare group of radio galaxies in which differing Fanaroff \& Riley morphologies (FR I/II) are observed for each of the two lobes. While they potentially provide insights into the formation of lobe structure, particle acceleration, and the FR dichotomy, previous work on HyMoRS has mainly been limited to low-resolution studies, searches for new candidates, and milliarcsecond-scale VLBI observations of the core region.
In this paper, we use new multi-array configuration Very Large Array (VLA) observations between 1 and 8 GHz to determine the morphology of HyMoRS on arcsecond scales and perform the first well-resolved spectral study of these unusual sources. We find that while the apparent FR I lobe is centre-brightened, this is the result of a compact acceleration region resembling a hotspot with a spectrum more consistent with an FR II (``strong-flavour'') jet. We find that the spectra of the apparent FR I lobes are not similar to their classical counterparts and are likely the result of line-of-sight mixing of plasma across a range of spectral ages. We consider possible mechanisms that could lead to the formation of HyMoRS under such conditions, including environment asymmetry and restarted sources, concluding through the use of simple modelling that HyMoRS are the result of orientation effects on intrinsically FR II sources with lobes non-parallel to the inner jet.
\end{abstract}


\begin{table*}
\centering
\caption{List of target sources, properties, and observations}
\label{targets}
\begin{tabular}{cccccccccccccc}
\hline
\hline
Source&R.A.&Dec.&$z$&Ref.&$S_{1.4}$&$S_{4.9}$&LAS&\multicolumn{3}{c}{L-Band ToS (mins)}&\multicolumn{3}{c}{C-Band ToS (mins)}\\
Name&(J2000)&(J2000)&&&(mJy)&(mJy)&(arcsec)&A&B&Total&B&C&Total\\
\hline
J1154$+$513&11h 53m 46.43s&+51d 17m 04.1s&0.31&1&495&137&44.70&90&60&150&90&60&150\\
J1206$+$503&12h 06m 22.39s&+50d 17m 44.3s&1.13&2&241&75&51.54&90&60&150&90&60&150\\
J1313$+$507&13h 13m 25.78s&+50d 42m 06.2s&0.56&1&277&84&58.61&90&60&150&90&60&150\\
J1315$+$516&13h 14m 38.12s&+51d 34m 13.4s&0.63&1&93&51&77.28&90&60&150&90&60&150\\
J1348$+$286&13h 47m 51.58s&+28d 36m 29.6s&0.74&3&241&117&47.66&90&60&150&90&60&150\\
\hline
\end{tabular}
\vskip 5pt
\begin{minipage}{17.8cm}
\small `Source Name' are the IAU names of the galaxies discussed in this paper. `R.A.' and `Dec.' columns list position in J2000 coordinates, taken from \citet{gawronski06}, `Redshift' provides the photometric redshift to the source as given by the `Ref.' column where: (1) \citet{york2000}; (2) \citet{richards2009}; (3) \citet{munoz2013}. `LAS' is the angular size of the source along its major axis, measured from the Faint Images of the Radio Sky at Twenty-centimetres (FIRST, \citealp{white97}) survey. 'L(C)-Band ToS' provides the time on source for each band and configuration in minutes.
\end{minipage}
\end{table*}

%\textcolor{red}{\textbf{Dummy line so latex will compile (reference cuts over 2 pages)}}

\section{Introduction}
\label{intro}

\subsection{Radio galaxy morphologies}
\label{rgintro}

Powerful radio galaxies can broadly be split into 2 morphological classes: \citet{fanaroff74} class I (FR I) and II (FR II) which were defined according to ``the ratio of the distance between the regions of highest brightness on opposite sides of the central galaxy or quasar, to the total extent of the source''. FR Is are characterised by their twin jet morphology, diffuse plumes, and have a surface brightness dominated by the inner jet region (ratio $< 0.5$; 'centre brightened'). These jets are thought to initially have highly relativistic bulk flow speeds that then decelerate on scales of tens of kpc to around $\lesssim 0.2c$ which is thought to be the primary cause of particle acceleration (e.g. \citealp{laing99}). In contrast, the jets of FR IIs are thought to be highly relativistic out to large distances (up to Mpc scales) and so are typically either not observed or are one-sided. These jets terminate in a shock, usually located at the extremities of the source, forming a compact region of emission known as a hotspot, which dominates the surface brightness (ratio $> 0.5$; `edge brightened'). As these hotspots advance through the external medium the shock accelerated plasma is left behind (possibly with some back flow) to form their characteristic radio lobes.

Although the FR I/II classification of radio galaxies is morphological, a break in radio luminosity is also observed (e.g. \citealp{owen94}). FR Is are generally the lower powered of the two classes having 178 MHz luminosity densities below $2 \times 10^{25}$ W Hz$^{-1}$ sr$^{-1}$ \citep{fanaroff74} with FR IIs residing above this threshold. Although recent studies have called this luminosity divide into question \citep{mingo19} there is strong evidence of differing evolution between the two classes (e.g. \citealp{wall80}), particularly with respect to the jet structure, particle acceleration, and particle content on large scales (e.g. \citealp{scheuer74, laing02, croston18}). While a dichotomy therefore still exists between the two radio galaxy types the underlying cause is not clear and has been the subject of much debate (e.g. \citealp{sheuer96, kaiser97}). Three solutions are generally favoured for explaining this dichotomy: 1) The FR I morphology is due to the entrainment of thermal plasma close to the core region (e.g. \citealp{laing07}). 2) The jet power and local environment determine how quickly a jet becomes decollimated (e.g. \citealp{gopal96}). 3) There is a fundamental difference in the nature of the central engine and/or the composition of the jets (electron-proton vs. electron-positron pairs, e.g. \citealp{celotti97}).

While classical FR I and FR II morphologies comprise the majority of the radio galaxy population, in \citeyear{gopal00} \citeauthor{gopal00} discovered a rare group of hybrid morphology sources (HyMoRS), in which differing FR morphologies are observed in each of the two lobes. These galaxies, of which only around 30 are currently known, provide the opportunity to probe two differing morphologies that are a result of just one central engine, potentially providing a unique insight into the cause of the dichotomy between the two FR classes. Despite this potential wealth of information, studies of HyMoRS have until recently remained limited to survey searches (e.g. \citealp{banfield15, kapinska17}) and investigations on milliarcsecond-scales using Very Long Baseline Interferometry (VLBI; e.g. \citealp{ceglowski13}). Where studies have used arcsecond resolution data (e.g. \citealp{gawronski06}, herein G06), they were archival, narrowband observations in a single array configuration, meaning the diffuse lobes are resolved out due to lack of \emph{UV} coverage on short spacings. The morphology and mechanisms that cause the formation of HyMoRS therefore remains relatively unexplored and non-intrinsic effects, such as orientation, cannot be ruled out in many cases.

\subsection{The radio spectrum of HyMoRS}
\label{specageintro}

A detailed analysis of the shape and spatial distribution of the spectrum of radio galaxies can provide key insights into their underlying physics, particularly when considered on resolved scales. A limited spectral study of HyMoRS was performed by G06, in which a gradient in the spectrum along the length of the lobes was observed, similar to what is seen in standard FR I and II sources. However, this study was hindered by observations being available at only two frequencies. A more recent study by \citet{gasperin17} at low frequencies of a newly discovered hybrid morphology source provides improved frequency coverage (323-1519 MHz) and an insight into the low-energy electron population, but is unable to provide the resolution required to determine the small scale structure, spectral curvature, or the most diffuse emission at GHz frequencies. Therefore while the observed large-scale spectral gradients are likely to be real, the observations were unable to reliably determine any small scale variations or any curvature present within the spectrum.

In theory, for an electron population radiating via synchrotron emission in a fixed magnetic field with an initial energy distribution described by a power law such that $N(E) = N_0 E^{-\delta}$ the energy losses scale as $\tau = \frac{E}{dE/dt} \propto 1/E \propto 1/\nu^{2}$, leading to a preferential cooling of higher energy electrons. In the absence of any further particle acceleration, this produces a spectrum that becomes increasingly curved over time (spectral ageing) and the fitting of such models to well constrained spectra, particularly to the most diffuse emission where the oldest plasma is likely to be located, is able to provide us with the characteristic age of a source. In addition, the derived model parameters are also able to provide key insights in to the underlying nature of a source. For example, the injection index that describes the initial electron energy distribution at the point of acceleration has been shown to be much steeper for FR II galaxies \citep{harwood13,harwood15, harwood17a} than for their FR I counterparts (e.g. \citealp{heesen14, heesen18}). There are therefore clear, testable predictions that can be addressed by this form of analysis with modern, broadband width observations, which will ultimately aid in determining the underlying mechanisms that drive the formation of HyMoRS.

\subsection{Outstanding questions addressed in this paper}
\label{aims}

It has been suggested by \citet{gopal00} that HyMoRS, simply by their existence, are highly indicative that the FR I/II dichotomy is a result of radio galaxy environment, rather than any intrinsic property of the central engine or jet; however, without detailed knowledge of the underlying physics of these sources neither these or any other possibilities can be excluded. By confirming the morphology and performing a detailed spectral study on small spatial scales for a given source, one is potentially able to derive key information about what is driving these powerful outflows. Using new Jansky Very Large Array (VLA) observations to provide high fidelity images over a wide frequency range (1 to 8 GHz) that will recover both the compact and extended emission of a sample of 5 hybrid morphology objects discovered by G06 (Table \ref{targets}), we intend to address three key questions:

\begin{enumerate}
\item What is the morphology of the extended emission on well resolved scales?\\
\item What is the spectrum of HyMoRS on well resolved scales and how is this spatially distributed?\\
\item What mechanism gives rise to hybrid morphology objects?\\

\end{enumerate}
\vspace{-2mm}

In Section \ref{method} we give details of target selection, data reduction and the analysis undertaken. Section \ref{results} presents our results and in Section \ref{discussion} we discuss these findings in the context of the aims outlined above. Throughout this paper, we define the spectral index such that $S \propto \nu^{-\alpha}$ and use a concordance model in which $H_0=71$ km s$^{-1}$ Mpc$^{-1}$, $\Omega _m =0.27$ and $\Omega _\Lambda =0.73$ is used \citep{spergel03}.

\begin{figure*}
\centering
\includegraphics[angle=0,height=6.8cm]{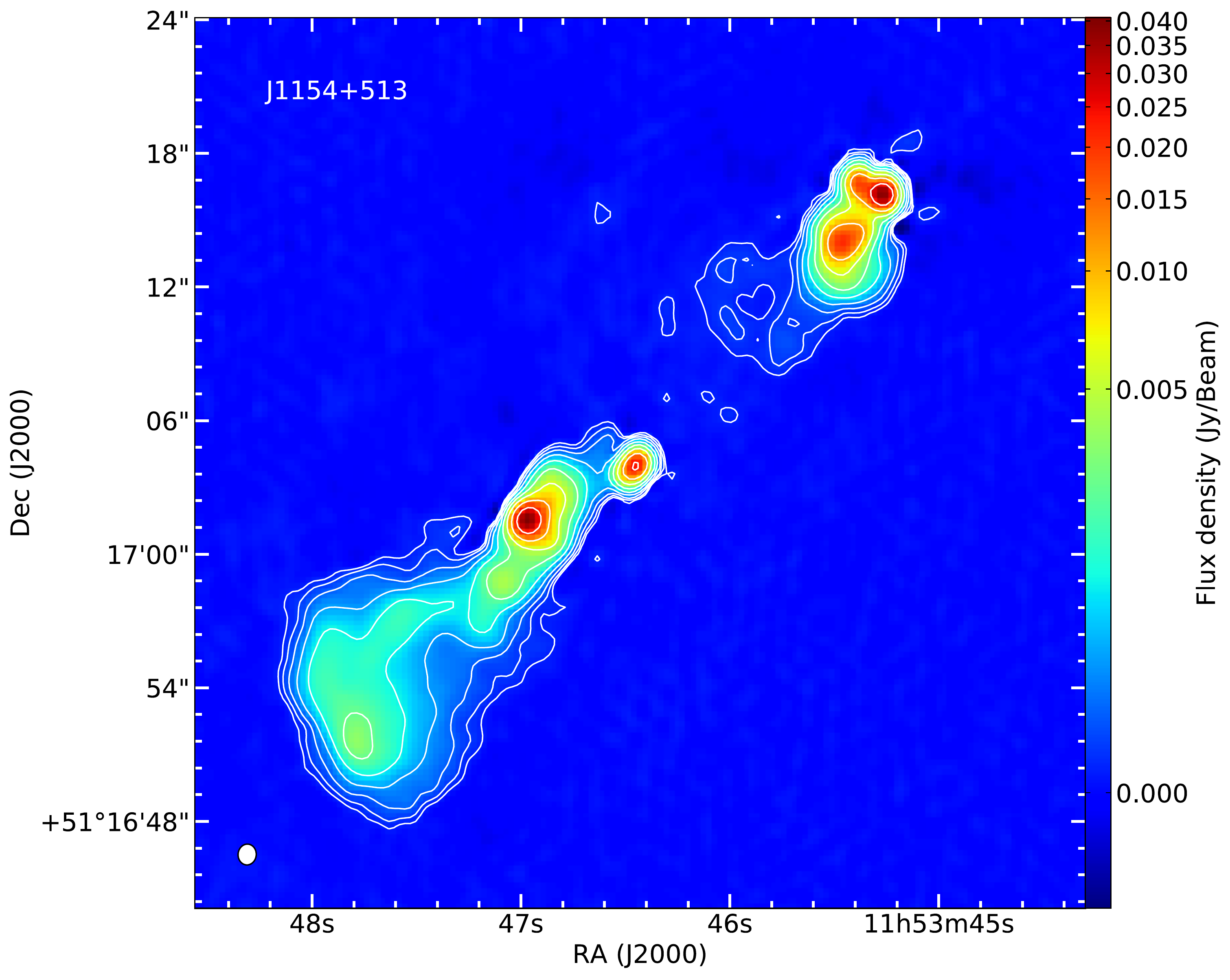}
\includegraphics[angle=0,height=6.8cm]{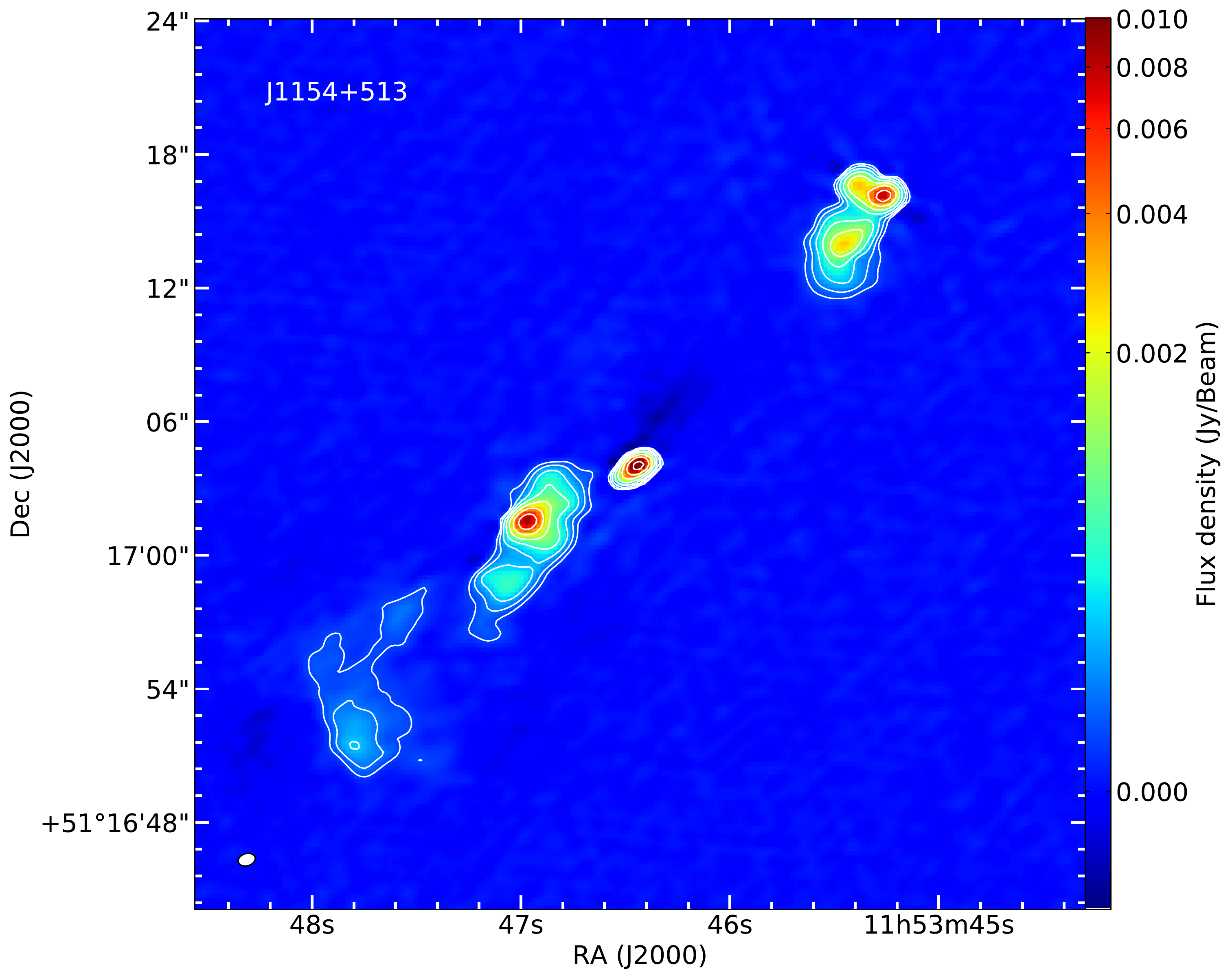}\\
\vspace{0.35cm}
\includegraphics[angle=0,height=6.9cm]{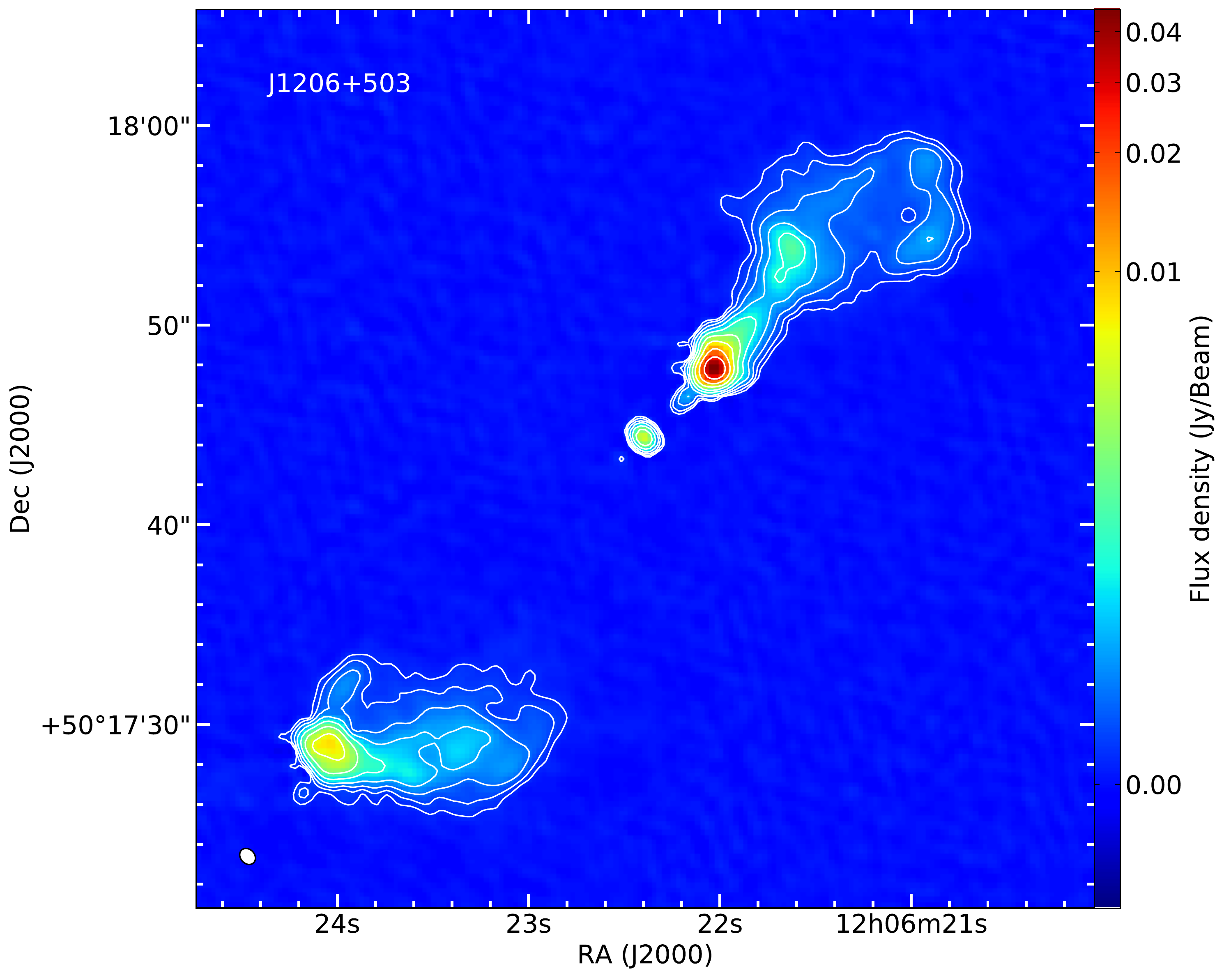}
\includegraphics[angle=0,height=6.9cm]{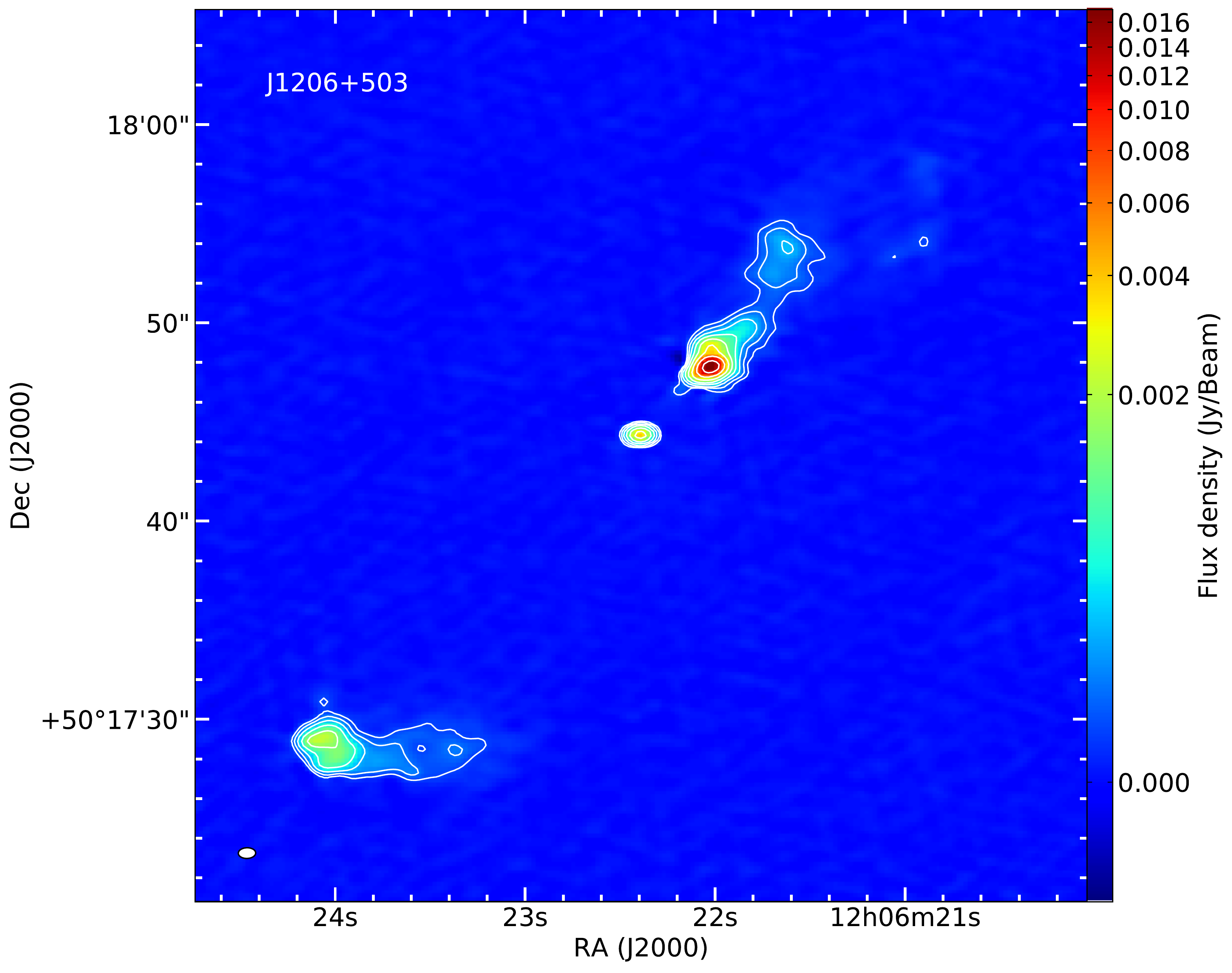}\\
\vspace{0.35cm}
\includegraphics[angle=0,height=6.9cm]{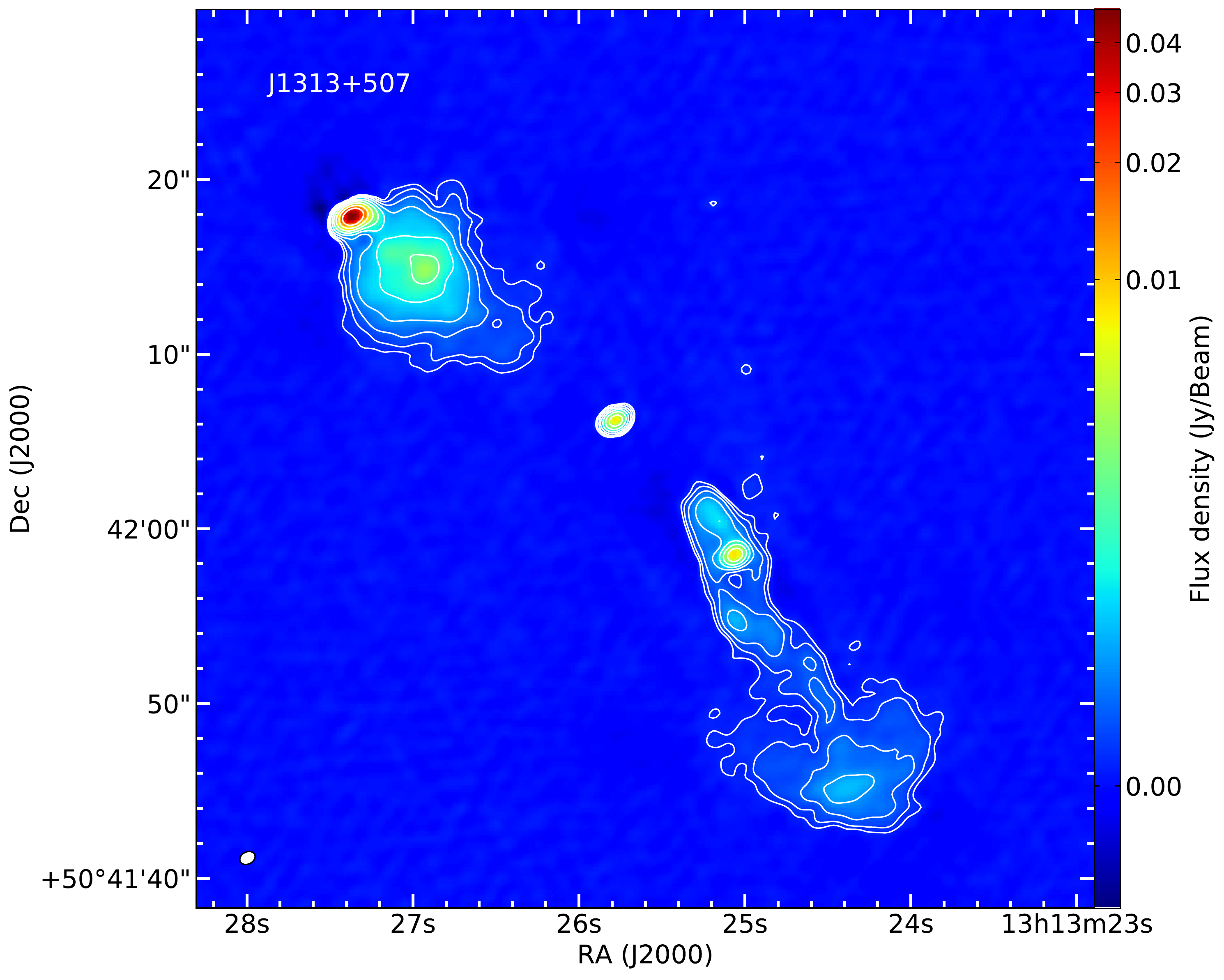}
\includegraphics[angle=0,height=6.9cm]{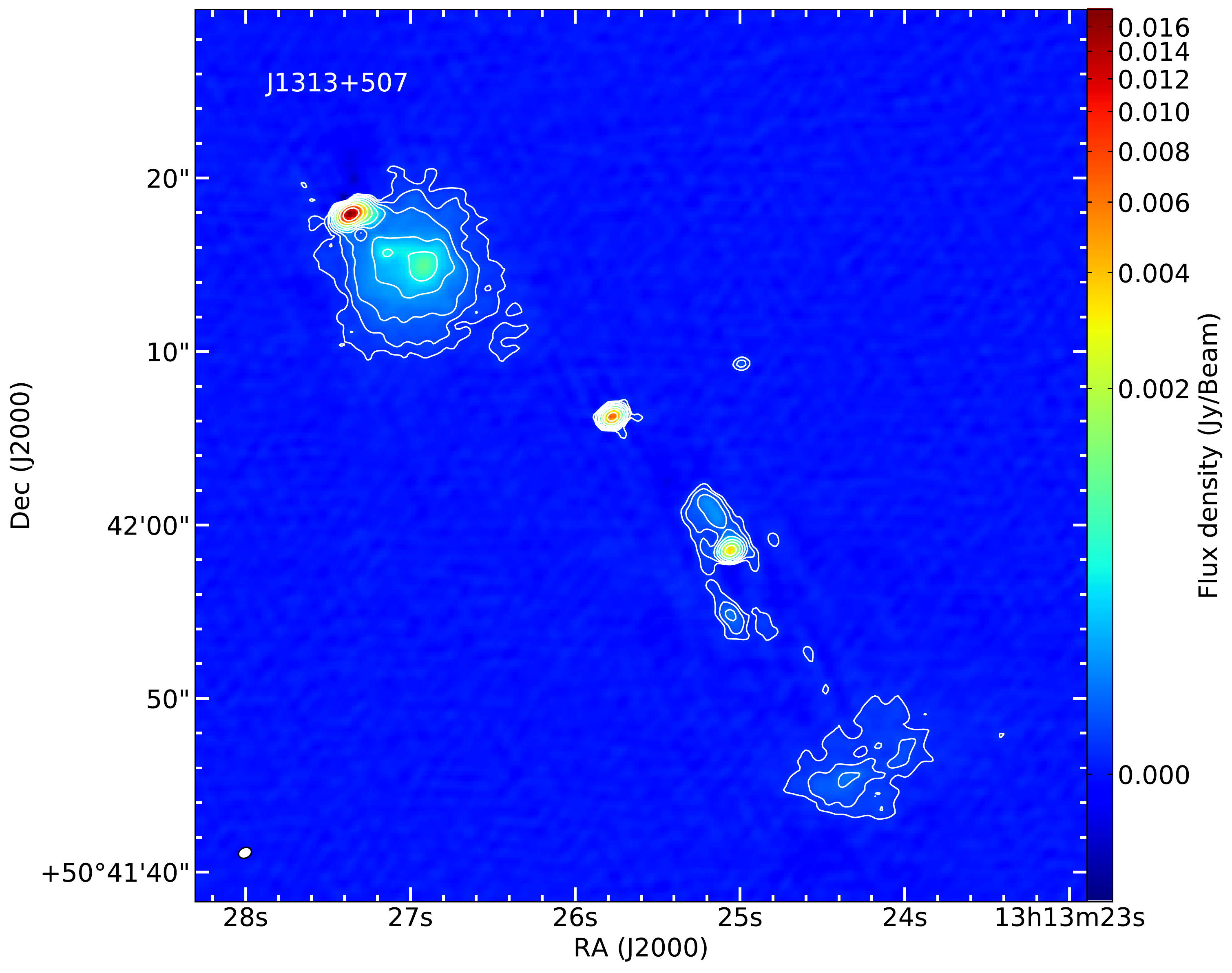}
\vspace{2.5cm}
\end{figure*}

\begin{figure*}
\centering
\includegraphics[angle=0,height=6.9cm]{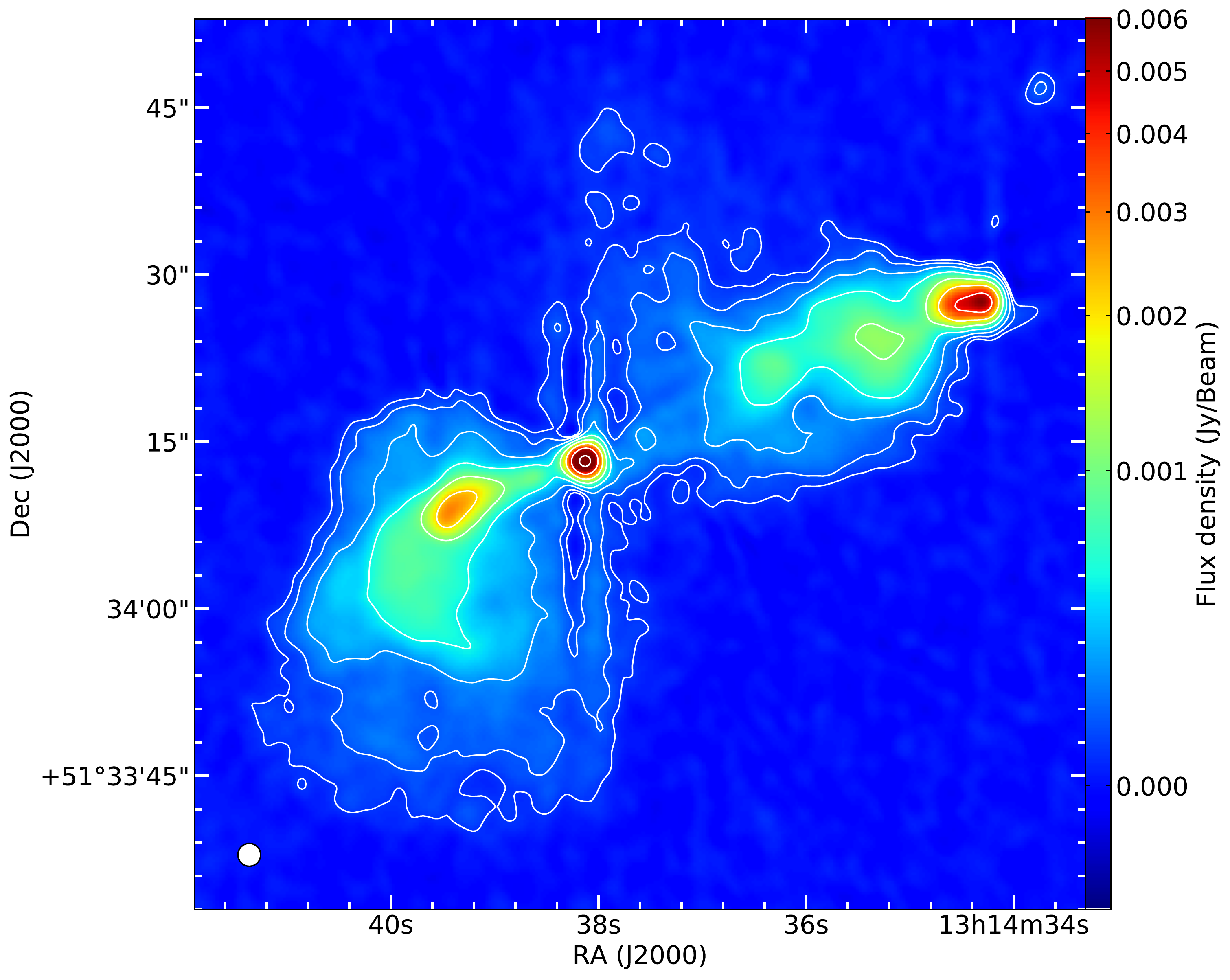}
\includegraphics[angle=0,height=6.9cm]{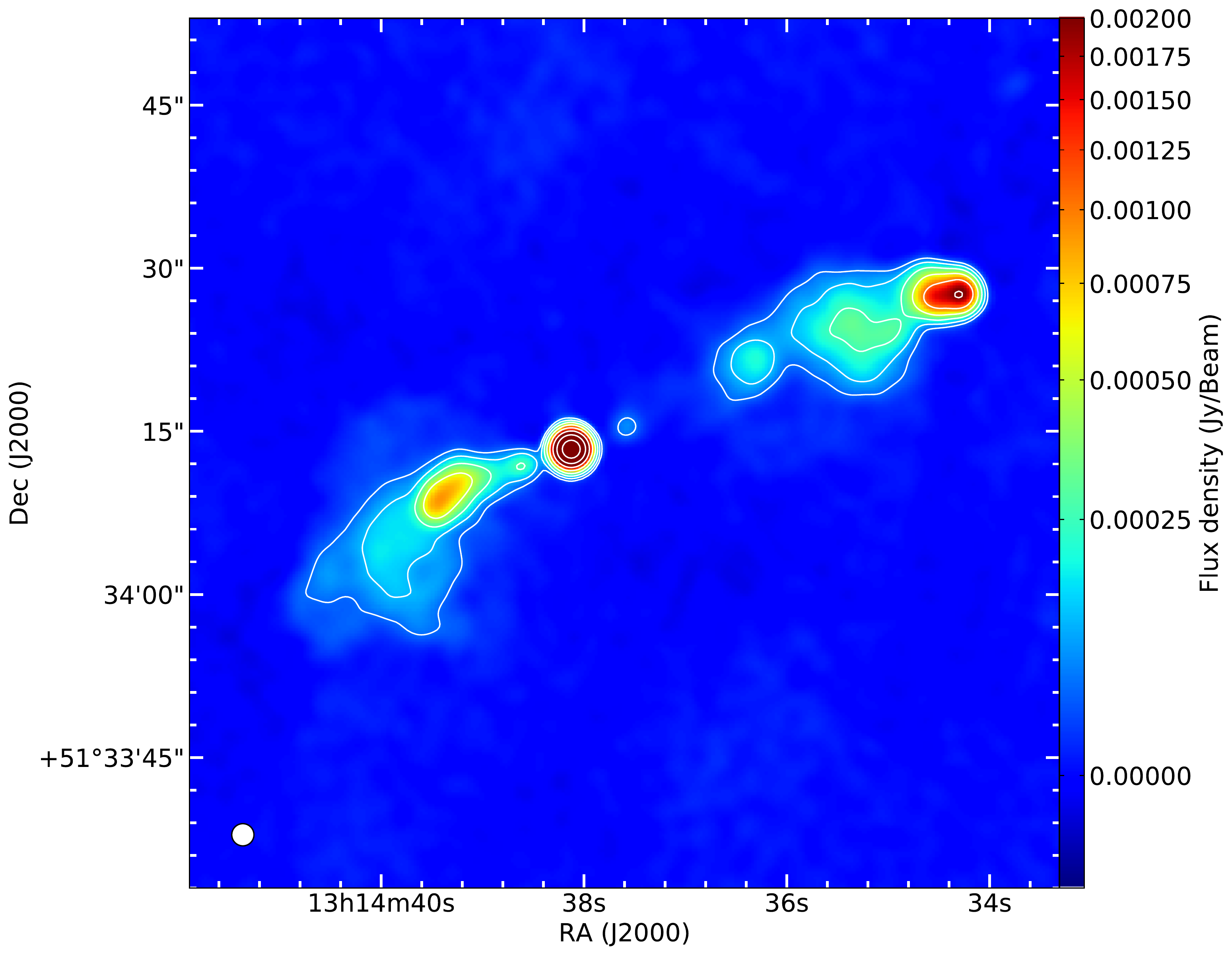}\\
\vspace{0.35cm}
\includegraphics[angle=0,height=6.9cm]{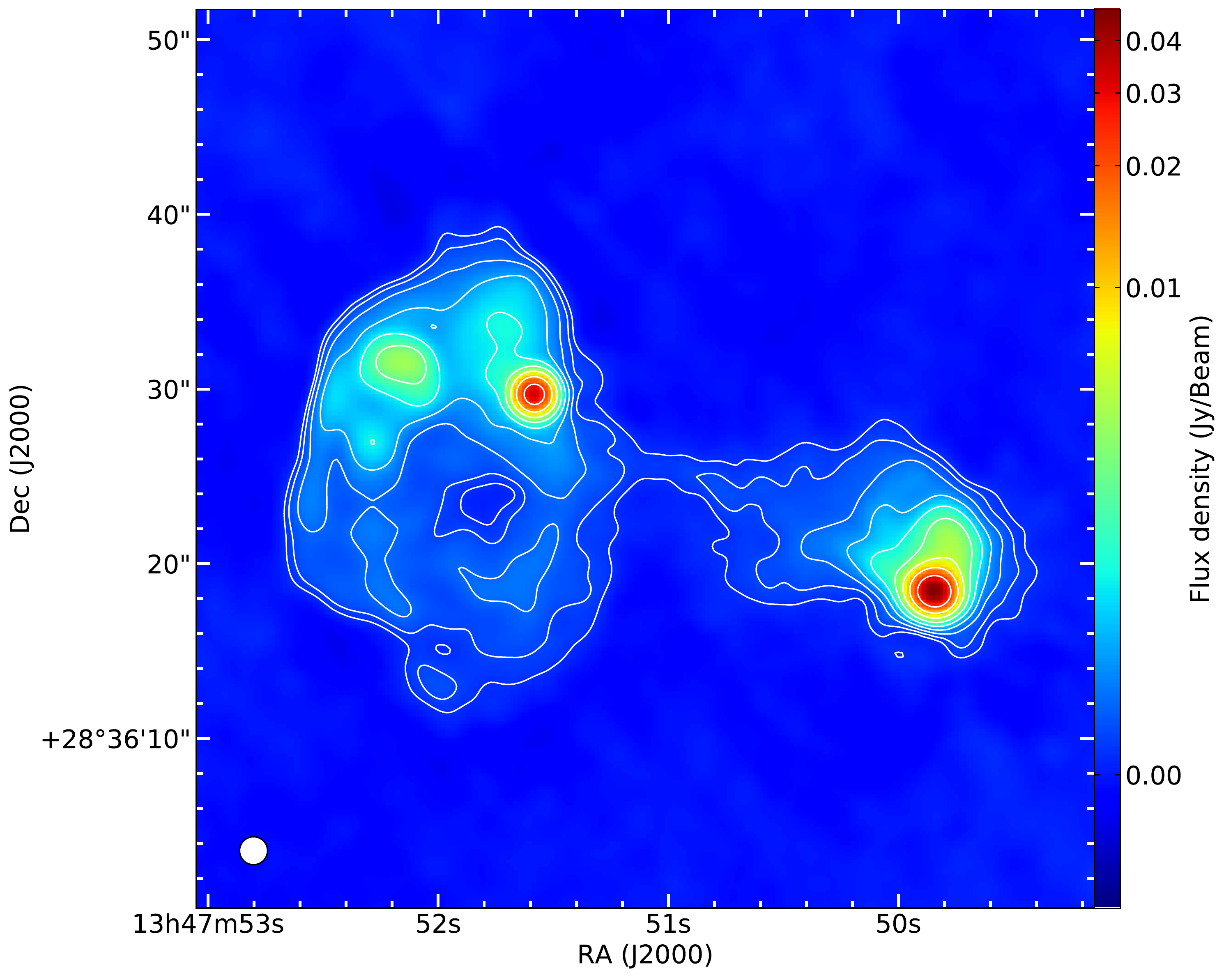}
\includegraphics[angle=0,height=6.9cm]{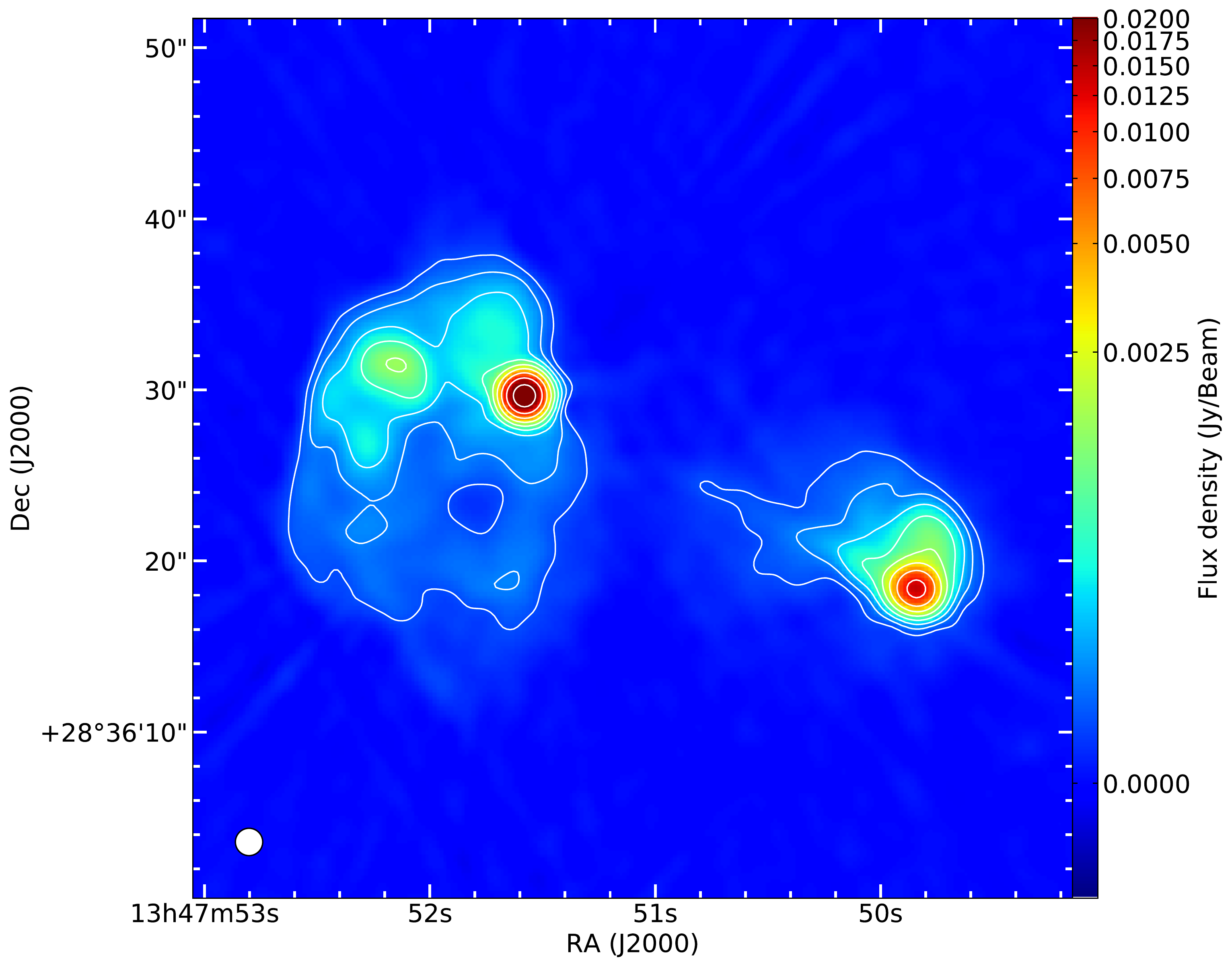}
\caption{Full bandwidth radio maps at 1.52 (left) and 6.00 (right) GHz of J1154+513, J1206+503, J1313+507, J1315+516, and J1348+286 (top to bottom) imaged using multiscale CLEAN and \emph{nterms} = 2. Noise levels and properties are shown in Table \ref{combinedsummary}. Contours levels are set such that $S_{cont} = 5\sigma_{rms} \times 2^{n}$.  In order to highlight structure in the extended emission, the maximum colour scale values have been set to match the brightest emission in the lobes.}
\label{intensitycombined}
\end{figure*}

\section{Data reduction and spectral analysis}
\label{method}

\subsection{Target selection and observations}
\label{targetselection}

In order to address the questions outlined above, the sample of 5 HyMoRS (Table \ref{targets}) discovered by G06 in the FIRST survey were observed with the VLA in multiple array configurations and frequency bands. As the dynamics of these rare sources are likely to be complex, this sample is ideal for addressing our scientific aims as:

\begin{itemize}
\item Investigations using narrowband, single configuration VLA observations have already been undertaken, meaning they provide the best known candidates for being true hybrid sources.\\
\item There is a clear disparity between the extended emission observed in the FIRST survey images and those at higher resolutions. The resolved morphology of these sources' extended emission is therefore still unknown.\\
\item VLBI observations of the sample \citep{ceglowski13} provides additional information about the inner jets, which may prove vital in understanding their large scale morphology.
\end{itemize}

The ability of the VLA to provide broad bandwidth, high sensitivity observations in a variety of array configurations makes it the ideal instrument for the study of the large scale structure of HyMoRS. As spectral curvature due to ageing is most easily observable at a few GHz (\citealp{alexander87}; \citealp{alexander87a}; \citealp{carilli91}; \citealp{perley97}; \citealp{hardcastle01}), the VLA C-band ($4.0-8.0$ GHz) was a natural choice for the required observations with complementary L-band observations ($1.0-2.0$ GHz) also taken to constrain the low-frequency spectrum and spectral ageing model parameters (see Section \ref{spectralanalysis}). In order to ensure that both the compact and diffuse structure was recovered, observations were made using the A and B configurations at L-band, and B and C configurations at C-band frequencies. This provided matched $UV$ coverage, a largest angular scale of 120 and 240 arcsec at L- and C-band respectively, and a resolution on the order of 1 arcsecond at both frequencies meaning that all 5 sources are well resolved. All observations used can be found under VLA project reference 16A-391 with a summary given in Table \ref{targets}.

\subsection{Calibration and imaging}
\label{datareduction}

As the data are standard continuum observations, initial calibration was performed using the \casa VLA pipeline\footnote{\url{https://science.nrao.edu/facilities/vla/data-processing/pipeline}} in the standard manner. The calibrated target sources were then split off and any remaining RFI flagged using a combination of the \casa \textsc{rflag} task and manual inspection.

In order to combine the various array configurations, the longer baseline observations (A and B configurations at L- and C-band respectively) were self-calibrated in phase and \textsc{clean}ed to convergence using the multiscale clean algorithm \citealp{cornwell08, rau11}). For the imaging the multiscale deconvolution was performed on scales of 0, 5, 10, and 20 times the cell size, which was set to one-fifth of the beam size. As the bandwidth of the observations was fractionally large, the \textsc{casa} multi-frequency synthesis (MFS) \textsc{clean} parameter $nterms=2$ \citep{rau11}, which scales the flux by a spectral index value fitted over the observed bandwidth, was used in line with standard practice. The calibrated longer baseline data where then used as a model to cross-calibrate the shorter baseline observations. The data were then concatenated and further phase only self-calibration cycles were performed, culminating in one round in both phase and amplitude.

The L-band A-configuration for J1315+516 required some changes to the method described above due to self-calibration resulting in the data between 1.5 and 2 GHz being entirely flagged. The raw, uncalibrated data were therefore revisited with additional flagging being carried out and the bandpass, flux, and phase calibrated manually but this resulted in only a minor improvement. The B-configuration data did not suffer from this same problem therefore the A-configuration data without any self calibration was concatenated with the self-calibrated B-configuration data and one additional round of phase only self-calibration applied using the B configuration model.

Both J1348+286 and J1315+516 have several other bright sources within the L-band primary beam that required cleaning in order to decrease the noise values. This was achieved using the ``outlier fields" parameter of the \casa \textsc{clean} task to specify additional phase centres to image following the standard procedure outlined in section 5.3.18.1 if the \casa cookbook\footnote{\url{https://casa.nrao.edu/casa_cookbook.pdf}}. This significantly reduces the required size of the final images which would have required a significantly larger image size. Bright outlier sources were identified by searching the NVSS catalogue within a radius of $30\,$arcminutes of the phase centres and selecting sources with $1.4\,$GHz flux densities in NVSS greater than $\sim 50\,$mJy. These outlier fields were deconvolved jointly with the images of the target sources using the same clean parameters. The outlier imaging was not required for the C-band data due to the much smaller primary beam size at the higher frequencies decreasing the uncorrected flux densities of the outlier sources. 

To perform the resolved spectral analysis detailed in Section \ref{spectralanalysis}, multiple images matched in terms of resolution, cell size, and $UV$ coverage are required over the frequency space to constrain any curvature preset within the spectrum. After initial exploratory imaging, we found the lowest noise images with minimal artefacts were produced by a robust value of 0 and a common resolution of 1.3 arc seconds with a cell size of 0.26 arc seconds (one fifth the resolution). In theory, a larger number of quasi-monochromatic observations are preferable; however, this can only be achieved at the cost of sensitivity. As the surface brightness of HyMoRS differs significantly between the lobes 2 imaging strategies were adopted. For the images used in the fitting of the higher surface brightness FR II lobes, the L-band observations were imaged between 1 and 1.5 GHz using contiguous 64 MHz spectral windows (SPW), increasing to 2 SPW's (128 MHz) between 1.5 and 2 GHz. At C-band frequencies, where the spectrum is generally much steeper, the data was split in to 256 MHz chunks (4 SPWs) to recover the most diffuse emission. For the images used in the fitting of the FR I lobes where the plume-nature of the lobes means that the surface brightness is much lower in the most diffuse regions of plasma, we retain the L-band imaging strategy but combined the full C-band bandwidth to create a single deep image. As for spectral fitting it is crucial that we are recovering the same structure across all images, and that fitting is limited by the worse quality image, only SPW's that contained high quality data in both configurations were included.

In addition to the images used for spectral fitting, full bandwidth images were also created with central frequencies of 1.5 and 6 GHz for all sources to fully explore the morphology at the maximum possible resolution and recover the faintest emission. As the full bandwidth images do not require the $UV$ coverage to be matched for all subbands, as is the case for those used in the spectral fitting, images were made using all SPW's and cleaned using MFS with $nterms=2$ on scales of 0, 5, 10, and 20 times the cell size. The resulting images are shown in Figure \ref{intensitycombined} with a summary of the images given in Table \ref{imagingparameters}.

\subsection{Spectral Analysis}
\label{spectralanalysis}

\subsubsection{Image alignment}
\label{imagealignment}

To perform the well-resolved spectral analysis used in this paper, accurate image alignment is critical. Frequency dependent shifts introduced during calibration and imaging mean that to achieve accurate alignment, we used the Gaussian fitting method described by \citet{harwood13, harwood15} that has previously proved robust for this form of analysis.

Regions around the core of each source were first selected using \casaviewer and saved in pixel coordinates. Gaussians were then fitted to each image using the \casa \textsc{imfit} task and the mean coordinates determined to provide a reference pixel for the images to be aligned. The difference between the peak of the Gaussian and the reference pixel were then determined and the \python \textsc{scipy} library's interpolation method used to shift each image into alignment. The accuracy of the alignment was then checked refitting the Gaussians which resulted in a discrepancy on the order of only 0.001 pixel. The effects of misalignment are therefore unlikely to have a significant effect on our analysis.

While this method of image alignment has been performed on a range of past projects, it has previously been a manual process. In preparation for the large number of images that will require accurate alignment in future and forthcoming surveys, we have therefore automated this process in the form of a publicly available \python script\footnote{\url{http://www.askanastronomer.co.uk/brats/downloads/alignment.py}}.

\begin{table}
\centering
\caption{Summary of imaging parameters}
\label{imagingparameters}
\begin{tabular}{lccccccc}
\hline
\hline

Source&\multicolumn{4}{c}{Bandwidth (MHz)}&Cell&Res.&Outlier\\
&A&B&C&D&(arcsec)&(arcsec)&fields\\

\hline
J1154$+$513    &    64   &   128   &   256   &   256   &   0.26   &   1.3   &   No   \\
J1206$+$503    &    64   &   128   &   256   &   256   &   0.26   &   1.3   &   No   \\
J1313$+$507    &    64   &   128   &   256   &   256   &   0.26   &   1.3   &   No   \\
J1315$+$516    &    128  &   128   &   \multicolumn{2}{c}{Full band}   &   0.42   &   2.1   &   Yes   \\
J1348$+$286    &    \multicolumn{2}{c}{Full band} &   256    &   256   &   0.32   &   1.6   &   Yes   \\

\hline
\end{tabular}

\vskip 5pt
\begin{minipage}{8.5cm}
Summary of parameters used for the spectral age modelling images described in Section \ref{datareduction}. `Bandwidth' is the bandwidth used for each image between A: 1--1.5 GHz; B: 1.5--2 GHz; C: 2--4 GHz; D: 4--8 GHz. `Cell' and `Res' list the common cell size and resolution used for each source in arcseconds. `Outlier fields' denotes whether outlier fields were used during imaging.
\end{minipage}
\end{table}

\begin{table*}
\centering
\caption{Summary of full bandwidth images}
\label{combinedsummary}
\begin{tabular}{lcccccc}
\hline
\hline

Source&\multicolumn{2}{c}{Resolution (arcsec)}&\multicolumn{2}{c}{RMS Noise ($\micro$Jy beam$^{-1}$)}&\multicolumn{2}{c}{$S_{core}$ (mJy beam$^{-1}$)}\\
&1.52 GHz&6.00 GHz&1.52 GHz&6.00 GHz&1.52 GHz&6.00 GHz\\

\hline
J1154$+$513    &    $0.94 \times 0.82$   &   $0.79 \times 0.55$   &   $17.9$   &   $9.10$   &   $34.4$   &   $18.8$   \\
J1206$+$503    &    $0.88 \times 0.71$   &   $0.88 \times 0.54$   &   $19.8$   &   $8.39$   &   $5.89$   &   $4.29$  \\
J1313$+$507    &    $0.94 \times 0.82$   &   $0.79 \times 0.55$   &   $15.7$   &   $7.37$   &   $8.42$   &   $7.23$   \\
J1315$+$516    &    $2.1 \times 2.1$     &   $2.1 \times 2.1$     &   $14.6$   &   $3.58$   &   $11.4$   &   $12.9$   \\
J1348$+$286    &    $1.6 \times 1.6$     &   $1.6 \times 1.6$     &   $19.4$   &   $3.84$   &   $32.1$   &   $30.6$   \\

\hline
\end{tabular}

\vskip 5pt

\end{table*}

\subsubsection{Spectral fitting}
\label{spectralfitting}

As described in Section \ref{specageintro}, a detailed analysis of spectral curvature present within radio galaxy lobes can provide key insights in to the underlying mechanics of a source. In order to address the questions laid out in Section \ref{aims}, we therefore used the Broadband Radio Astronomy ToolS (\textsc{brats}\footnote{\url{http://www.askanastronomer.co.uk/brats}}; \citealp{harwood13, harwood15}) software package, which provides a suite of tools for the spectral analysis of radio sources on resolved scales, to explore the spectral structure of the sample. The underlying functionality of this software has been discussed in detail elsewhere \citep{harwood13, harwood15}, and so we do not repeat that process here, but we instead provide a summary of the functions and parameters relating specifically to this investigation.

To determine the thermal noise for each image, a background region was first defined for each field in a blank area of sky well away from the target and any other bright emission using \casaviewer. The data were then loaded into \brats and the `\emph{setregions}' command was used to define regions to which the model will be fitted on a pixel by pixel basis, assuming an on-source noise multiplier of 3 to account for the increased RMS noise arising from uncertainty in the modelling of extended emission that occurs during imaging \citep{harwood13, harwood15}. We assumed the standard flux calibration errors for L- and C-band VLA observations of 2 per cent \citep{perley13}. These values were used throughout for determining both region selection and the statistical values associated with the model fitting. As in HyMoRS the morphology of each lobe is different, it is plausible that the parameters used to model the spectra may also vary. This process was therefore performed for each lobe and modelled individually. In all cases the core region which is not expected to be described by models of spectral ageing was excluded.

In order to determine the characteristic age of the observed plasma, the Tribble model of spectral ageing \citep{tribble93} was fitted to the observations using the \brats `\emph{fitjptribble}' command. This model is based on the canonical Jaffe and Perola model of spectral ageing (JP model, \citealp{jaffe73}), which assumes a single injection of particles with an energy distribution described by a power law that are then subject to radiative losses. The Tribble model uses the standard JP model assumptions that pitch angles are isotropic on short timescales relative to the radiative lifetime and that the magnetic field strength is constant over the lifetime of the source, but attempts to more realistically model the magnetic field structure by assuming a Gaussian random field, allowing electrons to diffuse across regions of varying field strength. In the weak field, high diffusion (i.e. free-streaming) case, the spectrum can be modelled by integrating the standard JP losses over a Maxwell-Boltzmann distribution. While computationally expensive, the Tribble model has been shown to provide an improved goodness-of-fit over the traditionally used JP model for radio galaxy lobes (e.g. \citealp{harwood17a}) whilst maintaining the more physically plausible assumptions it makes compared to the often better fitting model of \citep{kardashev62} and \citep{pacholczyk70} (KP model). It is therefore ideal for use in the moderately sized sample investigated here. The Tribble model has been discussed at length previously and so we do not repeat that process here but instead direct the interested reader to work by \citet{hardcastle13} and \citet{harwood13} who provide a full derivation and comparison between the different model types.

While the magnetic field strength of the sources would ideally be determined via synchrotron/inverse-Compton fitting (e.g. \citealp{croston04,harwood16}), this is only possible for a limited number of sources where high resolution, high sensitivity X-ray data are available. We have therefore taken the standard assumption of equipartition determined using the \textsc{synch} code of \citet{hardcastle98}, corrected by a factor of $0.4 B_{eq}$; the median value derived from a large sample of radio galaxies where the magnetic field strength has been determined via synchrotron/inverse-Compton fitting \citep{ineson17}. While it has so far only been possible to determine such a correction for FR II radio galaxies, and values are known to range between $0.3$ and $1.3 B_{eq}$ (e.g. \citealp{croston05}), it is unlikely to have a significant impact on the the conclusions of this paper and provides a better estimate than an uncorrected equipartition value. We also note that in a similar manner to variations in the magnetic field strength, the redshifts for our sample are photometric and so a change in distance could alter the spectral age significantly; however, as within this paper we are primarily interested in the relative, rather than absolute, ages within a source this is unlikely to have an impact on our results.

One of the key parameters which must also be determined when fitting models of spectral ageing is the power law describing the initial electron energy distribution, known as the injection index. Where the initial distribution is described such that \begin{equation}\label{initialdist}N(E) = N_{0}E^{-\delta}\end{equation} a spectrum which has not been subject to losses will be observed as \begin{equation}\label{alphainject}\alpha_{inj} = \frac{\delta - 1}{2}\end{equation} Thus if lobe emission is observed at low enough frequencies such that radiative losses are negligible, and/or any curvature in the spectrum can be accounted for, it is possible to determine the injection index empirically. We therefore determined the injection index for the lobes of each source using the \textsc{brats} `\emph{findinject}' command which sets the injection index as a free parameter and uses $\chi^{2}$ minimisation to determine which injection index best describes the observed emission, allowing any curvature within the spectrum to be accounted for (see \citealp{harwood13} for a detailed discussion). A grid search was performed using this method with a step size of 0.01 between 0.50 (the physical minimum limit on first-order Fermi acceleration) and a reasonable upper limit of 1.0. For sources where a minimum had not yet been reached, the bounds were extended accordingly.

Once all of the parameters described above had been determined, a final model fitting run was performed for each source assuming standard Lorentz factor values of $\gamma_{min} = 10$ and $\gamma_{max} = 1 \times 10^{6}$ (\citealp{carilli91, hardcastle98, godfrey09, harwood16}) and the relevant images, data, and statistical values exported for analysis.

\section{Results}
\label{results}

\subsection{Morphology}
\label{morphology}

The overall morphology of the sample shown in Figure \ref{intensitycombined} are in good agreement with the original investigation by G06, displaying a clear difference between the two lobes in all cases confirming their apparent hybrid morphology classifications. While some of the most diffuse emission is not recovered in the combined images, this is due to the higher resolution and frequencies used compared to the original study. This is made evident from the contours of Figure \ref{ageingmaps} where lower resolution images have been used at GHz frequencies for the purposes of model fitting. We note that an unusual north/south structure is also observed close to the core in the L-band image of J1313+507 but, as this feature is not seen at higher frequencies or in the original study by G06, is most likely an imaging artefact. We therefore exclude this region from any analysis performed in this paper. Given the good agreement with previous studies we are therefore confident that the observed structure of the lobes for our sample is robust, in many cases revealing additional emission due to the improved \emph{UV} coverage and increased sensitivity.

While the morphology on the largest scales is therefore in agreement with previous studies, the improved resolution and image fidelity reveals a number of previously unobserved structures. On these improved scales the FR II lobes remain morphologically as one would expect in all cases; however, the same cannot be said for the apparent FR I lobes. Most notably distinct disconnects between the core and the peak lobe emission (where particle acceleration is thought to occur) exist in all sources, resembling inner hotspots rather than the extended brightening observed in archetypal FR I sources such as the plumed 3C31 \citep{heesen18} or the lobed 3C296 \citep{hardcastle97}. Even for J1315+516, which represents the best example of a classical FR I morphology in our sample, we note the acceleration region remains relatively compact and emission connecting to the core is much weaker that one would expect for a typical FR I source, with peak brightness located well away from the core region. Interestingly, a previously undiscovered second disconnect is also observed in J1154+513 between the inner regions surrounding the peak emission and the emission towards the outer edge of the lobe. We also note that for J1206+503 the improved sensitivity of the lower resolution GHz modelling images reveals previously unseen emission connecting the bright inner core to the more diffuse outer regions at C-band frequencies, matching that observed at L-band, but still lacks any jet-like emission close to the core.

Due to the sensitivity limits of the study by G06, little was previously known about the high resolution morphology of J1348+286. From the lower resolution maps of G06 the morphology appears to be that of a hybrid; however, from Figure \ref{intensitycombined} we see that the peak emission in the apparent FR I lobe is located at the edge of the sources with the majority of the diffuse emission located south of the core. The brightness distribution of this diffuse emission is clumpy with bright regions spread throughout the lobe and a notable ``hole'' with a significant brightness deficit located centrally in the lobe. Multiple calibration and imaging methods were used to confirm the robustness of the images, with little variation in the observed structure. This unusual morphology is discussed in Section \ref{orientation}, but note here that J1348+286 can be unified under the same scheme as the other sources in our sample.

One surprising feature revealed by the improved resolution of the images presented here is the presence of secondary peaks close to the end of the apparent FR I lobes in some sources. This is most prevalent in J1313+507 with 3 brightened regions along the length of the lobe terminating in a mushroom-shaped large-scale morphology. Similar features are also observed in J1206+503 and J1154+513, with the later having a particularly bright peak located close to the end of the lobe. Such features are absent from both J1315+516 and J1348+286 suggesting that they may be a common, but not a required, observational feature of hybrid morphology sources.

\subsection{Spectral modelling}
\label{specage}

\subsubsection{Modelling parameters}
\label{modelparams}

The results of the injection index determination described in Section \ref{spectralfitting} are summarised in Table \ref{bestinject}. The sources J1154+513 and J1206+503 follow a $\chi^2$ curve similar to most other radio galaxies, but note that both lobes take values more commonly associated with FR IIs ($\alpha_{inj} > 0.6$; \citealp{harwood13, harwood15, harwood17a}).  The derivation of uncertainties on these values and their application to spectral age modelling have been described in detail in previous work (\citealp{avni76, harwood15}) and so we do not repeat that process here, but it is worth noting that the errors given in Table \ref{bestinject} are those of the model fit and do not account for additional physical or observational effects. Once such process is that of the superposition of spectra is likely to have a significant impact on the apparent FR I lobe (see Section \ref{orientation}) and so the uncertainties are likely to be significantly higher than the values stated. Regardless, even assuming a significantly higher error estimate of $\delta\alpha_{inj} = 0.05$ results in the injection index for the FR I lobe lying within the FR II regime. While morphologically similar to these sources, the injection index of J1313+507 is more inline with what one would expect for a truly hybrid source with one lobe taking steeper, FR II type values and the other a flatter injection index commonly associated with FR Is ($\alpha_{inj} \approx 0.5$; \citealp{young05, laing13}).

In contrast, both lobes of J1315+516 and J1348+286 are observed to have flatter FR I type injection indexes, with values $\alpha_{inj} < 0.6$. While it is not uncommon for FR II lobes to take such values, and might be expected given the differing morphologies (Figure \ref{intensitycombined}) compared to the three sources discussed above, we note that for both sources one lobe goes below the commonly assumed lower limit of $\alpha_{inj} = 0.5$. For J1315+516 this difference is small and is discussed further in Section \ref{particleacceleration}, but for J1348+286 a minimum is still not observed despite extending our search range suggesting it falls well below $\alpha_{inj} = 0.4$. While a flatter injection index will increase the spectral age of the plasma, this will only have a significant impact for the eastern lobe of J1348+286. We therefore opted to use the physical limit for first-order Fermi acceleration of $\alpha_{inj} = 0.5$ during model fitting in all cases where flatter values are derived.

Using these injection index values, along with the dimensions and total flux measured from the full bandwidth 1.5 GHz images, the magnetic field strength for the lobes of each source were derived (Table \ref{bfield}). Typical nearby radio galaxies are observed to have equipartition field strengths on the order of ~1 nT, ranging from sub-nT values for remnant sources (e.g. Blob1, \citealp{brienza16}) to higher field strengths for bright FR II sources (e.g. 4 nT for Cygnus A, \citealp{vries18}). Accounting for the increased redshift of our sample, which is thought to approximately scale as $(1+z)^{0.8}$ \citep{krolik91}, the equipartition values which range from 0.95 nT (J1348+286, eastern lobe) to 7.42 nT (J1154+513, western lobe) are consistent with other radio galaxies. As described in Section \ref{alphainject}, we assume the magnetic field strength of the lobes to be $0.4 B_{eq}$ when performing the spectral age model fitting.

\begin{table}
\centering
\caption{Best fitting injection indices}
\label{bestinject}
\begin{tabular}{lccccc}
\hline
\hline

Source&Lobe&FR type&Injection&\multicolumn{2}{c}{Error}\\
&&&Index&+&-\\

\hline
J1154$+$513     &   West   &    II   &   1.01    &   0.01    &   0.01\\
                &   East   &    I    &   0.94    &   0.01    &   0.01\\
J1206$+$503	    &   West   &    I    &   0.65    &   0.01    &   0.01\\
                &   East   &    II   &   0.80    &   0.01    &   0.02\\
J1313$+$507	    &   West    &    I    &   0.49    &   0.01    &   0.01\\
                &   East   &    II   &   0.65    &   0.01    &   0.01\\
J1315$+$516     &   West    &    II   &   0.46    &   0.01    &   0.02\\
                &   East    &    I    &   0.62    &   0.01    &   0.01\\
J1348$+$286     &   West    &    II   &   0.63    &   0.01    &   0.01\\
                &   East    &    I    &   $<$0.4  &   --      &   --  \\

\hline
\end{tabular}

\vskip 5pt
\begin{minipage}{8.5cm}
Best fitting injection indices for the HyMoRS sample as detailed in Section \ref{spectralfitting}. Errors are determined using the methods of \citet{avni76} and its application to spectral ageing models by \citet{harwood15}.
\end{minipage}

\end{table}

\begin{table*}
\centering
\caption{Source dimensions and magnetic field strengths}
\label{bfield}
\begin{tabular}{cccccccccccc}
\hline
\hline
Source&Lobe&FR type&1.52 GHz&Length&Width&Length&Width&Volume&Geometry&Equipartition&Magnetic field\\
name&&&flux (Jy)&(arcsec)&(arcsec)&(kpc)&(kpc)&($\times 10^{23}$\,m$^3$)&&strength (nT)&strength (nT)\\
\hline
J1154$+$513&West&II&2.64&10.5&5.2&64.0&31.6&5.4&C&7.42&2.97\\
&East&I&3.72&18.9&4.5&113.0&27.1&22.0&C&4.80&1.92\\
J1206$+$503&West&I&1.95&18.6&4.9&153.5&40.3&21.1&C&3.44&1.38\\
&East&II&1.02&12.6&5.1&185.6&58.4&19.5&C&4.16&1.66\\
J1313$+$507&West&I&1.24&22.5&7.1&153.4&48.3&22.0&C&1.55&0.62\\
&East&II&3.13&14.0&8.3&109.7&65.2&39.6&C&2.11&0.84\\
J1315$+$516&West&II&4.04&38.5&19.4&262.9&132.4&65.2&C&1.72&0.69\\
&East&I&4.11&31.2&31.8&212.6&217.1&860&E&0.99&0.40\\
J1348$+$286&West&II&3.69&27.1&9.2&197.8&67.0&12.5&C&3.00&1.20\\
&East&I&2.95&19.0&29.2&138.6&213.2&53.4&E&0.95&0.38\\
\hline
\end{tabular}
\vskip 5pt
\begin{minipage}{17.8cm}
\small `Source Name' are the IAU names of the galaxies discussed in this paper. `FR type' lists the lobe morphology, with `1.52 GHz flux', `Length', and `Width' as measured from the 1.52 GHz full bandwidth images to the 3 sigma contours. Note the dimensions are those used to determine the magnetic field strength and may not fully extend to the core. `Volume' shows the derived volume of the lobe assuming either a cylindrical (C) or ellipsoid (E) as noted in the `Geometry' column. `Magnetic field strength' lists the value used in all modelling assuming $B = 0.4B_{eq}$, where $B_{eq}$ is given in the `Equipartition strength' column.
\end{minipage}
\end{table*}

\subsubsection{Spectral fitting}
\label{spectralmodelling}

\subsubsection{J1154+513}
\label{j1154specage}

The results of the model fitting for J1154+513 are shown in Figure \ref{ageingmaps} with the corresponding statistics and maximum age shown in Table \ref{fittingresults}. The eastern lobe is well fitted by the spectral ageing model; however, the distribution of ages is unusual compared to what one would expect for an FR I type source. While the inner regions display low-age plasma, consistent with being the primary site of particle acceleration as one would expect, multiple hotspot-like low age regions are also observed in the outer lobes. While low age regions of plasma are found both at the edge of the lobe and the core in other FR I sources (e.g. figure 11 of \citealp{laing11}), these are normally more extended structures and often connected to the core by a visible FR I type jet, compared to the compact and disconnected hotspot-like regions observed here. The inner and outer regions of plasma also appear to be separated by a high age band of emission at the midpoint of the source. While the most eastern of these low-age emission regions is likely due to edge effects as observed in previous ageing studies (e.g. \citealp{harwood13}), the remaining three features correlate well with both the brightening in the intensity map (Figure \ref{intensitycombined}) and with these features tentatively observed in the spectral index map of the original study (figure 1 of G06). We are therefore confident that these structures are intrinsic to the source.

In contrast to the eastern lobe, the western lobe of J1154+513 is poorly fitted by the spectral ageing model being rejected at the $>$99 per cent confidence level. From the $\chi^{2}$ map (Figure \ref{ageingmaps}) one can see that while the hotspot region provides a reasonable fit, the more diffuse regions result in particularly high $\chi^{2}$ values. This is particularly prominent in plasma to the west of the second peak of emission which may provide some clue as to the cause. We discuss this feature along with the possible cause of the usual eastern spectral structure further in Section \ref{particleacceleration}.

\subsubsection{J1206+503}
\label{j1206specage}

The spectral structure of J1206+503, at least initially, appears to take a more classically plumed FR I distribution than that of J1154+513 with an age gradient increasing away from the assumed sites of particle acceleration\footnote{We note that this is not a requirement for an FR I classification with a spectral index increasing towards the core being observed in lobed FR Is e.g. \citet{laing11}.}. The eastern lobe is well fitted by the model and is distributed as one would expect for an FR II type source; however, the western FR I lobe is rejected at the $>$99 per cent confidence level. This is in contrast to the poorly fitted FR II lobe of J1154+513 suggesting that the goodness-of-fit is not type dependent.

While the western lobe of J1206+503 appears to display an archetypal age gradient for a plumed FR I that increases with distance from the core, it is interesting to note that the secondary peak of emission in the outer lobe is coincident with the most poorly fitted regions of plasma ($\chi^{2}$ map of Figure \ref{ageingmaps}). It is therefore possible that a second low age region of plasma is present, similar to that observed in J1154+513, but with a significantly steeper injection index. For a fixed set of modelling parameters, this would likely manifest itself in the ageing maps as an old but poorly fitted region of plasma such as is observed. We explore this possibility further in Section \ref{particleacceleration}.

\subsubsection{J1313+507}
\label{j1313specage}

Both the western and eastern lobes of J1313+507 are well fitted by the spectral ageing model. The eastern, FR II lobe has a classical distribution of ages with zero age emission at the hotspot with age increasing towards the core. As is the case with J1154+513 and J1206+503 the FR I lobe, although well fitted by the model, also displays a variety of non-typical spectral features. At the inner acceleration region, older emission appears both away from and \emph{towards} the core. This unusual spectrum has the highest $\chi^{2}$ values across all of the source and is rejected at the $>$99 per cent confidence level. Further independent observations would therefore be required to confirm this feature and so is excluded from consideration for the remainder of this paper; however, the unusual spectral structure of the outer western lobe is more robust. Similar to the FR I lobes of the sources described previously, we observe a secondary peak of emission near the tip of the lobe, where low age emission is once again observed. While from the lower resolution contours the most easterly of this zero age emission appears to be well offset from this brightening, one can see from the high resolution C-band images that a faint tertiary peak is present. It is interesting to note that once again a slight offset if observed between the peak emission and these zero age regions.

\subsubsection{J1315+516}
\label{j1315specage}

J1315+516 provides the best model fitting results of all the sources studied, with very few regions being rejected above the 90 per cent confidence level (Table \ref{fittingresults}). The spectral curvature of both lobes follows a distribution typically associated with FR I and FR II lobes with zero age emission close to the acceleration regions and an increasing age away from and towards the core respectively. Accounting for the significantly smaller maximum value compared to the other sources in the sample, the $\chi^{2}$ map shows a relatively even distribution of goodness-of-fit across the source consistent with the uncertainties one would expect from such observations.

While the age gradient of J1315+516 therefore supports it being an archetypal hybrid morphology source, the primary acceleration region is still observed to be disconnected from the core. This common feature suggests that although the large-scale morphology differs between sources, the mechanism for particle acceleration is similar and that, despite appearing centre brightened, an FR II type jet may be present.

\begin{figure*}
\centering
\includegraphics[angle=0,width=17.6cm]{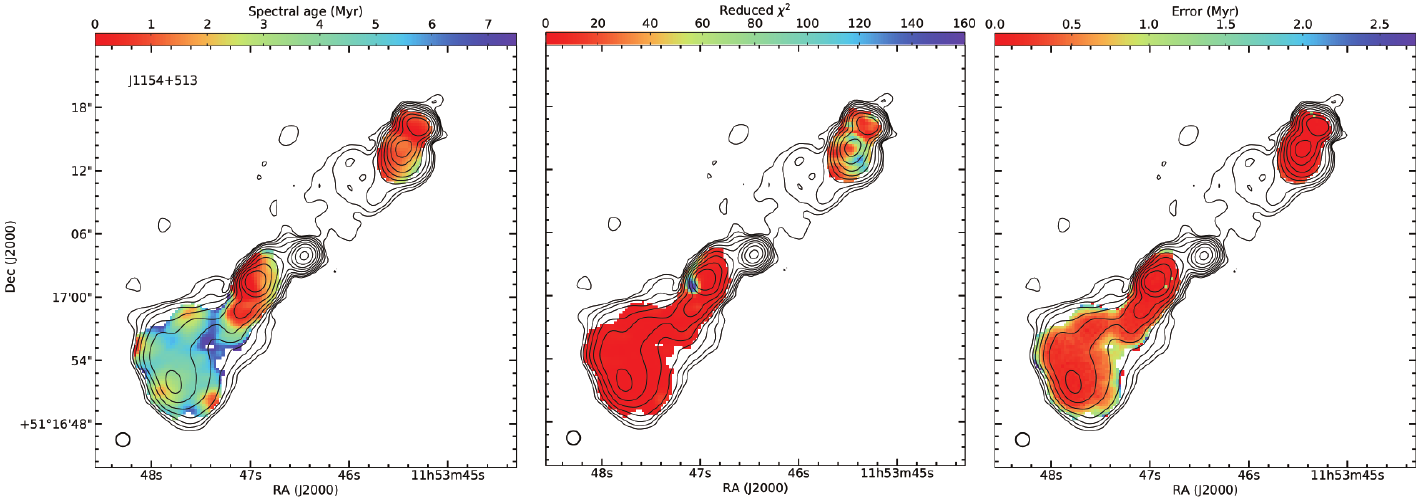}\\
\vspace{0.35cm}%
\includegraphics[angle=0,width=17.6cm]{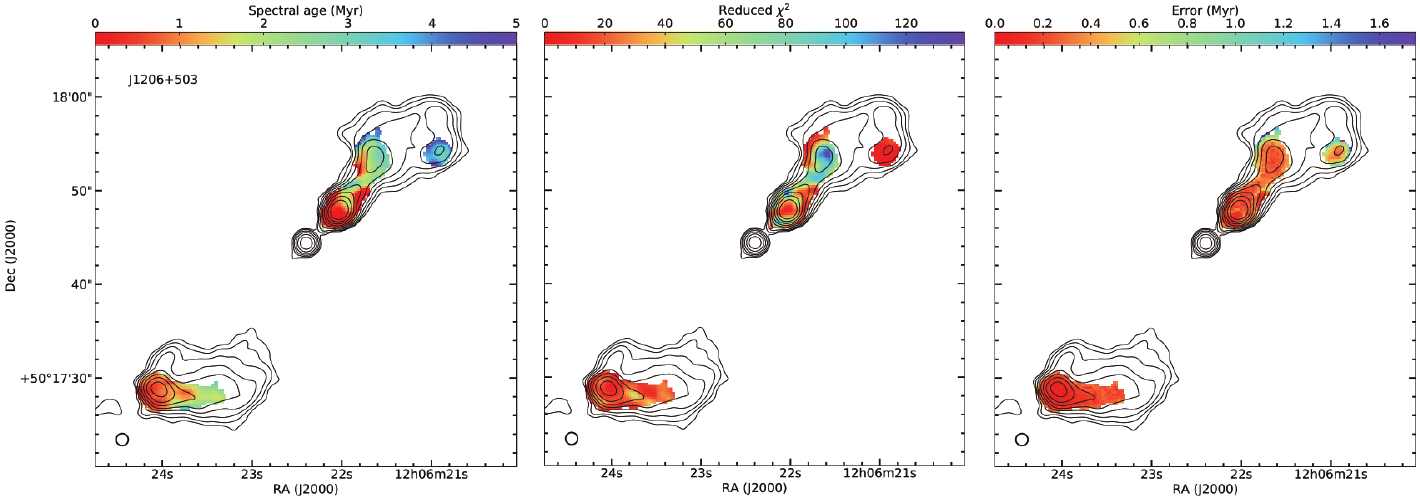}\\
\vspace{0.35cm}%
\includegraphics[angle=0,width=17.6cm]{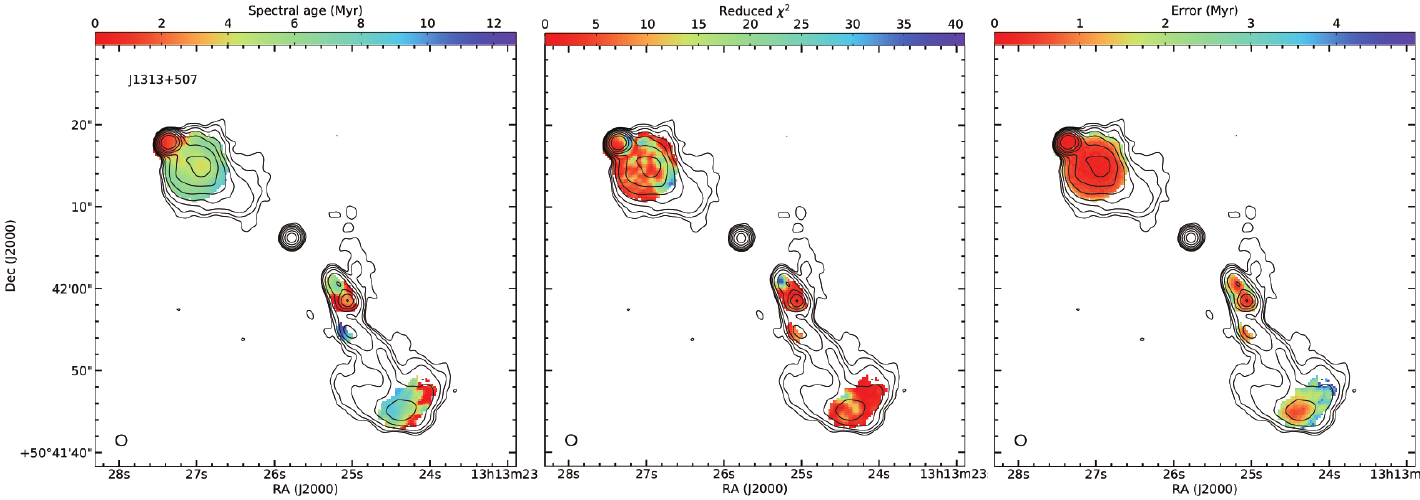}
\vspace{4.0cm}%
\end{figure*}

\begin{figure*}
\centering
\includegraphics[angle=0,width=17.6cm]{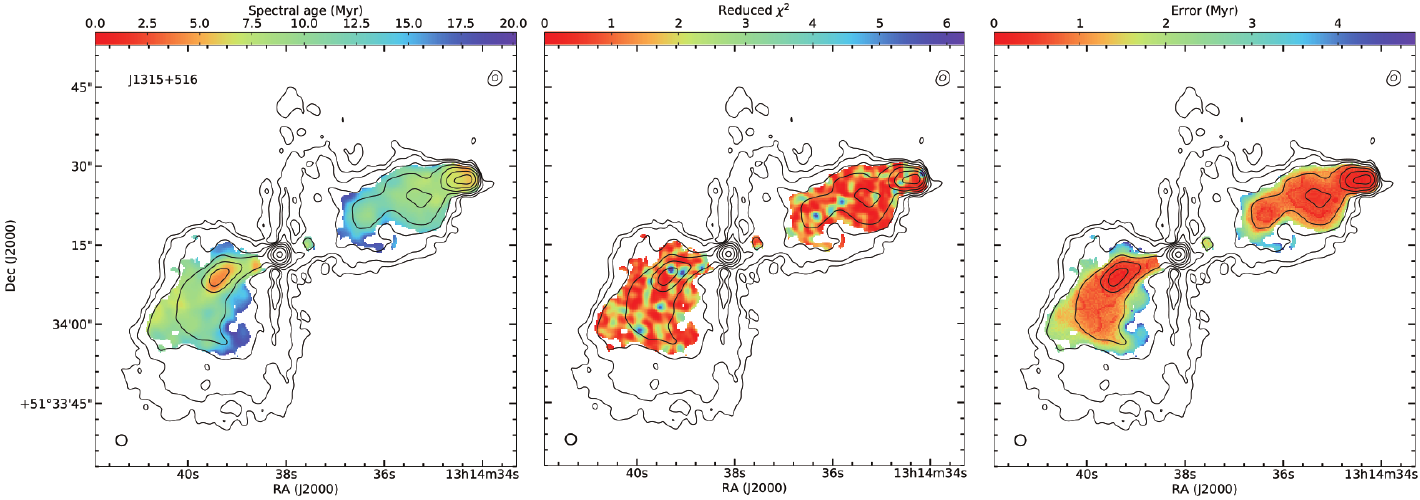}\\
\vspace{0.35cm}%
\includegraphics[angle=0,width=17.6cm]{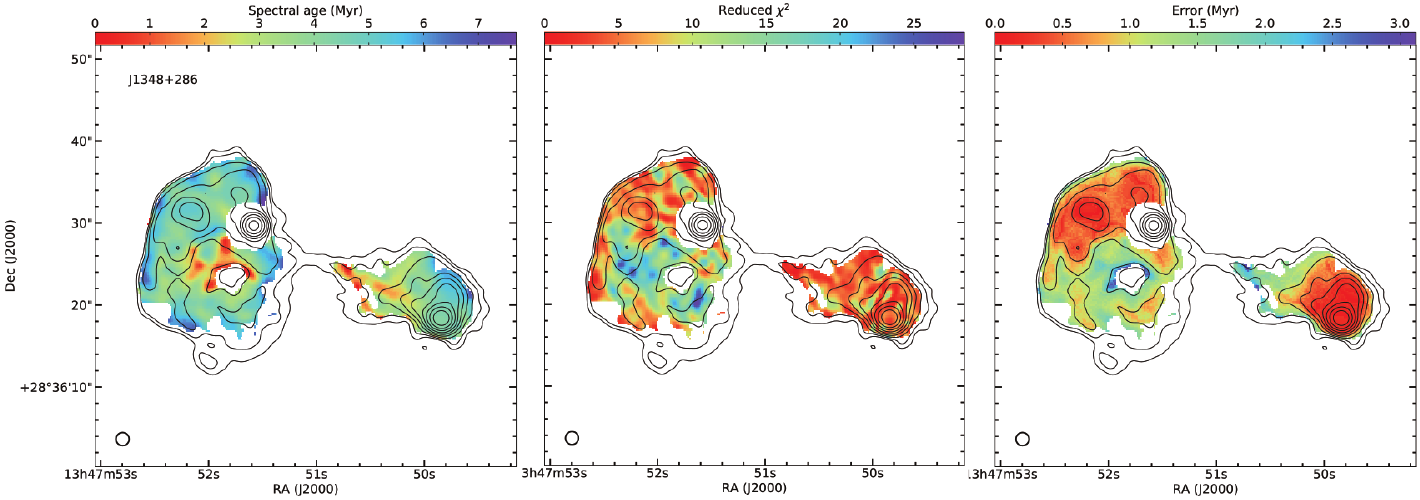}
\caption{Spectral ageing maps (left) of J1154+513, J1206+503, J1313+507, J1315+516, and J1348+286 (top to bottom) with corresponding $\chi^{2}$ (middle) and error maps (right). All maps have been overlaid with 1.5 GHz flux contours. Modelling parameters are shown in Tables \ref{bestinject} and \ref{bfield} with statistics and the maximum spectral ages given in Table \ref{fittingresults}.}
\label{ageingmaps}
\end{figure*}

\subsubsection{J1348+286}
\label{j1348specage}
The spectral structure of J1348+286 follows on from its unusual morphology described in Section \ref{morphology}. While the model provides a reasonable goodness-of-fit and we are unable to reject it even at the 68 per cent confidence level (Table \ref{fittingresults}), there is a notable lack of zero-age emission around the assumed sites of particle acceleration. The model in these regions is well fitted ($\chi^{2}$ map of Figure \ref{ageingmaps}) and so is unlikely to be a result of incorrect model parameters such as may be the case for certain regions of J1206+503. Some patchy regions of low-age emission are observed around the low-emission hole in the eastern lobe and coincident with the inner regions of the assumed jet path in the western lobe. Localised acceleration, such as turbulence, and jet/lobe interactions may explain such features, but we do not consider these to be the primary sites of particle acceleration as it is unlikely to be the source of the observed wider scale emission.

Due to the unusually steep spectrum of the primary acceleration region and its close proximity to the assumed core, composite maps using Pan-STARRS images \citep{flewelling16} and the 1.5 GHz radio contours were created. From Figure \ref{opticaloverlays} one can see that the core is coincident with a large, reddened, elliptical host. We are therefore confident that the location of the core is correct. The bright, flat-spectrum core ($\alpha_{core} \approx 0$), combined with the prominent hotspot in the western lobe implies that the jets are still currently active and so the lack of zero-age emission cannot be explained by J1348+286 being a remnant source. The hotspot like feature of the Eastern (FR I) lobe seen in intensity suggests that this usual spectrum is likely due to strong mixing of electron populations along the line-of-sight, which may also explain the unusual profile found when determining the injection index. We discuss this further in the context of the intrinsic nature of HyMoRS in Section \ref{particleacceleration}.

\section{Discussion}
\label{discussion}

The results presented in Section \ref{results} provide the first opportunity to explore the extended emission and spectrum of hybrid morphology radio galaxies on well-resolved scales. Understanding these observations is crucial if we are to determine the mechanisms and conditions that give rise to HyMoRS and gain insights into the wider radio galaxy formation process. In this section we discuss the observed properties of our sample and the underlying physics that cause the formation of these unusual objects.

\subsection{Particle acceleration}
\label{particleacceleration}

\subsubsection{Acceleration regions}
\label{accelerationregions}

While a variety of properties are observed to be different between radio galaxy types (most notably power) the classification of radio galaxies is purely morphological, defined as the centre-brightened FR Is and edge-brightened FR IIs. With modern observations it would therefore appear trivial to verify the classification of potential hybrid morphology sources; however, whether their morphologies are \emph{intrinsically} hybrid or only \emph{observationally} hybrid is key to determining their role and importance in the context of the wider population.

The brightening that defines FR I and IIs is the result of differing primary sites of particle acceleration within the lobe. For classical FR I sources such as 3C31, this is generally accepted to be due to an initially relativistic jet decelerating on scales of $\sim$10 kpc which results in a bright, extended ``flare'' that, at the resolutions discussed in this paper, appears to be connected to the core region (e.g. \citealp{laing02}). From Figure \ref{intensitycombined} it is apparent that this is not the case for the FR I lobes of our sample, with the primary particle accelerating regions for all sources being both compact and clearly detached from the core, similar to the hotspots found in FR II radio galaxies.

One of the key differences between the acceleration regions in FR I and II sources is the observed range of initial electron energy distribution values. While FR I sources are seen to have an injection index close to the limit of first-order Fermi acceleration ($\alpha_{inj} \approx 0.5$, \citealp{young05, laing13}) it is now well established through both GHz \citep{harwood13, harwood15} and MHz frequency observations \citep{harwood16, harwood17a} that FR IIs have a much wider, often steeper, range of values from $\alpha_{inj} = 0.5$ to $>1$. If the jets and particle acceleration mechanisms that form the lobes of hybrid sources are similar to their respective classical counterparts, one might expect to see similar injection index values. However, as noted in Section \ref{modelparams} even accounting for the caveats related to the uncertainties associated with the injection index and assuming a more conservative estimate, two of our sample (J1154+513 and J1206+503) have values that exceed the $\alpha_{inj} = 0.5$ to $0.6$ range found in FR I radio galaxies, taking values more commonly associated with FR IIs\footnote{Flatter injection index values ($\alpha_{inj}\approx0.5$) are relatively common in FR II sources, but the inverse (i.e. steeper values) cannot be said of FR Is.}.

For the two sources where $\alpha_{inj} \leq 0.6$ (the remaining source sitting on the border between the two extremes) the $\chi^2$ profile is relatively flat around the minimum values, only rising significantly above $\alpha_{inj} \approx 0.8$ (Appendix \ref{append_inj}). Previous studies of standard radio galaxies have generally produced a well-defined $\chi^{2}$ curve with a clear minimum when using a similar quality of data. One possible explanation is electron population mixing in which the observed plasma consists of a range of spectral ages leading to the underlying assumption that each region consists of a single age being violated. This in turn can lead to systematic errors in the derivation of both age and injection index, and possibly the poorly defined minima seen in the $\chi^2$ profiles. The most likely causes of such population mixing to be observed are intrinsic mixing due to e.g. turbulence, and empirical mixing due to the observers line-of-sight through the plasma.

Regardless of cause, the impact of varying the injection index values for the model fitting performed would be to systematically change the ages across the lobe of each source. Given the uncertainties in the dynamics of these sources (e.g. electron population mixing) such a shift is unlikely to significantly impact our findings as it is the relative age distribution, rather than absolute age, that is of interest. While such effects may therefore affect the measured ages, it is clear from the injection index values and unusual $\chi^2$ profile that the observed properties corresponding to the FR I and FR II lobes of hybrid morphology sources are not identical to their classical, edge-on radio galaxy counterparts. Given the overall morphology and initial electron distribution, our sample suggests that it is likely a ``strong-flavour'' FR II, rather than ``weak-flavour'' FR I, jet present in \textbf{both} lobes of hybrid sources despite being observed close to the core. Future VLBI observations of these regions would aid in confirming this scenario.

\begin{figure*}
\centering
\includegraphics[angle=0,height=6.9cm]{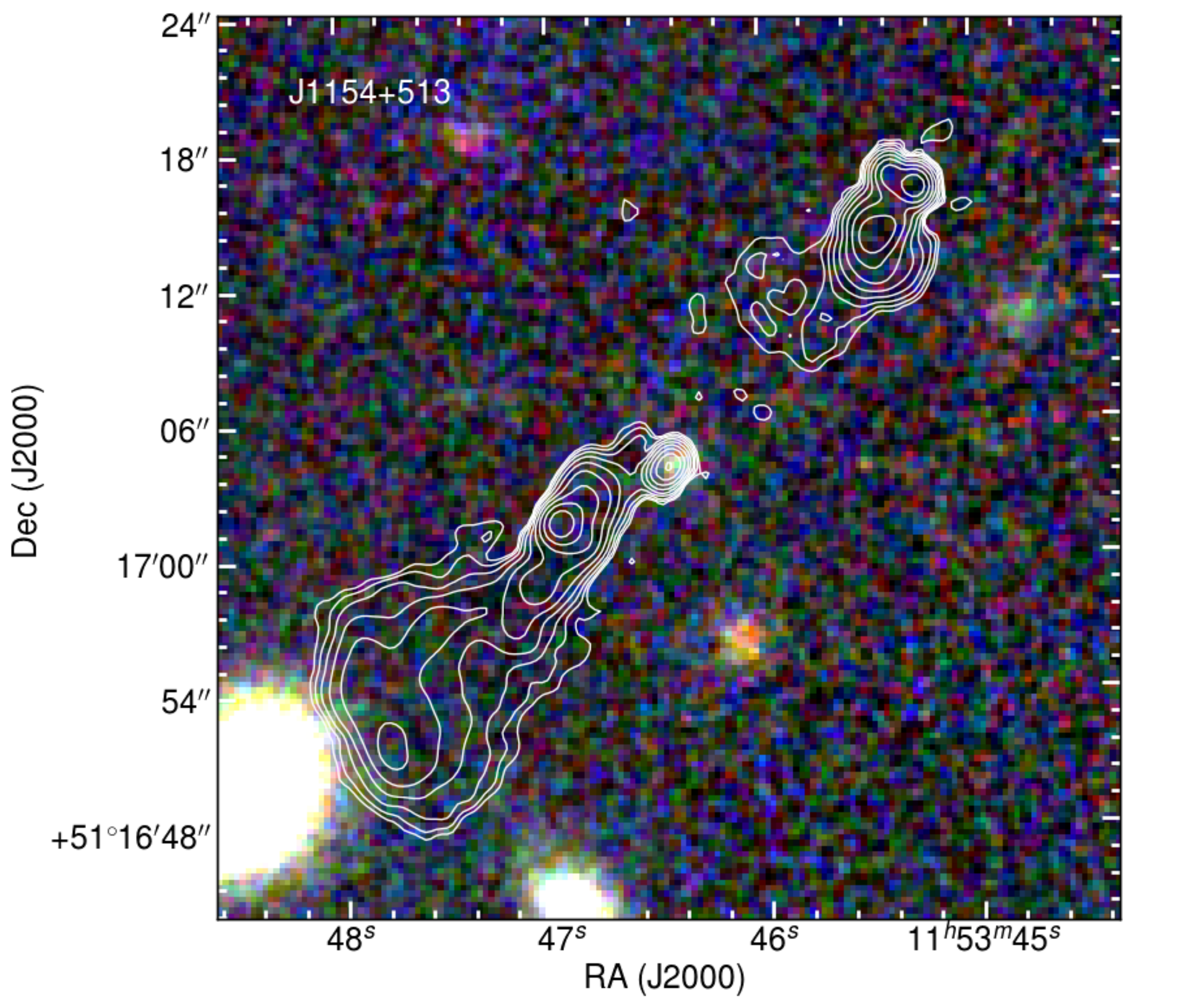}
\includegraphics[angle=0,height=6.9cm]{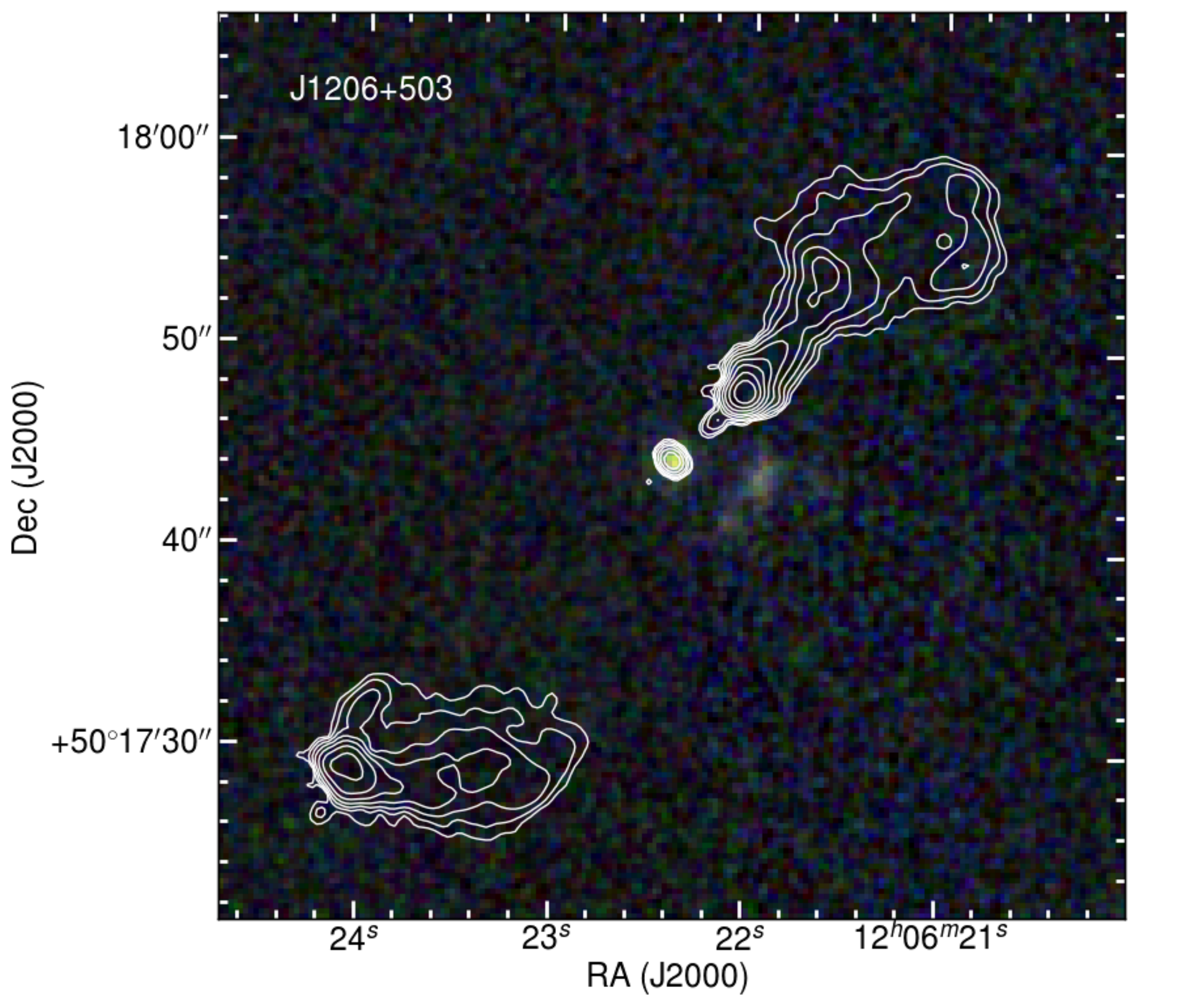}\\
\vspace{0.35cm}%
\includegraphics[angle=0,height=6.9cm]{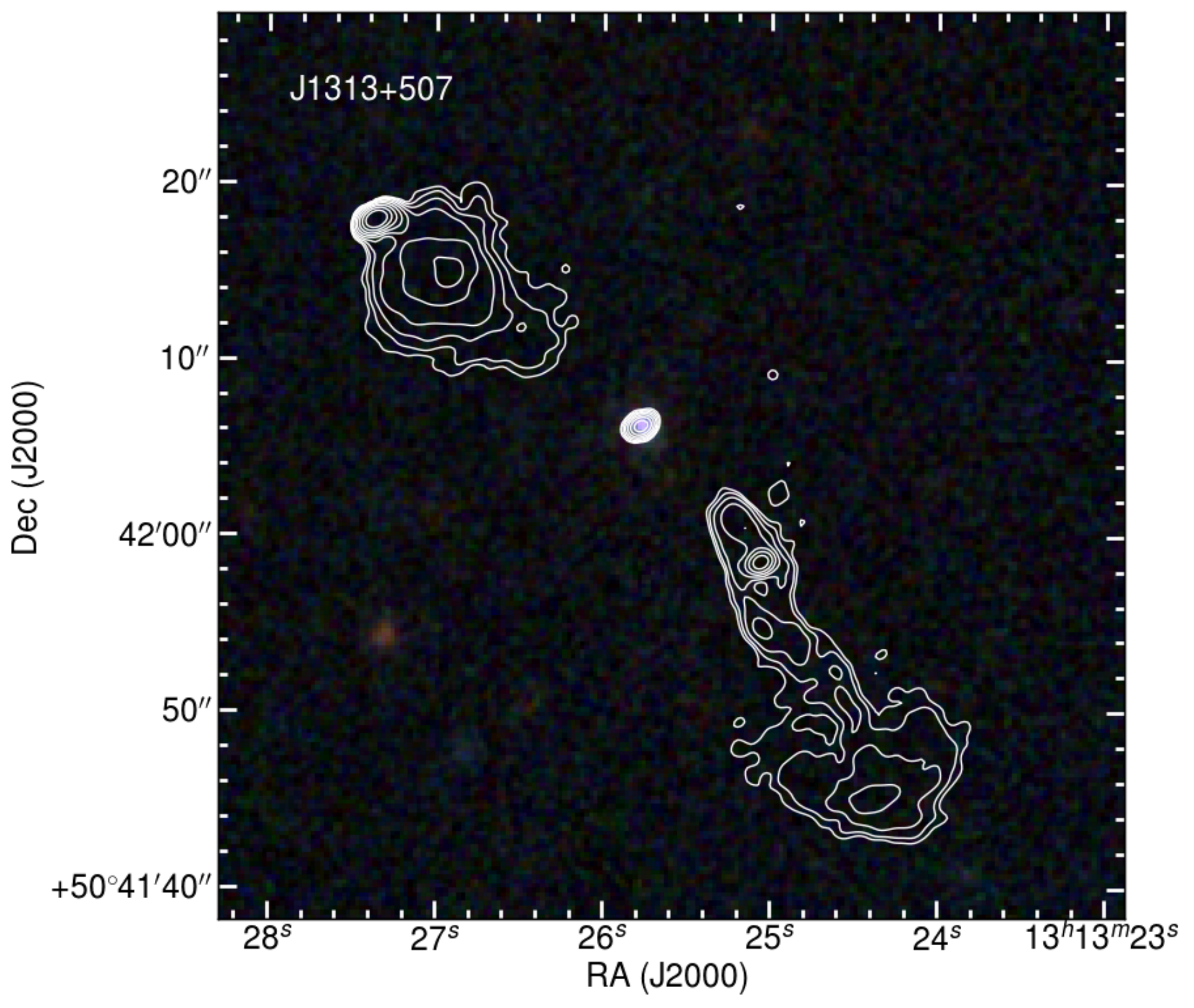}
\includegraphics[angle=0,height=6.9cm]{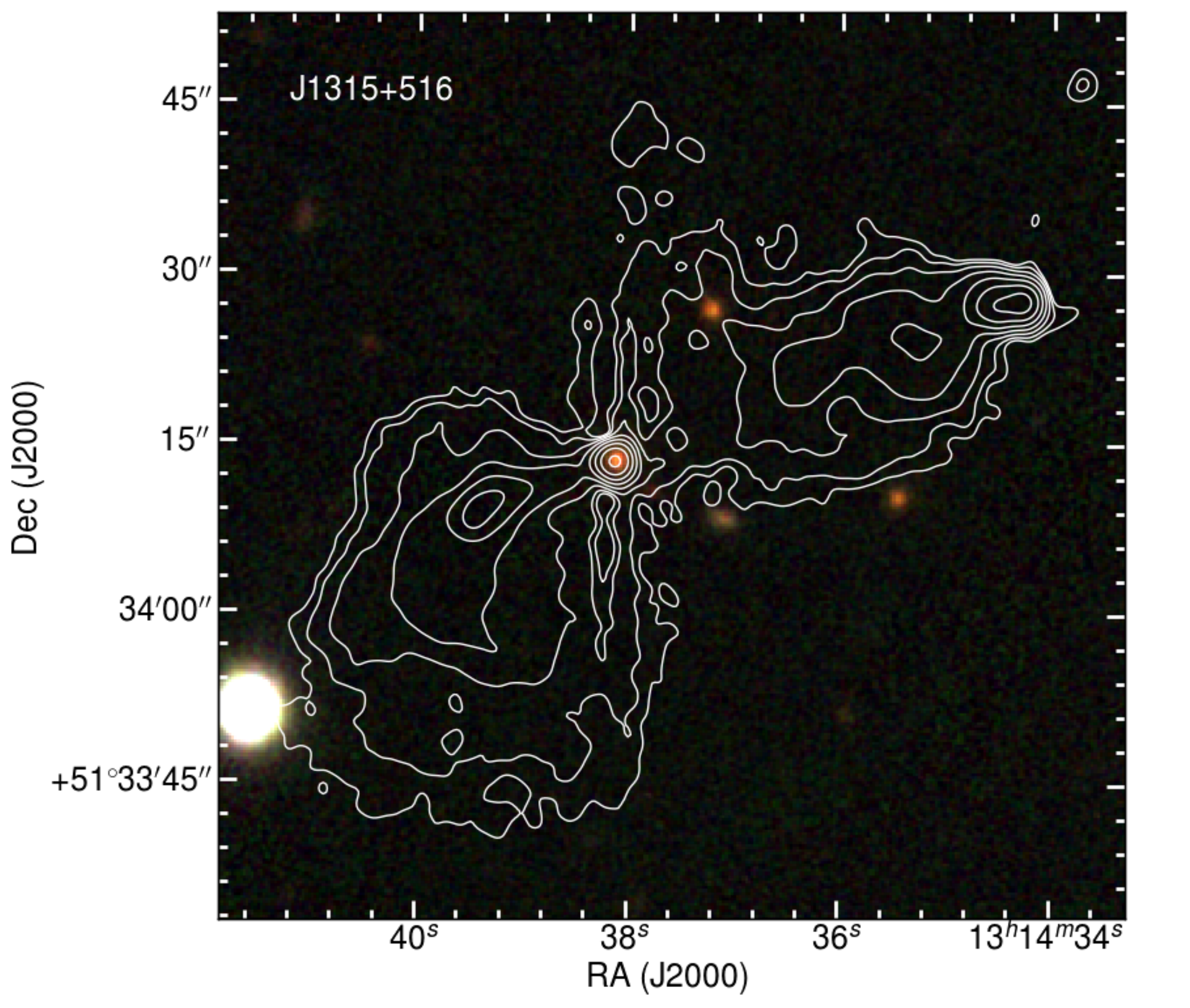}\\
\vspace{0.35cm}%
\includegraphics[angle=0,height=6.9cm]{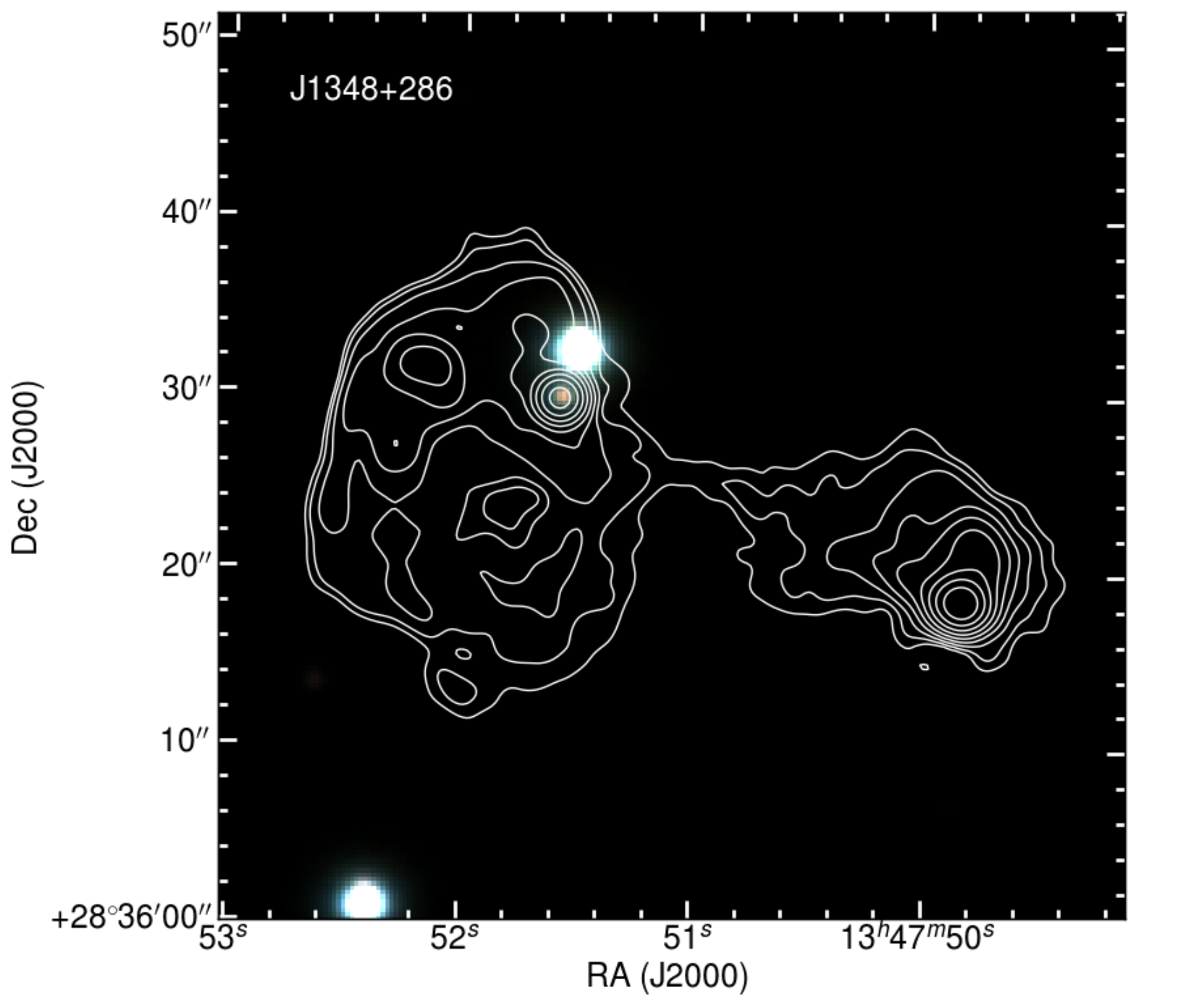}
\caption{L-band 1.5 GHz flux contours overlaid on optical RGB images. The optical image uses the $i$, $r$ and $g$ bands from the first data release of Pan-STARRS \citep{flewelling16}. The host galaxies are unequivocally identified as the sources that coincide with the location of the radio core. It is highly unlikely that the radio core and the galaxy are two unrelated sources seen in projection.}
\label{opticaloverlays}
\end{figure*}

\begin{table*}
\centering
\caption{Model Fitting Results}
\label{fittingresults}
\begin{tabular}{ccccccccccccccc}
\hline
\hline

Source&Lobe&FR&$\sum \chi^{2}$&Mean $\chi^{2}$&\multicolumn{5}{c}{Confidence bins}&Rejected&Median&Max age&+&-\\
name&&type&&&$<$ 68&68 - 90&90 - 95&95 - 99&$\geq$ 99&&Conf.&(Myr)&&\\

\hline
J1154$+$513 &	West	&   II  &	14885	&	40.89	&	34	    &	44	&	14	&	44	&	288 &	Yes	&	$>$ 99	&   5.40    &   0.60    &   0.53\\
	        &	East	&   I   &   10880	&	7.15	&	1243    &	126	&	29	&	40	&	83	&	No	&	$<$ 68	&   7.66    &   0.99    &   0.89\\
J1206$+$503   &	West	&   I   &	25398	&	41.98	&	106	    &	43	&	13	&	35	&	408 &	Yes	&	$>$ 99	&   5.50    &   0.94    &   0.69\\
	        &	East	&   II  &   7137	&	17.80	&	154     &	74	&	35	&	70	&	68	&	No	&	$<$ 68	&   4.01    &   0.49    &   0.41\\
J1313$+$507   &	East	&   II  &	8070	&	9.40	&	544	    &	117	&	53	&	64	&	81  &	No	&	$<$ 68	&   10.79   &   1.61    &   1.80\\
	        &	West	&   I   &   3113	&	5.01	&	398     &	105	&	35	&	46	&	37	&	No	&	$<$ 68	&   13.01   &   1.42    &   1.22\\
J1315$+$516   &	East	&   I   &	5828	&	1.37	&	4203    &	52	&	0	&	0	&	0   &	No	&	$<$ 68	&   19.49   &   4.50    &   2.56\\
	        &	West	&   II  &   4535	&	1.20	&	3743    &	23	&	3	&	1	&	0	&	No	&	$<$ 68	&   18.59   &   3.93    &   2.92\\
J1348$+$286   &	East	&   I   &	54418	&	10.14	&	2294    &	1237&	530	&	857	&	447 &	No	&	$<$ 68	&   7.81    &   0.96    &   0.84\\
	        &	West	&   II  &   11852	&	5.20	&	1954    &	270	&	35	&	18	&	1	&	No	&	$<$ 68	&   7.19    &   0.87    &   0.89\\
\hline
\end{tabular}

\vskip 5pt
\begin{minipage}{17.8cm}
`Model' refers to the spectral ageing model fitted to the target listed in the `Source' column.  $\sum \chi^{2}$ lists the sum of $\chi^{2}$ overall all regions with an equivalent mean value shown in the `Mean $\chi^{2}$' column. `Confidence Bins' lists the number of regions for which their $\chi^{2}$ values falls with the stated confidence range. `Rejected' lists whether the goodness-of-fit to the source as a whole can be rejected and `Median Conf.' the confidence level at which the model can be rejected.
\end{minipage}
\end{table*}

\subsection{What causes hybrid sources?}
\label{whatarehymors}

\subsubsection{Strongly asymmetric environment}
\label{asymmetricenvironment}

Arguably one of the most widely supported explanations for the formation of hybrid morphology sources based on previous studies is the presence of a strongly asymmetric environment. While no direct measurements of the environment are available, the findings of \citet{ceglowski13}, who use VLBI measurements to determine there is no discernible difference between the two inner jets, means such an environment is the most straight forward explanation. Under such conditions, the jet on one side is disrupted as it passes through a dense environment where as the other punches relatively unimpeded through a much weaker medium to form the FR I and FR II lobes respectively. However, the presence of FR II type hotspots revealed by the higher resolution observations presented in this paper make this scenario significantly less likely.

\citet{ceglowski13} find that the inner region of the sample displays FR II type jets on both sides of the core. We note that due to an error in the size scale calculated in the original paper, the observed jet lengths are on scale of 100-200 pc at the typical redshifts of the sample -- rather than the 1-10 pc originally stated -- a scale on which powerful FR I sources are typically observed to have well-collimated jets on both sides of the nucleus. Without further information, this makes such jets indistinguishable from those of FR IIs, but nevertheless limits the distance at which any jet disruption can occur to $\gtrsim 100$ pc.

Given the apparent compact nature of the acceleration regions, this leads to two possible scenarios if the newly discovered hotspots are to be formed intrinsically: the sources reside in relatively typical, albeit asymmetric, environment densities but the jet power is low; or the jet powers are typical of those found in FR II sources, but the environment densities are very high on one side. The luminosity of our sample $L_{178}\sim10^{27}$ W Hz$^{-1}$ sr$^{-1}$ all lie well above the 178 MHz FRI/II divide of $L_{178}\sim10^{25.5}$ W Hz$^{-1}$ sr$^{-1}$ \citep{fanaroff74} suggesting the jet powers of hybrid objects are similar to those of classical FR IIs, making the low power case unlikely. Under the high power, high density scenario, the presence of hotspots suggests that the jets retain their FR II type characteristics out to larger distances and are not disrupted or subject to recollimation shocks as would be the case for weak-flavour FR I jets. To create the conditions required to form a hybrid morphology source in this way, a steep density gradient would be needed to occur on scales of tens of kpc across the core region. This is a physically unlikely scenario for the large scale environment. Hence without strong empirical evidence of such conditions existing, we find this explanation unconvincing. While such an absolute difference in the large-scale external medium is therefore unlikely, a case for a localised difference could still be made in the case of restarted sources. We discuss this possibility further in the next section.

\subsubsection{Restarted sources}
\label{restarted}

One way in which the problems associated with a strong density asymmetry in the large-scale environment may be resolved is if it is only the environment local to the lobes which displays such an imbalance. Such differences are significantly easier to produce as they can occur on much shorter timescales as a result of the processes within the radio galaxy itself. One possibility for creating such conditions is the collapse of the lobe region cleared by an episode of previous active galactic nucleus (AGN) activity on one side of the source. In such a scenario, an initial phase of activity forms a classical radio galaxy before the central engine eventually switches off, leaving an adiabatically expanding low density remnant lobe. If, due to the external pressure, one side of the lobe was to collapse being replaced by the denser external environment, a second episode of AGN activity would result in significantly differing advance speeds for the newly formed acceleration regions.

From Figure \ref{intensitycombined} we see that three of our sample (J1154+513, J1206+503, J1313+507) show a bimodal brightness distribution in the FR I type lobe, with the bright primary acceleration region close to the core, a reduction in brightness towards the middle of the source, and a secondary peak of emission near the edge of the lobe. There is therefore some evidence to suggest that the sources within our sample could plausibly result from multiple episodes of activity; however, the morphology of the remaining two sources and the spectral age distribution of the entire sample is less simple to explain in the context of a restarted AGN.

The morphology of J1315+516 and J1348+286 show no signs of such a bimodal brightness distribution, with the morphology of J1315+516 being particularly unlike most restarted sources. While this does not completely exclude the possibility of being restarted, the spectral age distribution of the full sample (Figure \ref{ageingmaps}) shows little evidence of remnant emission. While J1154+513 does show a distinctive jump in characteristic age between the inner and outer regions of the FR I lobe, the age distribution of the remaining four sources is consistent with a single episode of activity. Strong mixing of the remnant and current electron populations would therefore be required to explain the observed distribution. Given that a condition of forming the lobe asymmetry through restarted sources is a high local density in the particle acceleration region, the strong flow of plasma towards the edge of the lobe that would be required to form such conditions seems unlikely. Combined with their relatively young age ($T_{avg} \approx 10$ Myr) meaning that the duty cycle and collapse of the cavity cleared by a previous episode of emission must occur on relatively short timescales\footnote{Mixing of fresh plasma with remnant emission would increase the steepening of the spectrum, meaning the characteristic ages would be an upper limit under such conditions.}, therefore leads us to believe that restarted sources are not the cause of hybrid morphology objects.

\subsubsection{Orientation}
\label{orientation}

One possibility yet to be considered is the effects of orientation that can significantly impact the observed morphology and spectrum of a source. As discussed in Sections \ref{accelerationregions} and \ref{restarted}, electron population mixing due to line-of-sight can significantly alter both the observed curvature and spatial distribution of the spectrum but, in the case of orientation, without the need for tightly constrained conditions intrinsic to the source.

Using the code of Martin Krause (private communication) which is based on the relativistic aberration formulas of \citet{gower82} and assuming both lobes have an intrinsically FR II morphology, the simplest case of directing one jet towards the observer can comfortably account for difference in distance between the two acceleration regions through light travel time effects. However, even with precession it cannot fully account for the lobes extending well beyond the acceleration regions or for spatial offsets in which the two lobes and core are not aligned (e.g. J1154+513). Adding jet precession can to some extent improve on these problems, particularly in the context of spatial offsets, but again cannot fully account for the lobes' extended morphology.

One way in which the observed morphology could be formed in the case of two intrinsically FR II type lobes is if the inner jet, hotspots, and diffuse lobe emission are not parallel. This can be achieved either through the heavy bending of the jet itself, the direction of the lobes being at an angle to a lightly-bent jet, or some combination of the two. Such bent-jet radio galaxies are more commonly observed for FR I type morphologies (known as wide angle tail, WATs, in their FR I form) and are thought to be the result of the lobes being swept back by the relative motion of the surrounding medium in a dense environment, such as those found in clusters (e.g. \citealp{begelman79, sakelliou00}). Recent surveys have shown that radio galaxies possess a wide range of unusual morphologies (e.g. \citealp{shimwell19}) with bent-jet FR IIs having now being observed in the plane of the sky (e.g. J0827+464 \citealp{missaglia19}). For such sources, one would expect offsets between the hosts/radio cores with respect to the position of the lobes to be common, a feature supported for our sample by Figure \ref{opticaloverlays}, which unambiguously identifies the hosts as coincident with the bright core and offset from the projected position of the lobes. Such conditions are therefore plausible for our sample of rare hybrid objects.

\begin{figure*}
\centering
\includegraphics[angle=0,width=17.6cm]{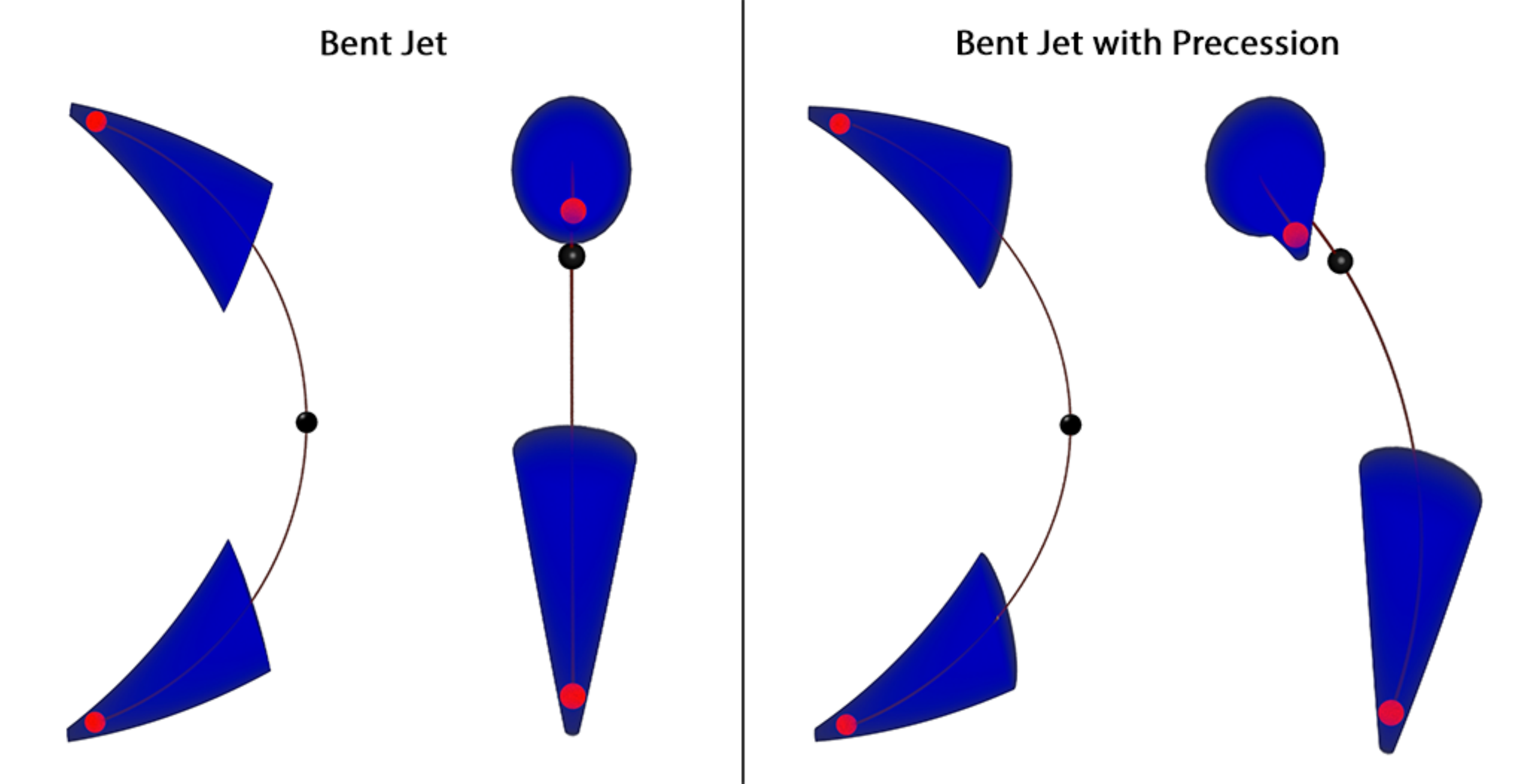}\vspace{10mm}//
\includegraphics[angle=0,width=15.6cm]{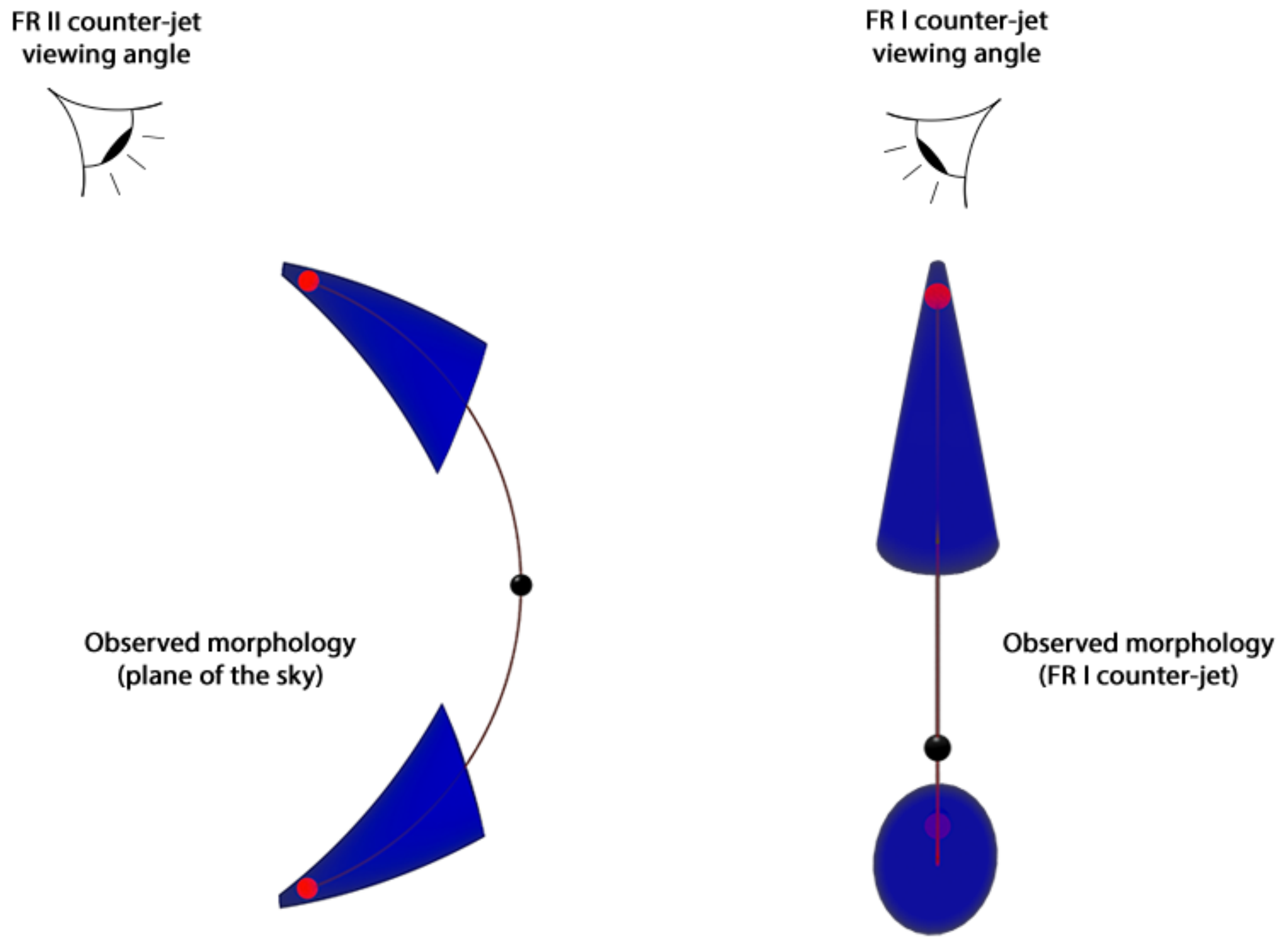}
\caption{Simple orientation models for hybrid morphology sources. Top: hotspots and core aligned (left) and offset (right) with the apparent FR I lobes orientated towards the observer. Bottom: Apparent FR I lobes orientated away from the observer where the "eye" symbol shows the viewing angles for the left hand model. Note precession also can be introduced in this orientation to provide the same hotspots/core offset effects as above.}
\label{orientationmodels}
\end{figure*}

The precession code discussed above cannot currently account for jet curvature or lobes not parallel to the jet; however, using a 3-dimensional toy model of a bent-jet FR II, orientating towards the line-of-sight can reproduce a morphology in which the lobe emission extends well beyond the hotspot (Figure \ref{orientationmodels}, left). Similarly, the spatial offset between the hotspots and core can also be replicated by introducing a simple precession to the jet (Figure \ref{orientationmodels}, right). While such models do not account for relativistic effects, using the precession code to replicate the jet angle for this model we find this is unlikely to have a significant impact on our results as only large angles to the line-of-sight ($>\!45\degree$) and only relatively minor precession are required to reproduce the observed morphology. These large angles to the line-of-sight also satisfy the criteria set by the VLBI observations of \citet{ceglowski13}, who limit the orientation of the inner jet to $>\!21\degree$, the point at which the counter-jet no longer becomes visible.

Such a straight-forward interpretation of the complex morphology of hybrid morphology sources not only provides a solution that is not bound by the tight environmental constraints discussed in Sections \ref{asymmetricenvironment} and \ref{restarted}, but can also explain the unusual lobe and acceleration region spectra and its spatial distribution. For such orientations, the spectrum of the acceleration regions of an intrinsically strong-flavour FR II jets are a superposition of freshly accelerated plasma and the aged electron populations present in the lobe. As discussed in Section \ref{accelerationregions}, such population mixing can explain the unusual $\chi^{2}$ profile when determining the injection index values as well as the particularly steep values such those observed in J1154+513 ($\alpha_{inj}>0.9$) and occurs naturally as as result of the apparent position of the hotspots relative to the lobes in such models.

In the same way, the spectrum of the lobes will be also affected by line-of-sight mixing. The extent to which the spectrum is affected will depend on the exact orientation and intrinsic morphology of the lobe, which cannot be determined here, but can range from a slight departure resulting in unreliable characteristic ages, to a complete rejection of standard spectral ageing models. The morphology of the lobes are unlikely to be as simple as the toy models presented here and are observed to have a range of shapes and smaller scale features such as clumpy emission and filaments (e.g. atlas of DRAGNs: Leahy, Bridle \& Strom\footnote{http://www.jb.man.ac.uk/atlas/}). Combined with the finding that the spatial distribution of spectral ages is often asymmetric along the jet axis (e.g. \citealp{harwood13, harwood15, harwood17a}), this means that the exact spectrum will depend on the geometry and dynamics of each lobe. For example, if a low-age region of plasma close to the edge of the lobe lies coincident with a much older, more extended region sharp age gradients may be observed such as appear in J1154+513. Similarly, the superposition of clumpy and/or filamentary structure in the lobes could lead to similar features in projection that do not manifest themselves spatially as would be expected if one were to assume the lobe is approximately in the plane of the sky. Such conditions could also potentially explain the secondary low age/intensity peaks in the outer regions of the FR I lobes detailed in Section \ref{spectralfitting} that are particularly prominent in J1154+513; however, simulations would be required to determine the viability of this scenario.

J1348+286 provides a particularly interesting case, with a clear lack of low-age plasma coincident with the intensity peak that is usually associated with the particle acceleration region. So far, we have only considered an orientation in which the counter-jet produces the FR II side of the source, with the jet pointing towards the observer resulting in the apparent FR I morphology. However, the reverse of this orientation can -- and indeed should -- also be present within the radio galaxy population. From Figure \ref{orientationmodels} we see that a model with the same starting conditions as previously used but with the viewing angle rotated $\!180\degree$ along the x-axis results in the same observational effect, but with the apparent FR I lobe being associated with the counter-jet. We critically note that the hotspot spectrum under such conditions can become severely affected by superposition with a large volume of heavily aged plasma, as it is viewed through a significant fraction of the lobe emission. This is likely to lead to a more diffuse, steep spectrum acceleration region such as is observed in J1348+286. Of the two sources in which both jets are clearly detected by \citet{ceglowski13} the counter-jet of J1348+286 is directed towards the FR I lobe, compared to J1154+513 in which it is directed towards the apparent FR II lobe, consistent with the orientation base interpretation described above.

While heavily bent jets are used in the toy models presented here, as noted above lobes lying at an angle to a much straighter jet can achieve the same effect. For the majority of our sample, determining the host galaxy properties is currently not possible due to their low surface brightness and the limited number of Sloan Digital Sky Survey (SDSS) bands available. However for two sources, J1315+516 and J1348+286, optical classifications have been made as part of SDSS Data Resease 15 \citep{aguado19}. These spectra suggest that both have a quasar host, hence are orientated more towards the observer than in the plane of the sky. The 3 arcsecond fibre size of SDSS means that these classifications should be viewed with some caution, particularly for J1348+286 where another bright galaxy is located nearby. Due to the degeneracy between viewing angle, lobe angle, and jet curvature it is therefore not possible with the data currently available to determine tight constraints on these free parameters, but the morphological and spectral effects remain similar. Follow-up Gemini observations will confirm the initial quasar classification as well as determine if this holds for the rest of the sample (Stroe, Harwood and Vernstrom, in prep).

A number of testable predictions can be made for this orientation based model to determine its validity, particularly when applied to larger samples. In the case of jets inclined towards the observer signatures of beaming should be apparent. One method for the detection of such effects is the ratio of core to total emission (core prominence) which is known to have a median value for the 3CRR objects \citep{laing83} at 1.4 GHz of $S_{core}/S_{tot} \sim 3 \times 10^{-4}$ \citep{mullin08}. From Tables \ref{targets} and \ref{combinedsummary} we see that our sample has core prominence values ranging from $2.4 \times 10^{-2}$ (J1206+503) to $1.3 \times 10^{-1}$ (J1348+286), well above of the 3CRR sample. As noted above, the lack of jet asymmetry seen in VLBI observations for most sources excludes very small angles to the line of sight, but such beaming effects suggest that moderate angles are present in the sample, consistent with an orientation based interpretation.

The aspect ratio of the lobes (length/width) should also provide a testable feature for an orientation based interpretation. For the bent-jet models presented in Figure \ref{orientationmodels}, the aspect ratio of the apparent FR I lobe should be systematically higher than the FR II side. From Figure \ref{intensitycombined} it is clear that for J1315+516 and J1348+286 such an aspect ratio difference between the two lobes exists; however, for the remaining three sources this is far less clear. Such ratio differences are affected by both the unknown viewing angle and intrinsic source morphology which may vary significantly in the presence of a dense environment and so the expected aspect ratio will likely be systematic over large samples, rather than applicable to individual sources. We discuss the wider population and surveys to which such tests will be applied further in Section \ref{population}.

Another testable feature of the bent-jet model is a distinct asymmetry in the polarisation between the two lobes. Under such conditions, the receding lobe should be more depolarized than its advancing counterpart due to emission having to pass through a larger amount of plasma in the intergalactic medium. The expected difference again relies heavily on the exact orientation, jet curvature, and lobe morphology of the source but should be present in observations of hybrid morphology objects. A detailed analysis of the polarization properties of our sample is outside the scope of this paper, but will be the subject of future investigation (Vernstrom, Harwood and Stroe, in prep). 

It is clear from the discussion above that quantifying the dependency between orientation, jet curvature, and lobe morphology is key to providing quantitative predictions and determining their intrinsic morphology. Simulations are therefore likely to be vital in determining the most likely parameters for the sources in our sample (Shabala \& Harwood, in prep) which can subsequently be tested on larger, more uniform samples of HyMoRS. However, regardless of the detailed parameters, the ability of the orientation models presented here to explain all of the unusual morphological and spectral features observed in our sample provides both the simplest and most complete explanation for the cause of the hybrid morphology sources in our sample.

\subsubsection{The hybrid morphology population}
\label{population}

While the orientation based explanation described above holds well for the sample presented here, there remains the question of whether this is true for the wider population of hybrid morphology objects. One simple metric for determining its viability is whether the population of bent-jet FR II radio galaxies as viewed in the plane of the sky if sufficient to support the number of HyMoRS given a homogeneous distribution of viewing angles. The largest published sample of HyMoRS is currently that of \citet{kapinska17} who present 25 new candidates found in the $\sim$$170000$ radio sources of Radio Galaxy Zoo data release 1, giving a candidate fraction of $0.015$ per cent. From the original 21 candidates found in FIRST by \citet{gawronski06} only 5 were confirmed through $4.9$ GHz follow-up observations suggesting that this fraction may be significantly lower; however, as both of these searches are based on the same survey they are likely to suffer from similar limitations. 

Due to the FIRST survey using only 3 minutes snapshots in the VLA B configuration, diffuse emission is often not observed. If the acceleration regions are the result of compact hotspots as suggested by the bent-jet model, a fraction of the sample will be missing significant emission from the lobes -- particularly from the apparent FR I lobe -- and the source may instead being classified as e.g. and asymmetric FR II, or in the most extreme cases as unassociated point sources. Such values should therefore be taken as a lower limit rather than definitive values. Similar problems occur when attempting to determine the total number of bent-jet FR II in the plane of the sky in such surveys. As the hotspots and core are necessarily misaligned, identification of these sources is difficult without sufficient diffuse emission to confidently associate the three components, meaning the total number of bent-jet FR II therefore currently remains unknown. The short baselines, low frequency, and resolution provided by LOFAR \citep{haarlem13} mean that the forthcoming LOFAR Two Meter Sky Survey (LoTSS) data release 2 will help to resolve these issues by providing a large sample of sources in which complex morphology objects can be confidently identified and will therefore be the subject of future work to test the prevalence of both bent-jet FR IIs and hybrid morphology sources (Harwood and Mingo, in prep).

While we are therefore currently unable to determine reliable statistics on the hybrid population, comparisons can be made to other known HyMoRS with respect to the robustness of the bent-jet model. Comparing the findings of \citet{kapinska17} we see that the majority have a similar morphology to our sample as observed in FIRST, with 19 of the 25 sources appearing to contain a compact acceleration region that is disconnected from the core (figures 1 and 2 of \citealp{kapinska17}). The remaining 6 sources lack either the resolution or extended emission required to determine if such conditions exist. While follow up observations would be required to robustly confirm such structure, preliminary exploration of the LOFAR Two Meter Sky Survey value added catalogue (LoTSS; \citealp{shimwell19, williams19}) using the automated morphological classification of \citet{mingo19} suggests that those sources where structure can be confidently identified contain similar morphological features to those presented here. It is therefore likely that this orientation based explanation can be extended to explain a significant fraction of hybrid population.

While the bent-jet model described above can therefore account for a large fraction of the known hybrid population, some alternative explanations, while not being directly applicable to our sample, cannot be ruled out and may be applicable when exploring larger (and potentially less biased) samples. One such possibility is a population of intrinsically FR II radio galaxies that are orientated close to the line-of-sight in which the inner jet outshines the near-side hotspot. Under such conditions the approaching jet would be classified as FR I, but is likely to result in a more extended acceleration region that may be more closely connected to the core. This explanation is unlikely to be the cause for the majority of our sample due to the compact nature of the acceleration regions and strong beaming is not found through VLBI observation \citep{ceglowski13} and so the a model in which the jet is more strongly bent would be favoured. However, a case could be made that such effects partially impact J1315+516 given the more extended acceleration region, the core prominence shown in Section \ref{orientation}, and the tenuous detection of a quasar spectrum for its host galaxy although either jet-bending and/or non-aligned lobes would still be required to explain the diffuse emission well beyond the jet termination point. For a larger sample of HyMoRS in which such conditions are suspected and optical data is available to confirm their host spectrum, such a scenario should be relatively straight forward to test as the observed asymmetry between core and acceleration regions should be well described by relativistic effects.

Another plausible scenario is one in which HyMoRS are observed at similar viewing angles to those proposed for the bent-jet model, but the lobes have been swept away from the core such as is observed in WATs creating a bent FR I type morphology on both sides of the source. WATs are known to have a wide range of unusual morphologies and are commonly observed to have acceleration regions disconnected from the core when in the plane of the sky (e.g. 3C465; \citealp{odea85}). Simple modelling of such a morphology in the same manner used in Section \ref{orientation} suggests that, while plausible, for an apparent FR II lobe to be observed requires the source to be at such an orientation that significant emission should be observed extending across the core region; a feature that not seen in our sample. However, this does not exclude the possibility of such sources and, given that a significant number of WATs are know to exist in the plane of the sky, should make up some fraction of the hybrid population when larger samples are explored.
 
While a number of other niche cases, such as remnant radio galaxies in an asymmetric environment, are also possible such explanations generally require a narrow set of conditions to be viable and are therefore not likely to make up a significant fraction of the overall population. We therefore conclude that while the exact fraction of each scenario cannot yet be determined, the orientation of an intrinsically FR II type jet -- whether heavily bent, a WAT, or inner jet dominated -- is likely to be the primary cause for the hybrid morphology population.

\section{Conclusions}
\label{conclusions}

In this paper we have presented the first resolved spectral study at radio frequencies of hybrid morphology sources (HyMoRS), recovering both compact and diffuse emission on arcsecond scales. Using the sample of \citet{gawronski06}, we have uncovered the kpc scale morphology, spectrum and underlying cause of these unusual sources. Addressing the primary science questions posed in Section \ref{specageintro} we find with respect to the morphology and spectrum of hybrid morphology objects that:

\begin{enumerate}
\item The morphology of the apparent FR I lobes are centre brightened as expected.
\begin{itemize}
\item[--] The centre brightening is the result of a compact, hotspot-like acceleration region well separated from the core.\\
\item[--] The morphology and spectrum suggest that the acceleration regions are the result of FR II (strong-flavour) jets.
\end{itemize}
\item The spatial distribution of the resolved spectrum is not similar to most classical radio galaxy lobes.
\begin{itemize}
\item[--] The unusual spectrum is most likely the result of line-of-sight electron population mixing.
\end{itemize}
\end{enumerate}

\noindent We also address point (iii) of our primary science questions, the underlying cause of hybrid morphology objects, concluding that:

\begin{itemize}
\item Hybrid morphology objects are most likely the result of orientation and are intrinsically FR II radio galaxies.\\
\item The underlying radio galaxies are expected to contain either bent jets, lobes lying at an angle to a straight jet, or some combination of the two.\\
\item The orientation models presented can explain both the morphological and spectral features observed in both this paper, and previous investigations.\\
\item The apparent FR I lobe can result from either the approaching or counter-jet, which have distinct features in both their intensity and spectrum.\\
\item Other models considered, including large scale environmental asymmetry and restarted sources, cannot account for all spectral features observed and require a restrictive range of conditions to form the required morphology.
\end{itemize}

In order to determine the detailed parameters and conditions under which hybrid morphology sources exist, further investigations will be required, including simulations and optical observations to determine host galaxy properties and AGN type. Environmental and clustering information -- gained from for example X-ray observations -- will also be key in confirming our non-parallel jet/lobe interpretation. Regardless of the detailed mechanics of HyMoRS, the orientation interpretation presented here currently provides the best explanation for the cause of hybrid morphology objects with the ability to provide a plausible explanation for all of the observed spectral and morphological features.

\section*{Acknowledgements}
\label{acknowledgements}
We would like to thank Stas Shabala, Martin Hardcastle, Judith Croston and Beatriz Mingo for useful discussions throughout the course of this project and the anonymous referee for their constructive comments which have helped improve the paper. JH gratefully acknowledges support of the University of Hertfordshire's Vice Chancellor's research fellowship. AS gratefully acknowledges support of a Clay Fellowship. We wish to thank staff of the NRAO Jansky Very Large Array of which this work makes heavy use. The National Radio Astronomy Observatory is a facility of the National Science Foundation operated under cooperative agreement by Associated Universities, Inc. This work has made use of the University of Hertfordshire Science and Technology Research Institute high-performance computing facility.

\bibliographystyle{mnras}
\bibliography{hymors}

\newcommand{\noop}[1]{}
\begin{thebibliography}{}
\makeatletter
\relax
\def\mn@urlcharsother{\let\do\@makeother \do\$\do\&\do\#\do\^\do\_\do\%\do\~}
\def\mn@doi{\begingroup\mn@urlcharsother \@ifnextchar [ {\mn@doi@}
  {\mn@doi@[]}}
\def\mn@doi@[#1]#2{\def\@tempa{#1}\ifx\@tempa\@empty \href
  {http://dx.doi.org/#2} {doi:#2}\else \href {http://dx.doi.org/#2} {#1}\fi
  \endgroup}
\def\mn@eprint#1#2{\mn@eprint@#1:#2::\@nil}
\def\mn@eprint@arXiv#1{\href {http://arxiv.org/abs/#1} {{\tt arXiv:#1}}}
\def\mn@eprint@dblp#1{\href {http://dblp.uni-trier.de/rec/bibtex/#1.xml}
  {dblp:#1}}
\def\mn@eprint@#1:#2:#3:#4\@nil{\def\@tempa {#1}\def\@tempb {#2}\def\@tempc
  {#3}\ifx \@tempc \@empty \let \@tempc \@tempb \let \@tempb \@tempa \fi \ifx
  \@tempb \@empty \def\@tempb {arXiv}\fi \@ifundefined
  {mn@eprint@\@tempb}{\@tempb:\@tempc}{\expandafter \expandafter \csname
  mn@eprint@\@tempb\endcsname \expandafter{\@tempc}}}

\bibitem[\protect\citeauthoryear{{Aguado} et~al.,}{{Aguado}
  et~al.}{2019}]{aguado19}
{Aguado} D.~S.,  et~al., 2019, \mn@doi [\apjs] {10.3847/1538-4365/aaf651},
  \href {https://ui.adsabs.harvard.edu/abs/2019ApJS..240...23A} {240, 23}

\bibitem[\protect\citeauthoryear{Alexander}{Alexander}{1987}]{alexander87a}
Alexander P.,  1987, MNRAS, 225, 27

\bibitem[\protect\citeauthoryear{Alexander \& Leahy}{Alexander \&
  Leahy}{1987}]{alexander87}
Alexander P.,  Leahy J.~P.,  1987, MNRAS, 225, 1

\bibitem[\protect\citeauthoryear{Avni}{Avni}{1976}]{avni76}
Avni Y.,  1976, ApJ, 210, 642

\bibitem[\protect\citeauthoryear{{Banfield} et~al.,}{{Banfield}
  et~al.}{2015}]{banfield15}
{Banfield} J.~K.,  et~al., 2015, \mn@doi [\mnras] {10.1093/mnras/stv1688},
  \href {https://ui.adsabs.harvard.edu/#abs/2015MNRAS.453.2326B} {453, 2326}

\bibitem[\protect\citeauthoryear{{Begelman}, {Rees}  \& {Blandford}}{{Begelman}
  et~al.}{1979}]{begelman79}
{Begelman} M.~C.,  {Rees} M.~J.,   {Blandford} R.~D.,  1979, \mn@doi [\nat]
  {10.1038/279770a0}, \href
  {https://ui.adsabs.harvard.edu/abs/1979Natur.279..770B} {279, 770}

\bibitem[\protect\citeauthoryear{Brienza et~al.,}{Brienza
  et~al.}{2016}]{brienza16}
Brienza M.,  et~al., 2016, A\&A, 585, 29

\bibitem[\protect\citeauthoryear{Carilli, Perley, Dreher  \& Leahy}{Carilli
  et~al.}{1991}]{carilli91}
Carilli C.,  Perley R.,  Dreher J.,   Leahy J.,  1991, ApJ, 383, 554

\bibitem[\protect\citeauthoryear{{Ceg{\l}owski}, {Gawro{\'n}ski}  \&
  {Kunert-Bajraszewska}}{{Ceg{\l}owski} et~al.}{2013}]{ceglowski13}
{Ceg{\l}owski} M.,  {Gawro{\'n}ski} M.~P.,   {Kunert-Bajraszewska} M.,  2013,
  \mn@doi [\aap] {10.1051/0004-6361/201220544}, \href
  {https://ui.adsabs.harvard.edu/#abs/2013A\&A...557A..75C} {557, A75}

\bibitem[\protect\citeauthoryear{{Celotti}, {Padovani}  \&
  {Ghisellini}}{{Celotti} et~al.}{1997}]{celotti97}
{Celotti} A.,  {Padovani} P.,   {Ghisellini} G.,  1997, \mn@doi [\mnras]
  {10.1093/mnras/286.2.415}, \href
  {https://ui.adsabs.harvard.edu/#abs/1997MNRAS.286..415C} {286, 415}

\bibitem[\protect\citeauthoryear{Cornwell}{Cornwell}{2008}]{cornwell08}
Cornwell T.~J.,  2008, JSTSP, 2, 793

\bibitem[\protect\citeauthoryear{Croston, Birkinshaw, Hardcastle  \&
  Worrall}{Croston et~al.}{2004}]{croston04}
Croston J.~H.,  Birkinshaw M.,  Hardcastle M.~J.,   Worrall D.~M.,  2004,
  MNRAS, 353, 879

\bibitem[\protect\citeauthoryear{Croston, Hardcastle, Harris, Belsole,
  Birkinshaw  \& Worrall}{Croston et~al.}{2005}]{croston05}
Croston J.~H.,  Hardcastle M.~J.,  Harris D.~R.,  Belsole E.,  Birkinshaw M.,
  Worrall D.~M.,  2005, ApJ, 626, 733

\bibitem[\protect\citeauthoryear{{Croston}, {Ineson}  \&
  {Hardcastle}}{{Croston} et~al.}{2018}]{croston18}
{Croston} J.~H.,  {Ineson} J.,   {Hardcastle} M.~J.,  2018, \mn@doi [\mnras]
  {10.1093/mnras/sty274}, \href
  {https://ui.adsabs.harvard.edu/abs/2018MNRAS.476.1614C} {476, 1614}

\bibitem[\protect\citeauthoryear{{Fanaroff} \& {Riley}}{{Fanaroff} \&
  {Riley}}{1974}]{fanaroff74}
{Fanaroff} B.~L.,  {Riley} J.~M.,  1974, \mn@doi [\mnras]
  {10.1093/mnras/167.1.31P}, \href
  {https://ui.adsabs.harvard.edu/#abs/1974MNRAS.167P..31F} {167, 31P}

\bibitem[\protect\citeauthoryear{{Flewelling} et~al.,}{{Flewelling}
  et~al.}{2016}]{flewelling16}
{Flewelling} H.~A.,  et~al., 2016, arXiv e-prints, \href
  {https://ui.adsabs.harvard.edu/abs/2016arXiv161205243F} {p. arXiv:1612.05243}

\bibitem[\protect\citeauthoryear{{Gawro{\'n}ski}, {Marecki},
  {Kunert-Bajraszewska}  \& {Kus}}{{Gawro{\'n}ski} et~al.}{2006}]{gawronski06}
{Gawro{\'n}ski} M.~P.,  {Marecki} A.,  {Kunert-Bajraszewska} M.,   {Kus} A.~J.,
   2006, \mn@doi [\aap] {10.1051/0004-6361:20053996}, \href
  {https://ui.adsabs.harvard.edu/#abs/2006A\&A...447...63G} {447, 63}

\bibitem[\protect\citeauthoryear{{Godfrey} et~al.,}{{Godfrey}
  et~al.}{2009}]{godfrey09}
{Godfrey} L.~E.~H.,  et~al., 2009, \mn@doi [ApJ] {10.1088/0004-637X/695/1/707},
  695, 707

\bibitem[\protect\citeauthoryear{{Gopal-Krishna} \& {Wiita}}{{Gopal-Krishna} \&
  {Wiita}}{2000}]{gopal00}
{Gopal-Krishna} {Wiita} P.~J.,  2000, \aap, \href
  {https://ui.adsabs.harvard.edu/#abs/2000A\&A...363..507G} {363, 507}

\bibitem[\protect\citeauthoryear{{Gopal-Krishna}, {Wiita}  \&
  {Hooda}}{{Gopal-Krishna} et~al.}{1996}]{gopal96}
{Gopal-Krishna} {Wiita} P.~J.,   {Hooda} J.~S.,  1996, \aap, \href
  {https://ui.adsabs.harvard.edu/#abs/1996A\&A...316L..13G} {316, L13}

\bibitem[\protect\citeauthoryear{Gower, Gregory, Unruh  \& B.}{Gower
  et~al.}{1982}]{gower82}
Gower A.~C.,  Gregory P.~C.,  Unruh W.~G.,   B. H.~J.,  1982, ApJ, 262, 478

\bibitem[\protect\citeauthoryear{Hardcastle \& Krause}{Hardcastle \&
  Krause}{2013}]{hardcastle13}
Hardcastle M.~J.,  Krause M. G.~H.,  2013, MNRAS, 430, 174

\bibitem[\protect\citeauthoryear{Hardcastle \& Looney}{Hardcastle \&
  Looney}{2001}]{hardcastle01}
Hardcastle M.~J.,  Looney L.~W.,  2001, MNRAS, 320, 355

\bibitem[\protect\citeauthoryear{Hardcastle, Alexander, Pooley  \&
  Riley}{Hardcastle et~al.}{1997}]{hardcastle97}
Hardcastle M.~J.,  Alexander P.,  Pooley G.~G.,   Riley J.~M.,  1997, MNRAS,
  288, 859

\bibitem[\protect\citeauthoryear{Hardcastle, Birkinshaw  \& Worrall}{Hardcastle
  et~al.}{1998}]{hardcastle98}
Hardcastle M.~J.,  Birkinshaw M.,   Worrall D.~M.,  1998, MNRAS, 294, 615

\bibitem[\protect\citeauthoryear{{Harwood}, {Hardcastle}, {Croston}  \&
  {Goodger}}{{Harwood} et~al.}{2013}]{harwood13}
{Harwood} J.~J.,  {Hardcastle} M.~J.,  {Croston} J.~H.,   {Goodger} J.~L.,
  2013, \mn@doi [\mnras] {10.1093/mnras/stt1526}, \href
  {https://ui.adsabs.harvard.edu/#abs/2013MNRAS.435.3353H} {435, 3353}

\bibitem[\protect\citeauthoryear{{Harwood}, {Hardcastle}  \&
  {Croston}}{{Harwood} et~al.}{2015}]{harwood15}
{Harwood} J.~J.,  {Hardcastle} M.~J.,   {Croston} J.~H.,  2015, \mn@doi
  [\mnras] {10.1093/mnras/stv2194}, \href
  {https://ui.adsabs.harvard.edu/#abs/2015MNRAS.454.3403H} {454, 3403}

\bibitem[\protect\citeauthoryear{{Harwood} et~al.,}{{Harwood}
  et~al.}{2016}]{harwood16}
{Harwood} J.~J.,  et~al., 2016, \mn@doi [\mnras] {10.1093/mnras/stw638}, \href
  {https://ui.adsabs.harvard.edu/#abs/2016MNRAS.458.4443H} {458, 4443}

\bibitem[\protect\citeauthoryear{{Harwood} et~al.,}{{Harwood}
  et~al.}{2017}]{harwood17a}
{Harwood} J.~J.,  et~al., 2017, \mn@doi [\mnras] {10.1093/mnras/stx820}, \href
  {https://ui.adsabs.harvard.edu/#abs/2017MNRAS.469..639H} {469, 639}

\bibitem[\protect\citeauthoryear{{Heesen}, {Croston}, {Harwood}, {Hardcastle}
  \& {Hota}}{{Heesen} et~al.}{2014}]{heesen14}
{Heesen} V.,  {Croston} J.~H.,  {Harwood} J.~J.,  {Hardcastle} M.~J.,   {Hota}
  A.,  2014, \mn@doi [\mnras] {10.1093/mnras/stu043}, \href
  {https://ui.adsabs.harvard.edu/#abs/2014MNRAS.439.1364H} {439, 1364}

\bibitem[\protect\citeauthoryear{{Heesen} et~al.,}{{Heesen}
  et~al.}{2018}]{heesen18}
{Heesen} V.,  et~al., 2018, \mn@doi [\mnras] {10.1093/mnras/stx2869}, \href
  {https://ui.adsabs.harvard.edu/#abs/2018MNRAS.474.5049H} {474, 5049}

\bibitem[\protect\citeauthoryear{{Ineson}, {Croston}, {Hardcastle}  \&
  {Mingo}}{{Ineson} et~al.}{2017}]{ineson17}
{Ineson} J.,  {Croston} J.~H.,  {Hardcastle} M.~J.,   {Mingo} B.,  2017,
  \mn@doi [\mnras] {10.1093/mnras/stx189}, \href
  {https://ui.adsabs.harvard.edu/#abs/2017MNRAS.467.1586I} {467, 1586}

\bibitem[\protect\citeauthoryear{Jaffe \& Perola}{Jaffe \&
  Perola}{1973}]{jaffe73}
Jaffe W.,  Perola G.,  1973, A\&A, 26, 423

\bibitem[\protect\citeauthoryear{{Kaiser} \& {Alexander}}{{Kaiser} \&
  {Alexander}}{1997}]{kaiser97}
{Kaiser} C.~R.,  {Alexander} P.,  1997, \mn@doi [\mnras]
  {10.1093/mnras/286.1.215}, \href
  {https://ui.adsabs.harvard.edu/#abs/1997MNRAS.286..215K} {286, 215}

\bibitem[\protect\citeauthoryear{{Kapi{\'n}ska} et~al.,}{{Kapi{\'n}ska}
  et~al.}{2017}]{kapinska17}
{Kapi{\'n}ska} A.~D.,  et~al., 2017, \mn@doi [\aj] {10.3847/1538-3881/aa90b7},
  \href {https://ui.adsabs.harvard.edu/abs/2017AJ....154..253K} {154, 253}

\bibitem[\protect\citeauthoryear{Kardashev}{Kardashev}{1962}]{kardashev62}
Kardashev N.~S.,  1962, AJ, 6, 317

\bibitem[\protect\citeauthoryear{{Krolik} \& {Chen}}{{Krolik} \&
  {Chen}}{1991}]{krolik91}
{Krolik} J.~H.,  {Chen} W.,  1991, \mn@doi [\aj] {10.1086/115985}, \href
  {https://ui.adsabs.harvard.edu/\#abs/1991AJ....102.1659K} {102, 1659}

\bibitem[\protect\citeauthoryear{Laing \& Bridle}{Laing \&
  Bridle}{2002}]{laing02}
Laing R.~A.,  Bridle A.~H.,  2002, MNRAS, 336, 328

\bibitem[\protect\citeauthoryear{Laing \& Bridle}{Laing \&
  Bridle}{2013}]{laing13}
Laing R.~A.,  Bridle A.~H.,  2013, MNRAS, 432, 1114

\bibitem[\protect\citeauthoryear{Laing, Riley  \& Longair}{Laing
  et~al.}{1983}]{laing83}
Laing R.~A.,  Riley J.~M.,   Longair M.~S.,  1983, MNRAS, 204, 151

\bibitem[\protect\citeauthoryear{{Laing}, {Parma}, {de Ruiter}  \&
  {Fanti}}{{Laing} et~al.}{1999}]{laing99}
{Laing} R.~A.,  {Parma} P.,  {de Ruiter} H.~R.,   {Fanti} R.,  1999, \mn@doi
  [\mnras] {10.1046/j.1365-8711.1999.02548.x}, \href
  {https://ui.adsabs.harvard.edu/#abs/1999MNRAS.306..513L} {306, 513}

\bibitem[\protect\citeauthoryear{{Laing}, {Bridle}  \& {Canvin}}{{Laing}
  et~al.}{2007}]{laing07}
{Laing} R.~A.,  {Bridle} A.~H.,   {Canvin} J.~R.,  2007, in {Aschenbach} B.,
  {Burwitz} V.,  {Hasinger} G.,   {Leibundgut} B.,  eds, Relativistic
  Astrophysics Legacy and Cosmology - Einstein's Legacy. p.~445,
  \mn@doi{10.1007/978-3-540-74713-0_102}

\bibitem[\protect\citeauthoryear{{Laing}, {Guidetti}, {Bridle}, {Parma}  \&
  {Bondi}}{{Laing} et~al.}{2011}]{laing11}
{Laing} R.~A.,  {Guidetti} D.,  {Bridle} A.~H.,  {Parma} P.,   {Bondi} M.,
  2011, \mn@doi [\mnras] {10.1111/j.1365-2966.2011.19436.x}, \href
  {https://ui.adsabs.harvard.edu/abs/2011MNRAS.417.2789L} {417, 2789}

\bibitem[\protect\citeauthoryear{{Mingo} et~al.,}{{Mingo}
  et~al.}{2019}]{mingo19}
{Mingo} B.,  et~al., 2019, \mn@doi [\mnras] {10.1093/mnras/stz1901}, \href
  {https://ui.adsabs.harvard.edu/abs/2019MNRAS.488.2701M} {488, 2701}

\bibitem[\protect\citeauthoryear{{Missaglia}, {Massaro}, {Capetti}, {Paolillo},
  {Kraft}, {Baldi}  \& {Paggi}}{{Missaglia} et~al.}{2019}]{missaglia19}
{Missaglia} V.,  {Massaro} F.,  {Capetti} A.,  {Paolillo} M.,  {Kraft} R.~P.,
  {Baldi} R.~D.,   {Paggi} A.,  2019, \mn@doi [\aap]
  {10.1051/0004-6361/201935058}, \href
  {https://ui.adsabs.harvard.edu/abs/2019A&A...626A...8M} {626, A8}

\bibitem[\protect\citeauthoryear{{Mu{\~n}oz}, {Falco}, {Kochanek}, {Leh{\'a}r}
  \& {Mediavilla}}{{Mu{\~n}oz} et~al.}{2003}]{munoz2013}
{Mu{\~n}oz} J.~A.,  {Falco} E.~E.,  {Kochanek} C.~S.,  {Leh{\'a}r} J.,
  {Mediavilla} E.,  2003, \mn@doi [\apj] {10.1086/377077}, \href
  {http://adsabs.harvard.edu/abs/2003ApJ...594..684M} {594, 684}

\bibitem[\protect\citeauthoryear{{Mullin}, {Riley}  \& {Hardcastle}}{{Mullin}
  et~al.}{2008}]{mullin08}
{Mullin} L.~M.,  {Riley} J.~M.,   {Hardcastle} M.~J.,  2008, \mn@doi [\mnras]
  {10.1111/j.1365-2966.2008.13534.x}, \href
  {https://ui.adsabs.harvard.edu/abs/2008MNRAS.390..595M} {390, 595}

\bibitem[\protect\citeauthoryear{{O'Dea} \& {Owen}}{{O'Dea} \&
  {Owen}}{1985}]{odea85}
{O'Dea} C.~P.,  {Owen} F.~N.,  1985, \mn@doi [\aj] {10.1086/113802}, \href
  {https://ui.adsabs.harvard.edu/abs/1985AJ.....90..954O} {90, 954}

\bibitem[\protect\citeauthoryear{Owen \& Ledlow}{Owen \& Ledlow}{1994}]{owen94}
Owen F.~N.,  Ledlow M.~J.,  1994, in Bicknell G.~V.,  Dopita M.~A.,   Quinn
  P.~J.,  eds,  ASP Conference Series Vol. 54, \emph{`The first Stromlo
  symposium: The physics of active galaxies'}.

\bibitem[\protect\citeauthoryear{Pacholczyk}{Pacholczyk}{1970}]{pacholczyk70}
Pacholczyk A.~G.,  1970, Radio astrophysics. Nonthermal processes in galactic
  and extragalactic sources.
San Francisco, Freeman

\bibitem[\protect\citeauthoryear{Perley \& Butler}{Perley \&
  Butler}{2013}]{perley13}
Perley R.~A.,  Butler B.~J.,  2013, ApJ, 204, 19

\bibitem[\protect\citeauthoryear{Perley, R\"oser  \& Meisenheimer}{Perley
  et~al.}{1997}]{perley97}
Perley R.~A.,  R\"oser H.~J.,   Meisenheimer K.,  1997, A\&A, 328, 12

\bibitem[\protect\citeauthoryear{Rau \& Cornwell}{Rau \&
  Cornwell}{2011}]{rau11}
Rau U.,  Cornwell T.~J.,  2011, A\&A, 532, 71

\bibitem[\protect\citeauthoryear{{Richards} et~al.,}{{Richards}
  et~al.}{2009}]{richards2009}
{Richards} G.~T.,  et~al., 2009, \mn@doi [\apjs] {10.1088/0067-0049/180/1/67},
  \href {http://adsabs.harvard.edu/abs/2009ApJS..180...67R} {180, 67}

\bibitem[\protect\citeauthoryear{{Sakelliou} \& {Merrifield}}{{Sakelliou} \&
  {Merrifield}}{2000}]{sakelliou00}
{Sakelliou} I.,  {Merrifield} M.~R.,  2000, \mn@doi [\mnras]
  {10.1046/j.1365-8711.2000.03079.x}, \href
  {https://ui.adsabs.harvard.edu/abs/2000MNRAS.311..649S} {311, 649}

\bibitem[\protect\citeauthoryear{Scheuer}{Scheuer}{1974}]{scheuer74}
Scheuer P. A.~G.,  1974, MNRAS, 166, 513

\bibitem[\protect\citeauthoryear{{Scheuer}}{{Scheuer}}{1996}]{sheuer96}
{Scheuer} P.~A.~G.,  1996, in {Hardee} P.~E.,  {Bridle} A.~H.,   {Zensus}
  J.~A.,  eds, ~ Vol. 100, Energy Transport in Radio Galaxies and Quasars.
  p.~333

\bibitem[\protect\citeauthoryear{{Shimwell} et~al.,}{{Shimwell}
  et~al.}{2019}]{shimwell19}
{Shimwell} T.~W.,  et~al., 2019, \mn@doi [\aap] {10.1051/0004-6361/201833559},
  \href {https://ui.adsabs.harvard.edu/abs/2019A&A...622A...1S} {622, A1}

\bibitem[\protect\citeauthoryear{{Spergel} et~al.,}{{Spergel}
  et~al.}{2003}]{spergel03}
{Spergel} D.~N.,  et~al., 2003, \mn@doi [The Astrophysical Journal Supplement
  Series] {10.1086/377226}, \href
  {https://ui.adsabs.harvard.edu/#abs/2003ApJS..148..175S} {148, 175}

\bibitem[\protect\citeauthoryear{Tribble}{Tribble}{1993}]{tribble93}
Tribble P.,  1993, MNRAS, 261, 57

\bibitem[\protect\citeauthoryear{{Wall}}{{Wall}}{1980}]{wall80}
{Wall} J.~V.,  1980, \mn@doi [Philosophical Transactions of the Royal Society
  of London Series A] {10.1098/rsta.1980.0182}, \href
  {https://ui.adsabs.harvard.edu/#abs/1980RSPTA.296..367W} {296, 367}

\bibitem[\protect\citeauthoryear{{White}, {Becker}, {Helfand}  \&
  {Gregg}}{{White} et~al.}{1997}]{white97}
{White} R.~L.,  {Becker} R.~H.,  {Helfand} D.~J.,   {Gregg} M.~D.,  1997,
  \mn@doi [\apj] {10.1086/303564}, \href
  {https://ui.adsabs.harvard.edu/#abs/1997ApJ...475..479W} {475, 479}

\bibitem[\protect\citeauthoryear{{Williams} et~al.,}{{Williams}
  et~al.}{2019}]{williams19}
{Williams} W.~L.,  et~al., 2019, \mn@doi [\aap] {10.1051/0004-6361/201833564},
  \href {https://ui.adsabs.harvard.edu/abs/2019A&A...622A...2W} {622, A2}

\bibitem[\protect\citeauthoryear{{York} et~al.,}{{York}
  et~al.}{2000}]{york2000}
{York} D.~G.,  et~al., 2000, \mn@doi [\aj] {10.1086/301513}, \href
  {http://adsabs.harvard.edu/abs/2000AJ....120.1579Y} {120, 1579}

\bibitem[\protect\citeauthoryear{{Young}, {Rudnick}, {Katz}, {DeLaney},
  {Kassim}  \& {Makishima}}{{Young} et~al.}{2005}]{young05}
{Young} A.,  {Rudnick} L.,  {Katz} D.,  {DeLaney} T.,  {Kassim} N.~E.,
  {Makishima} K.,  2005, \mn@doi [ApJ] {10.1086/430107}, \href
  {http://adsabs.harvard.edu/abs/2005ApJ...626..748Y} {626, 748}

\bibitem[\protect\citeauthoryear{{de Gasperin}}{{de
  Gasperin}}{2017}]{gasperin17}
{de Gasperin} F.,  2017, \mn@doi [\mnras] {10.1093/mnras/stx210}, \href
  {https://ui.adsabs.harvard.edu/abs/2017MNRAS.467.2234D} {467, 2234}

\bibitem[\protect\citeauthoryear{{de Vries} et~al.,}{{de Vries}
  et~al.}{2018}]{vries18}
{de Vries} M.~N.,  et~al., 2018, \mn@doi [\mnras] {10.1093/mnras/sty1232},
  \href {https://ui.adsabs.harvard.edu/\#abs/2018MNRAS.478.4010D} {478, 4010}

\bibitem[\protect\citeauthoryear{{van Haarlem} et~al.,}{{van Haarlem}
  et~al.}{2013}]{haarlem13}
{van Haarlem} M.~P.,  et~al., 2013, A\&A, 556, A2

\makeatother
\end{thebibliography}

%\clearpage
%\newpage
\appendix

\section{Injection index fitting}
\label{append_inj}

\begin{figure*}
\centering
\includegraphics[angle=0,totalheight=5.9cm]{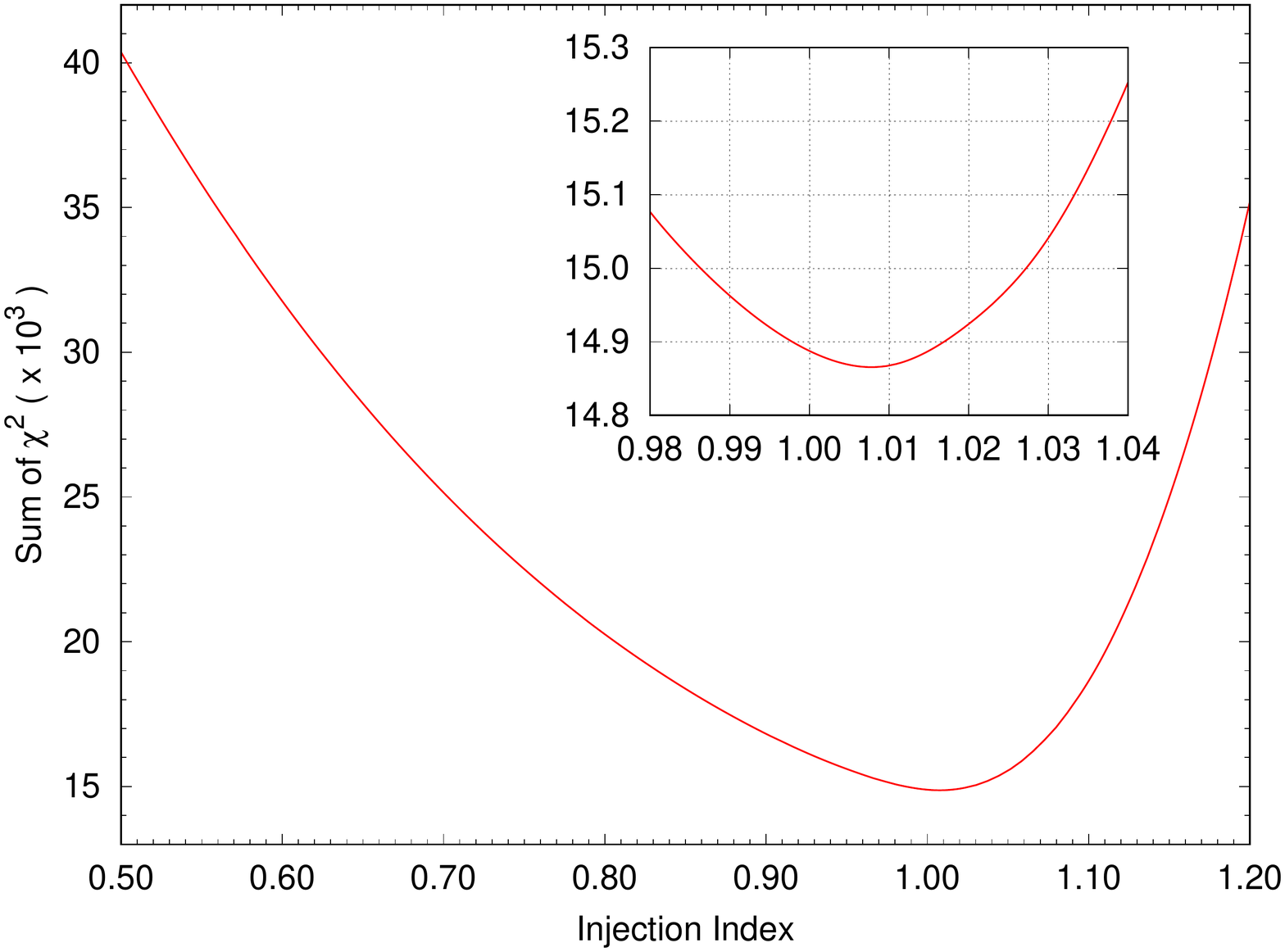}
\includegraphics[angle=0,totalheight=5.9cm]{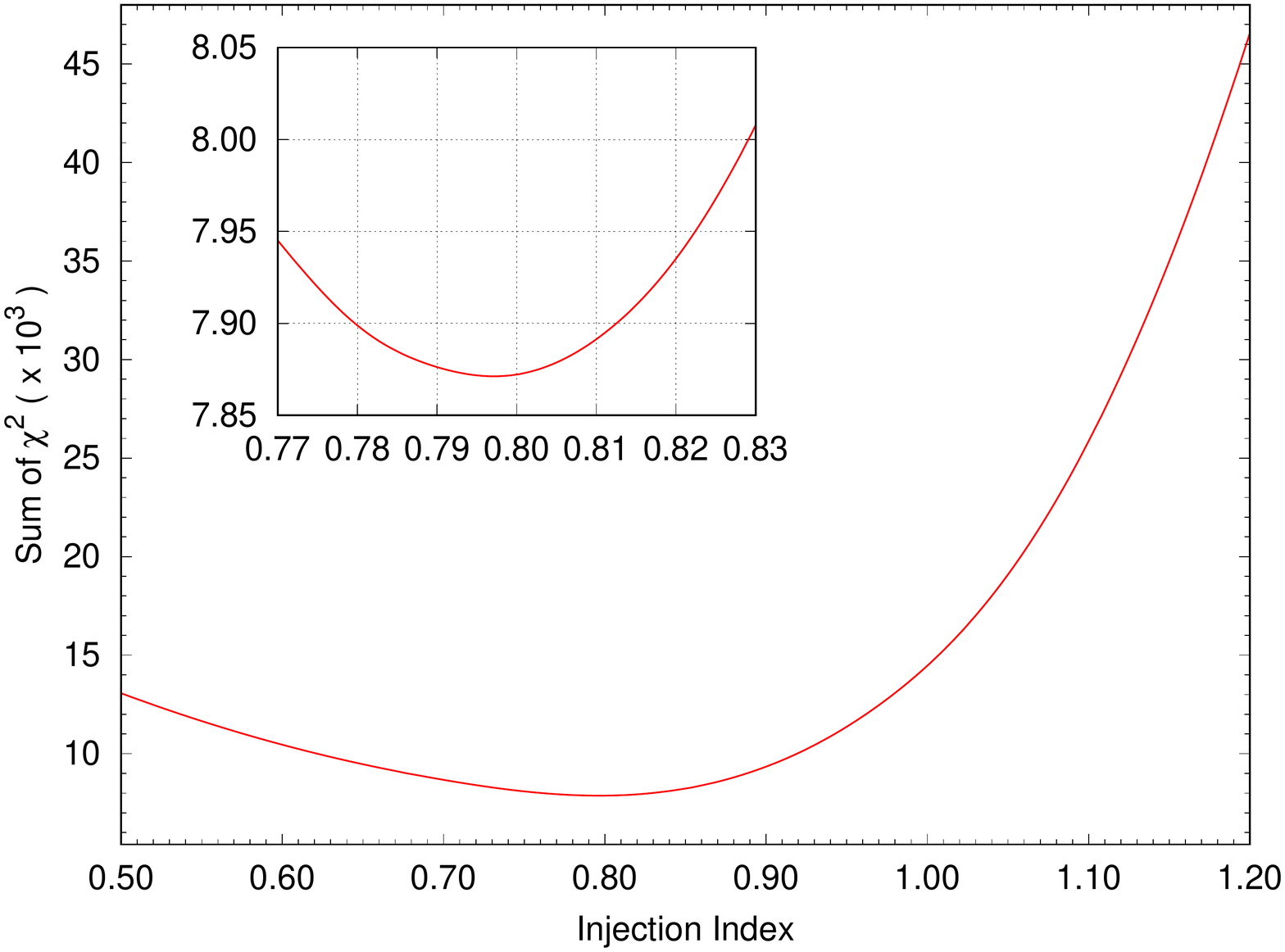}\\
\includegraphics[angle=0,totalheight=5.9cm]{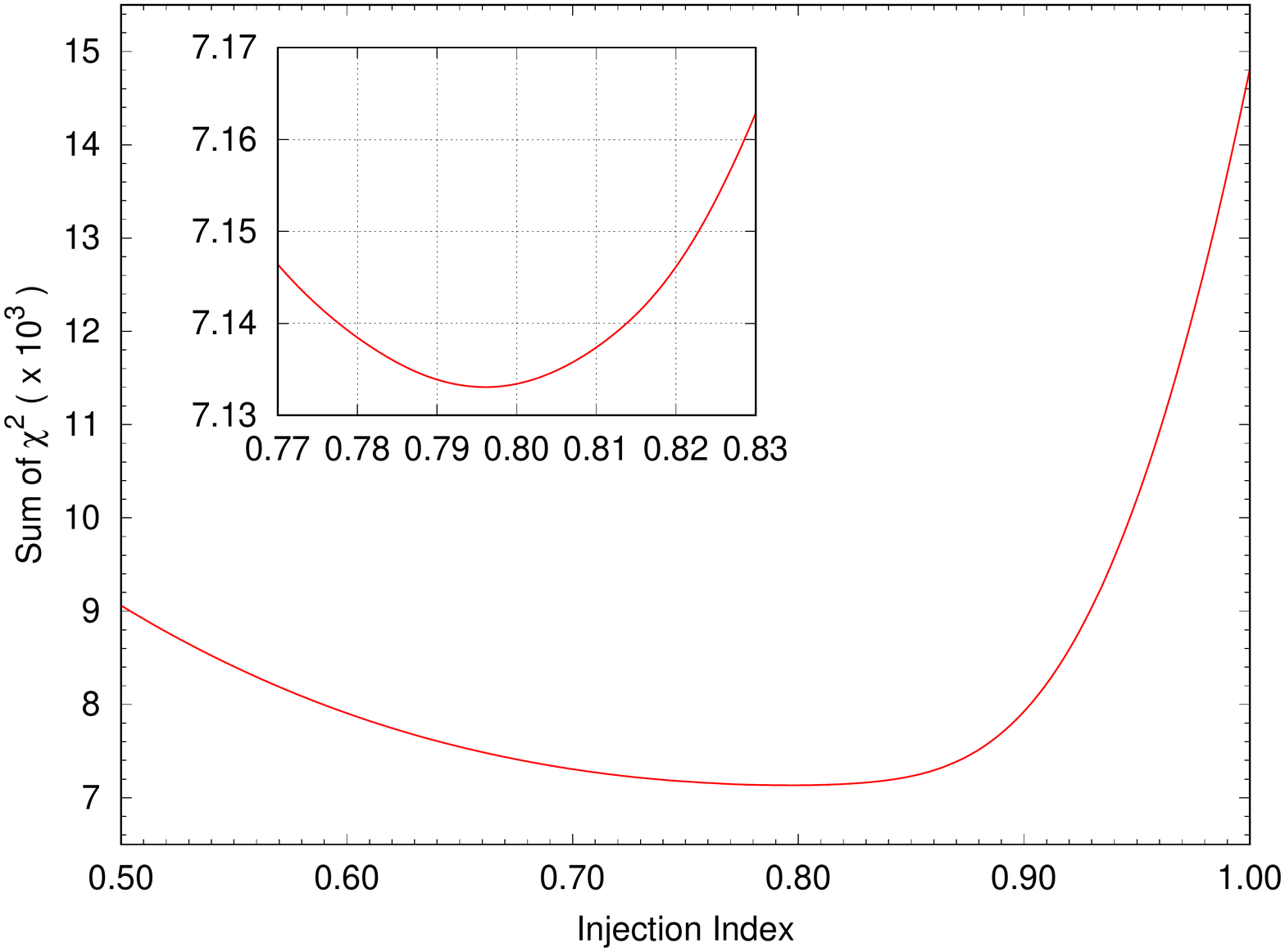}
\includegraphics[angle=0,totalheight=5.9cm]{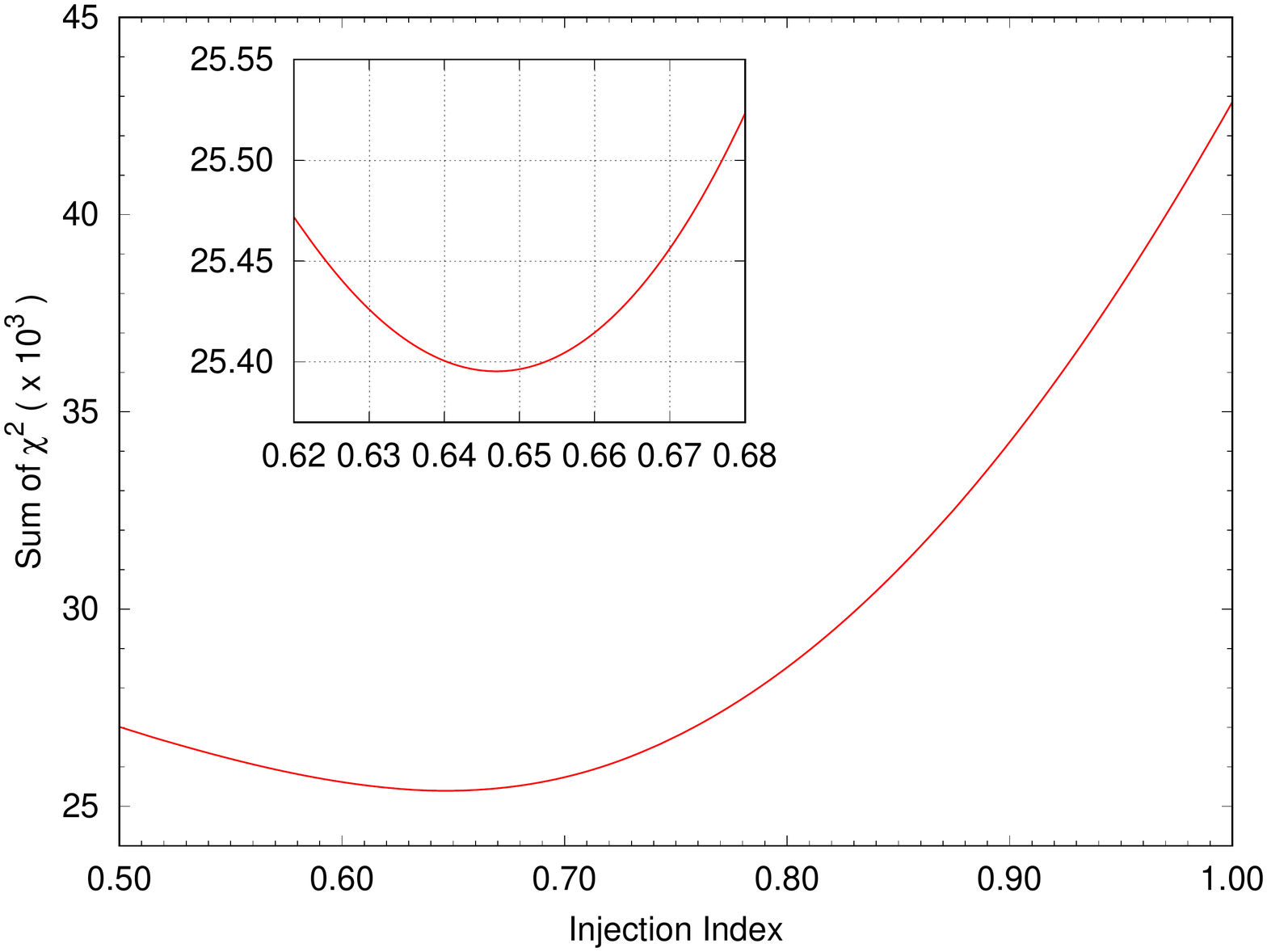}\\
\includegraphics[angle=0,totalheight=5.9cm]{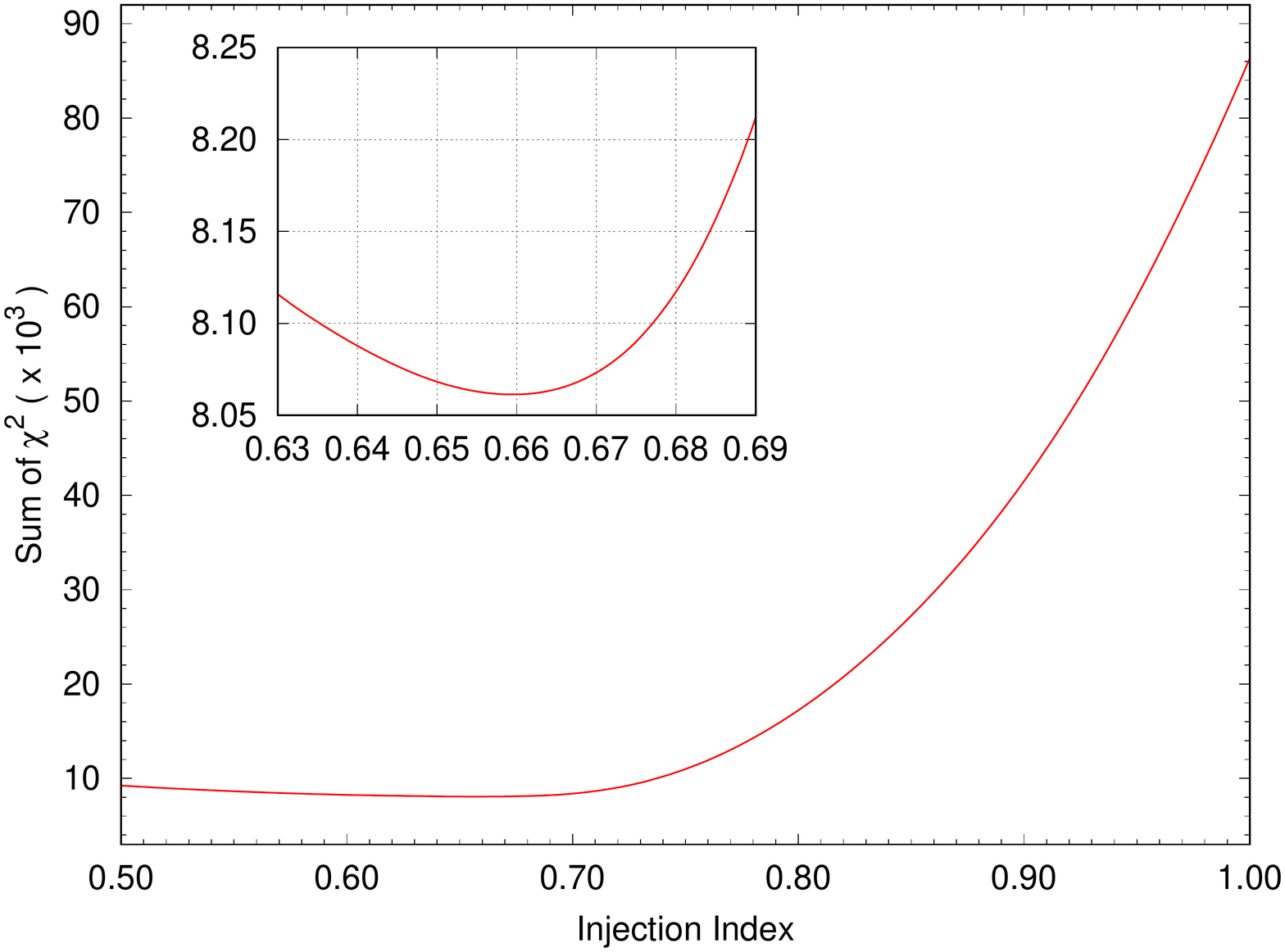}
\includegraphics[angle=0,totalheight=5.9cm]{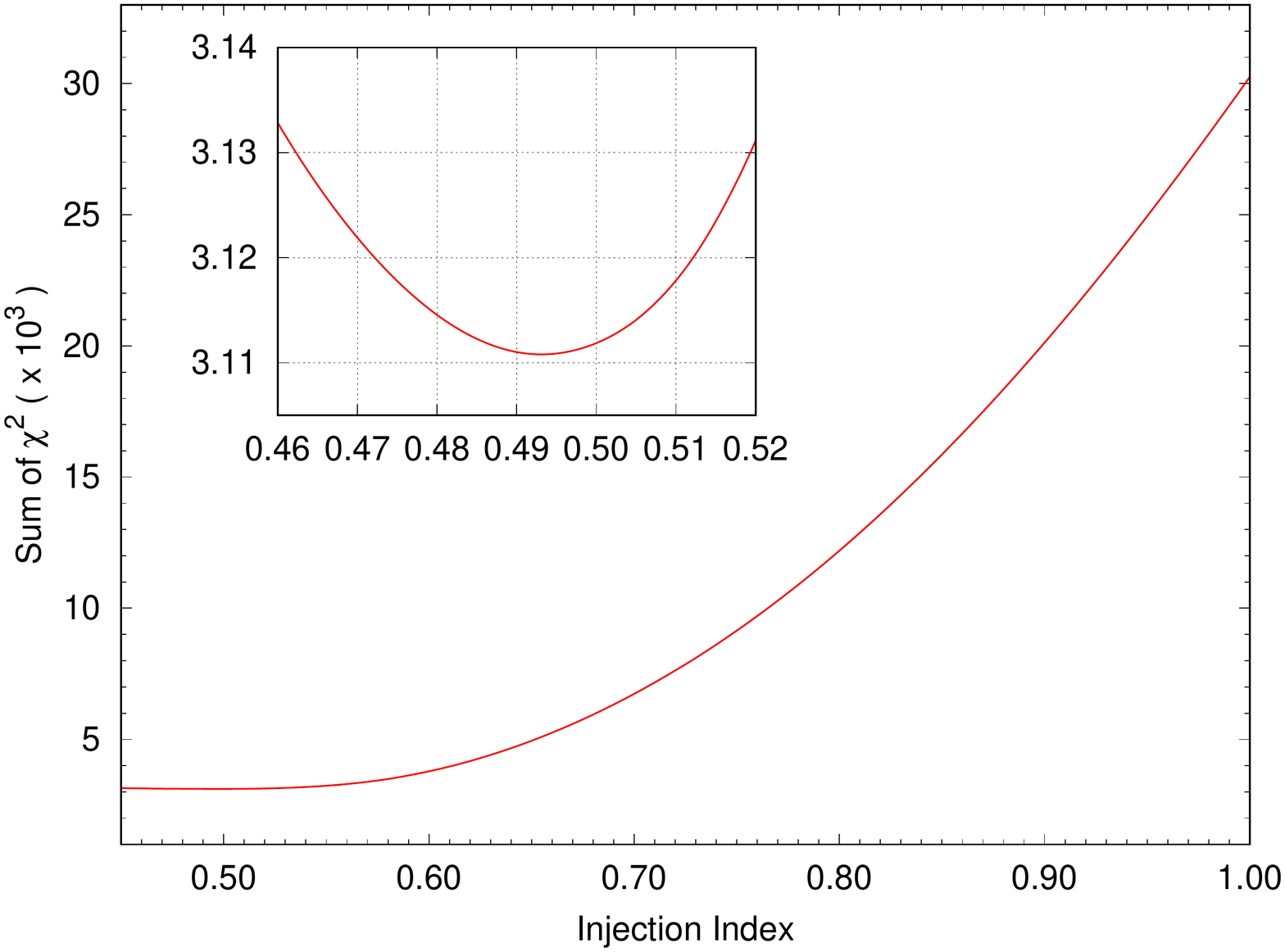}
\end{figure*}

\begin{figure*}
\centering
\includegraphics[angle=0,totalheight=5.9cm]{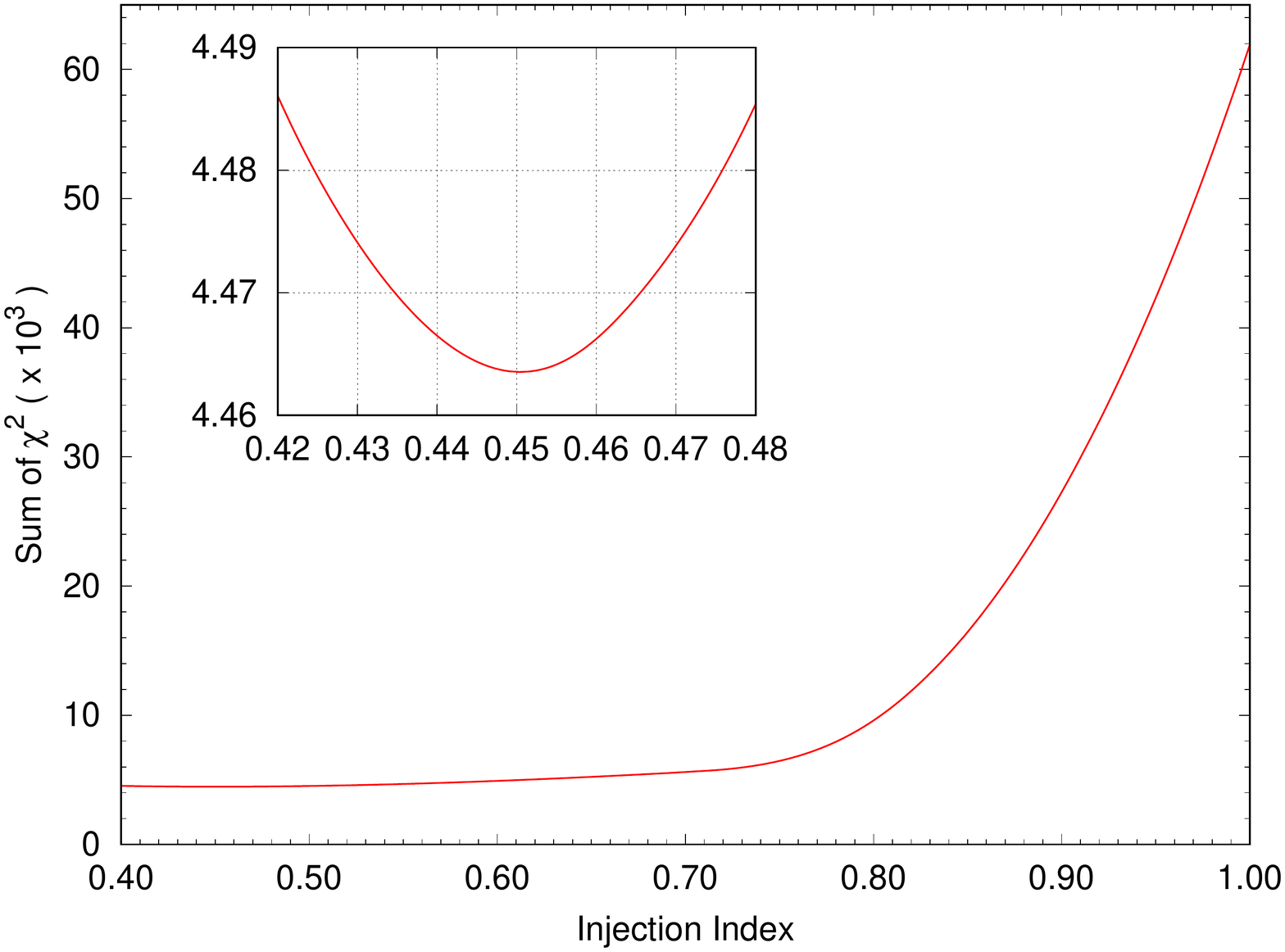}
\includegraphics[angle=0,totalheight=5.9cm]{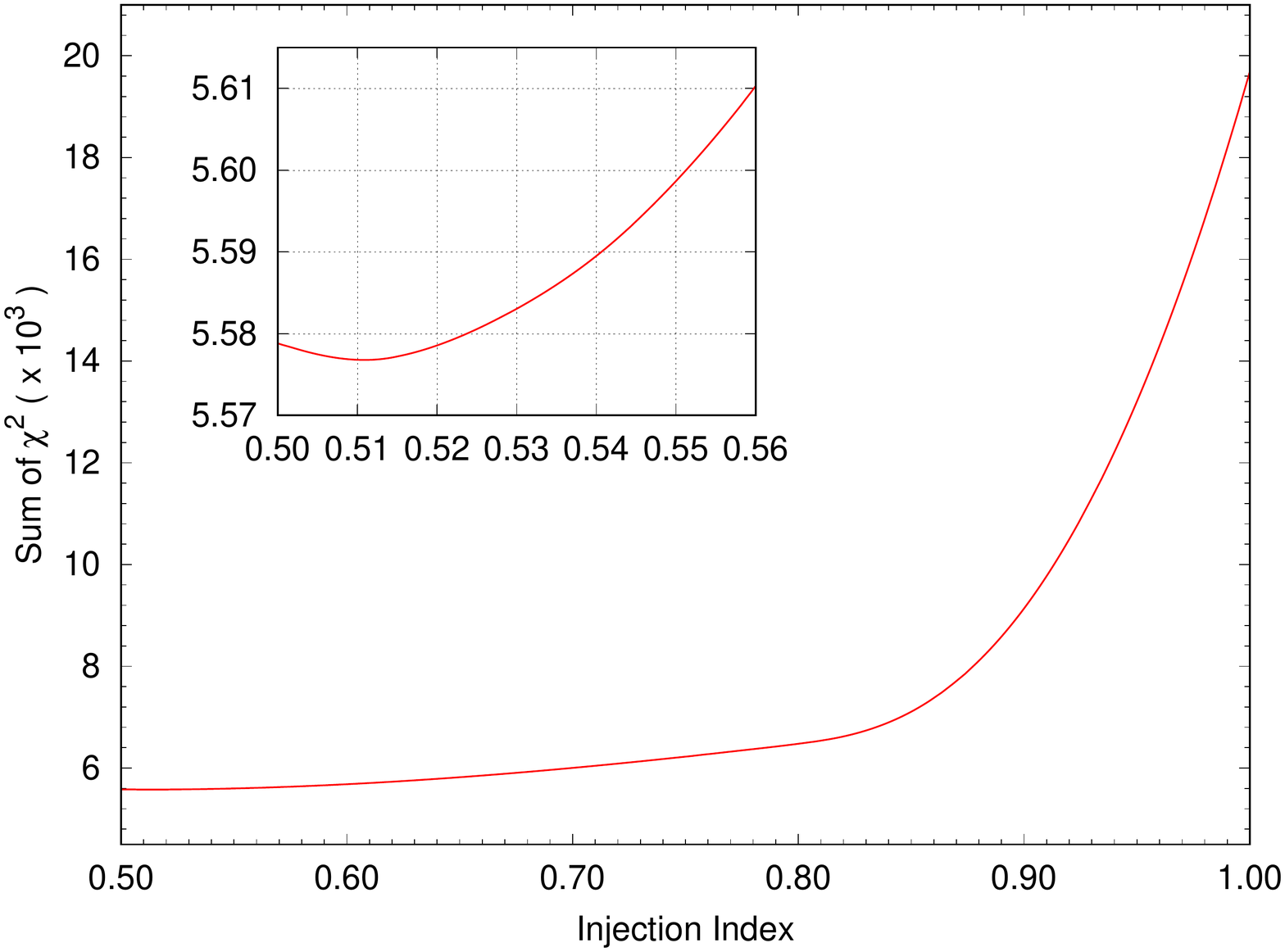}\\
\includegraphics[angle=0,totalheight=5.9cm]{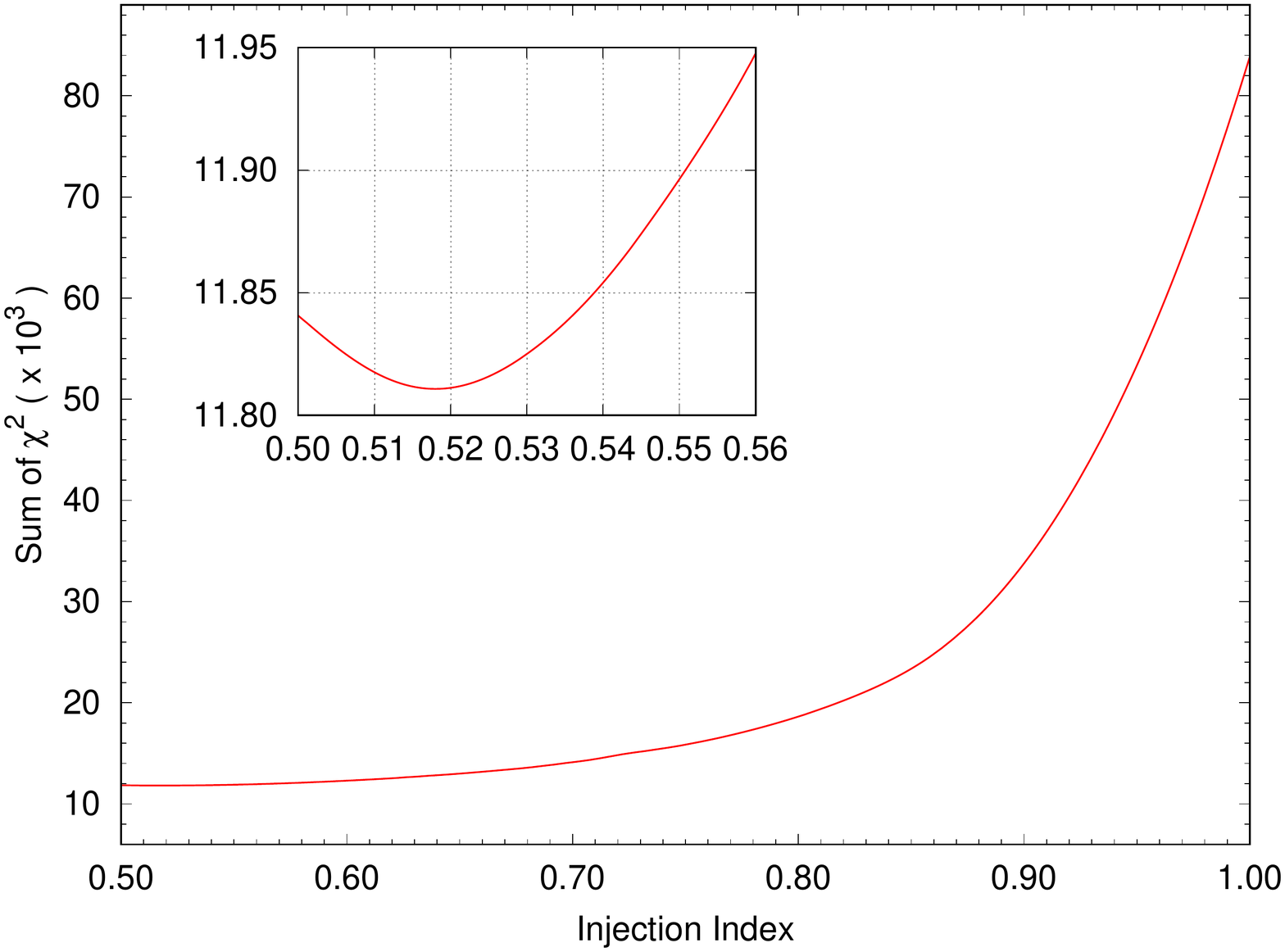}
\includegraphics[angle=0,totalheight=5.9cm]{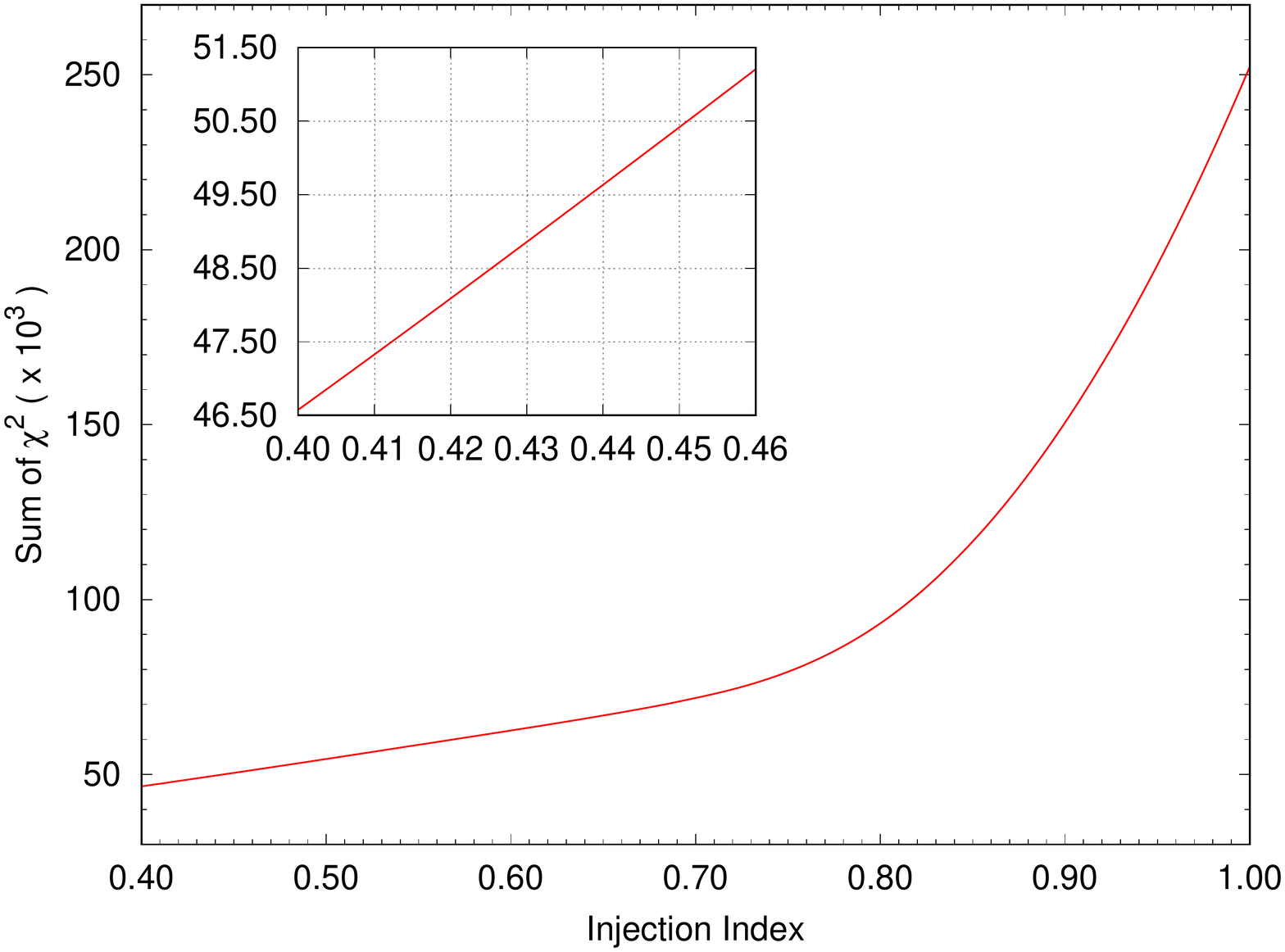}
\caption{$\chi^{2}$ values as a function of injection index. Plots zoomed to the minimum values are shown inset for each panel. Fitting was performed with an initial step of $0.05$ between $0.5$ and $1.0$ and a secondary step size of $0.01$ around the minimum values, extended to $0.4$ where appropriate. The data are fitted with natural cubic splines. As all points lie on the fitted spline they are excluded for clarity.}
\label{injectplots}
\end{figure*}

% Don't change these lines
\bsp	% typesetting comment
\label{lastpage}
\end{document}